\documentclass[aps,prl,twocolumn,groupedaddress,showpacs,notitlepage,nofootinbib]{revtex4-2}
\usepackage{hyperref} 
\hypersetup{colorlinks,linkcolor={blue},citecolor={blue},urlcolor={blue}}
\usepackage{xcolor}
\usepackage{graphicx}
\usepackage{subfigure}
\usepackage{amsmath}
\usepackage{braket}
\usepackage{soul} 
\usepackage[T1]{fontenc}
\usepackage[utf8]{inputenc}
\usepackage{float}
\usepackage{bm}
\usepackage[normalem]{ulem}
\usepackage{esint} 
\usepackage{wasysym} 
\usepackage{amssymb} 
\usepackage{MnSymbol}
\usepackage{enumitem}
\usepackage{CJKutf8} 
\usepackage{tabularx}  
\usepackage{makecell}  

\makeatletter
\def\maketitle{
\@author@finish
\title@column\titleblock@produce
\suppressfloats[t]}
\makeatother
\newcommand{\bk}{\boldsymbol{k}}
\newcommand{\br}{\boldsymbol{r}}

\newcommand{\bd}{\boldsymbol{d}}

\newcommand{\bs}{\boldsymbol}
\newcommand{\Cal}{\mathcal}

\newcommand{\CalG}{\mathcal{G}}

\newcommand\tinyhex{\vcenter{\hbox{\scalebox{0.8}{$\varhexagon$}}}} 
\newcommand\tinykag{\vcenter{\hbox{\scalebox{0.9}{$\davidsstar$}}}} 

\newcommand\tinyCW{\vcenter{\hbox{\scalebox{0.7}{$\circlearrowright$}}}} 

\begin{document}



\title{Fractional quantization in gapless flat bands with ill-defined topology and divergent quantum geometry}


\author{Wenqi Yang}
\thanks{Wenqi Yang and Dawei Zhai contributed equally to this work.}
\author{Dawei Zhai}
\email{dzhai@hku.hk}
\author{Wang Yao}
\email{wangyao@hku.hk}
\affiliation{New Cornerstone Science Laboratory, Department of Physics, The University of Hong Kong, Hong Kong, China}

\affiliation{HK Institute of Quantum Science \& Technology, The University of Hong Kong, Hong Kong, China}

\affiliation{State Key Laboratory of Optical Quantum Materials, The University of Hong Kong, Hong Kong, China}





\begin{abstract}
    Fractional quantum anomalous Hall (FQAH) effect, a zero-magnetic-field lattice analog of the fractional quantum Hall effect, offers a unique pathway toward fault-tolerant quantum computation and deep insights into the interplay of topology and strong correlations. The exploration has been successfully guided by the paradigm of ideal flat Chern bands, which mimic Landau levels in both band topology and local quantum geometry. Yet, given the boundless potential for Bloch bands in lattice systems, it remains a significant open question whether FQAH states can arise in scenarios fundamentally distinct from this paradigm. Here we turn to a class of gapless flat bands, featuring (i) ill-defined band topology, (ii) non-quantized Berry flux, (iii) divergent quantum geometry at singular band touchings, (iv) highly fluctuating and far-from-ideal quantum geometry across the Brillouin zone. Our exact diagonalization and density matrix renormalization group calculations unambiguously demonstrate an FQAH phase that is virtually independent of the interaction strength, persisting from the weak-interaction to the strong-interaction limit. We find the stability of the FQAH states does not uniquely correlate with the singularity strength. Instead, the many-body topological order can adapt to the singular and fluctuating quantum geometric landscape by spontaneously developing an inhomogeneous carrier distribution, while its quenching is accompanied by a drop in the occupation-weighted Berry flux. Our work reveals a profound interplay between local quantum geometry and many-body correlation, and significantly expands the exploration space for the FQAH effect and correlated phenomena in general.
\end{abstract}

\maketitle


\textit{\textcolor{blue}{Introduction---}}The fractional quantum anomalous Hall (FQAH) effect is the zero-magnetic-field lattice analog of the fractional quantum Hall effect, which has been experimentally discovered in twisted bilayer MoTe$_2$~\cite{FCIMoTe2Jiaqi2023,FCIMoTe2ShanJie2023,FCIMoTe2Park2023,FCIMoTe2PRX2023} and rhombohedral graphene/hBN heterostructures~\cite{lu2024fractional,JuLongFCI2025,LuXiaoboGrapheneFCI2025,GrapheneFCIPRX2025}.
Theoretical exploration of fractional Chern insulators on lattice models exhibiting the FQAH effect was initiated as early as a decade ago, with numerical predictions in various models of isolated flat Chern bands as the lattice correspondence of Landau levels (LLs)~\cite{tang2011high,neupert2011fractional,sheng2011fractional,regnault2011fractional,wang2011fractional,xiao2011interface,sun2011nearly}. Not all flat Chern bands can sustain FQAH states, and in screening for the suitable ones, mimicking the LL quantum geometry characterized by Berry curvature $\Omega(\bk)$ and Fubini-Study metric $\CalG(\bk)$
has served as a guideline~\cite{RoyPRB2012,RoyPRB2014,RoyNatCommun2015,FCIPositionMomentumDualityPRL2015,IdealFCIPRB2017,AshvinFCIPRR2020,WangjieIdealBandPRL2021,KahlerBandsPRB2021a,KahlerBandsPRB2021b,KahlerBandsPRB2021c,ChiralMultiGrapheneLedwith2022,ChiralMultiGrapheneWangJie2022,ChiralMultiGrapheneDongJunkai2023,WangjieIdealBandPRR2023,VortexablePRB2023,IdealChernCurveSpaceLL2023,IdealFlatStrainSunKaiPRL2023,VortexablePRL2025,WangjieHigherIdealBandPRX2025}. Uniform $\Omega(\bk)$ plus the trace condition $\text{tr}\,\CalG(\bk)=|\Omega(\bk)|$ will lead to the same Girvin-MacDonald-Platzman algebra of projected density operators in the lowest LL~\cite{GMPalgebra1986,RoyPRB2012,RoyPRB2014,RoyNatCommun2015}. More examples examined later suggest that uniformity of quantum geometry may not be essential, while fulfillment of the trace condition represents some idealness in stabilizing FQAH states~\cite{FCIPositionMomentumDualityPRL2015,AshvinFCIPRR2020,WangjieIdealBandPRL2021,KahlerBandsPRB2021a,KahlerBandsPRB2021b,KahlerBandsPRB2021c,ChiralMultiGrapheneLedwith2022,ChiralMultiGrapheneWangJie2022,ChiralMultiGrapheneDongJunkai2023,WangjieIdealBandPRR2023,VortexablePRB2023,IdealChernCurveSpaceLL2023,VortexablePRL2025,WangjieHigherIdealBandPRX2025}.
Various flat Chern bands with inhomogeneous but ideal quantum geometry---the so-called ideal Chern flat bands~\cite{WangjieIdealBandPRL2021,WangjieIdealBandPRR2023,AshvinFCIPRR2020,WangjieHigherIdealBandPRX2025,ChiralMultiGrapheneWangJie2022,ChiralMultiGrapheneLedwith2022,ChiralMultiGrapheneDongJunkai2023}, corresponding to the intersection of K\"ahler bands~\cite{KahlerBandsPRB2021a,KahlerBandsPRB2021b,KahlerBandsPRB2021c} and vortexable bands~\cite{VortexablePRB2023,VortexablePRL2025}---have been proposed, which offer favorable conditions for FQAH ground states under short-range interactions.

Recent studies have shown that FQAH effects extend well beyond the mimicry of LLs, from well-isolated Chern bands of far-from-ideal quantum geometry~\cite{FCIFarFromIdealPRL2024,FCIgradient2025}, to gapless critical Chern bands with band-touching points~\cite{WenqiPRL2025}, and even beyond the Chern band paradigm to isolated trivial flat bands~\cite{FCIC=0Lin2025,FCIC=0Lu2025,FCIC=0Bergholtz2025}. FQAH states in these non-Chern band contexts exhibit an inhomogeneous carrier distribution, where the preferentially occupied Brillouin zone (BZ) regions feature rather uniform quantum geometry satisfying the trace condition~\cite{FCIC=0Lin2025,FCIC=0Lu2025,FCIC=0Bergholtz2025}. These findings point to a broader principle for the emergence of FQAH states: it is not necessary to simultaneously have band topology and ideal quantum geometry. This, however, raises the next question: can FQAH states emerge when neither condition is present? We turn to the singular flat bands (SFBs)~\cite{SinguFlatClassificationPRB2019,SinguFlatAdvPhysX2021}, a context where both conditions can be absent. These exactly flat bands feature singular touchings to other bands around which Bloch states are discontinuous, leading to divergent quantum geometry around the touching points and a non-quantized Berry flux in the BZ. Their gapless nature further raises questions concerning the usual weak coupling paradigm of fractional quantum Hall physics, which requires the interaction strength $U$ to be well below the band gap $\Delta$ such that strong correlations are restricted to a single projected band. Notably, there exist examples of FQAH states in the strong coupling limit $U \gg \Delta$ at dilute fillings~\cite{InteractionExceedGapPRL2014,InteractionExceedGapPRB2015,FCIC=0Lu2025,WenqiPRL2025}.

\begin{figure*}[t]
	\centering
	\includegraphics[width=5in]{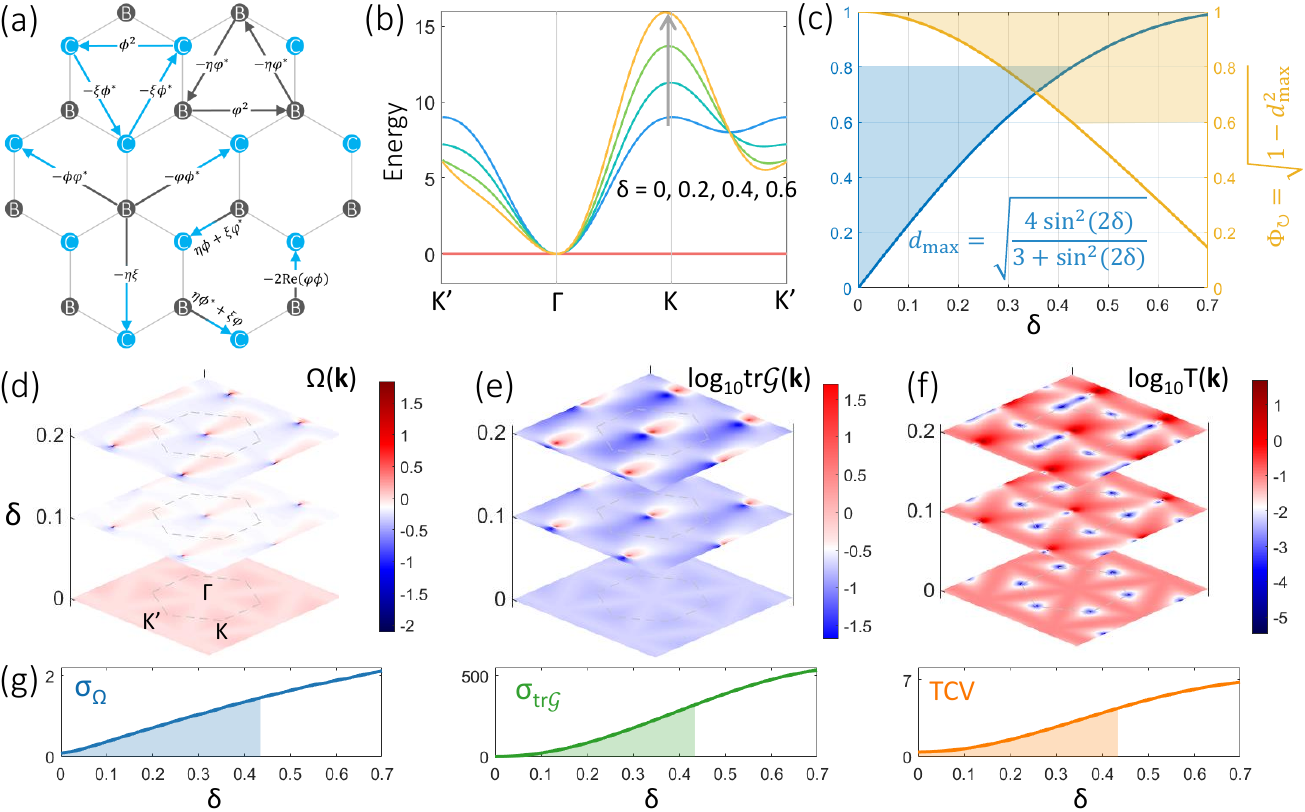}
	\caption{\textbf{Single-particle results of the honeycomb model.} (a) Schematics of the honeycomb model $\hat{H}_{\tinyhex}(\bk)$, where $\xi=2\cos\theta_+$, $\eta=2\cos\theta_-$, $\phi=e^{i\theta_+}$ and $\varphi=e^{i\theta_-}$. (b) Band structure of $\hat{H}_{\tinyhex}(\bk)$ for various $\delta$. (c) Evolution of $d_{\rm max}$ and Berry phase $\Phi_{\tinyCW}$ around the touching point with $\delta$. (d--f) Distribution of $\Omega(\bk)$, log$_{10}\text{tr}\,\CalG(\bk)$ and log$_{10}T(\bk)$ of the SFB in the $\bk$ space for a few $\delta$. (g) Quantum geometry fluctuation as $\delta$ is varied. The shaded areas covering $0\le\delta\lesssim0.43$ on the curves in panels (c) and (g) host FQAH effects.}
	\label{Fig:DiceSingleParticle}
\end{figure*}

Here we consider two distinct SFB models of honeycomb and kagome geometry respectively, featuring tunable quantum geometry divergence characterized by the maximum quantum distance $d_{\rm max}$ around the singular touching point and non-quantized Berry flux $\Phi_{\tinyCW}=\sqrt{1-d^2_{\rm max}}$ in the BZ. Over a broad range of singularity strength $d_{\rm max}$ and with an arbitrary-strength nearest-neighbor (NN) repulsion, an FQAH phase is demonstrated using exact diagonalization (ED) and density matrix renormalization group (DMRG) calculations. The two models display distinct many-body phase diagrams as $d_{\rm max}$ is varied, but a number of common features are observed in their FQAH phases. Away from the band touching, we show that $\text{tr}\,\CalG(\bk)$ of the SFB tends to repel interacting carriers away from its maxima, reminiscent of the observations in isolated flat bands~\cite{FCIQuantumMetricLiuZhaoPRL2020,FCIQuantumMetricPRR2023,FCIQuantumMetricYangBo2024,FCIC=0Lin2025,FCIC=0Lu2025,FCIC=0Bergholtz2025}. Around the singular touching point, we observe significant carrier occupation accompanied by finite band mixing. Notably, while the SFB's $\text{tr}\,\CalG(\bk)$ is divergent in the vicinity of the singular touching point, the two-band $\text{tr}\,\CalG(\bk)$ of the touching bands is vanishing. This points to the possibility that, near the touching point, two-band quantum geometry replaces one-band quantum geometry in determining the inhomogeneous occupation of interacting carriers in the BZ. The many-body gap of the FQAH phase does not uniquely correlate with the singularity strength, while its quenching is accompanied by a decrease in the occupation-weighted Berry flux. The persistence of the FQAH phase from the weak-interaction to strong-interaction limit in these contexts points to an intriguing adaptability of the many-body topological order to the singular and fluctuating quantum geometric landscape and ill-defined band topology.


\textit{\textcolor{blue}{Honeycomb model and its quantum geometry---}}The honeycomb lattice model, which is derived from a fluxed dice lattice model [see Supporting Information (SI)]~\cite{WenqiPRL2025}, is schematically shown in Fig.~\ref{Fig:DiceSingleParticle}(a).
In the orbital basis, the Hamiltonian reads $\hat{H}_{\tinyhex}(\bk)=
t\begin{pmatrix} 
|f|^2&gf^{*}\\
fg^{*}&|g|^2 
\end{pmatrix}$,
where $f=-2\cos\theta_-e_1 + e^{i\theta_-} e_2 + e^{-i\theta_-}  e_3$ and $g=2\cos\theta_+e_1^{*} - e^{i\theta_+} e_2^{*} - e^{-i\theta_+}  e_3^{*}$. Here $\theta_\pm=-\pi/3\pm\delta$, $e_i=e^{-i\bk\cdot\bd_i}$ and $\bd_{1,2,3}$ are the NN vectors. $\delta$ is the parameter used to tune the quantum geometry in this SFB. Importantly, $\delta\ne0$ leads to divergent and strongly fluctuating quantum geometry that intensifies as $\delta$ increases. The lattice constant and hopping amplitude $t$ are set to 1 throughout this work. $\hat{H}_{\tinyhex}$ has a zero-energy flat band with Bloch function $\psi_0=(g,\,-f)^{T}/\sqrt{|f|^2+|g|^2}$, and a dispersive band [Fig.~\ref{Fig:DiceSingleParticle}(b)]. The two bands have a quadratic band touching at the $\Gamma$ point, around which $\psi_0$ is discontinuous when $\delta\ne0$, rendering the flat band a SFB~\cite{SinguFlatClassificationPRB2019,SinguFlatAdvPhysX2021}. The singularity of the band touching can be measured by the maximal Hilbert-Schmidt distance $d_{\rm max}=\text{max}\sqrt{1-|\braket{\psi_0(\bk)|\psi_0(\bk')}|^2}$ defined on a vanishing circle centered at the $\Gamma$ point. Here $d_{\rm max}=\sqrt{\frac{4 \sin^2(2\delta)}{3 + \sin^2(2\delta)}}$, increasing from 0 to 1 monotonically with $\delta$ in the range $0\le\delta\le\pi/4$ [Fig.~\ref{Fig:DiceSingleParticle}(c) blue curve]. It is important to distinguish the case of $\delta=0$ ($d_{\rm max}=0$), which exhibits critical Chern band topology in this model~\cite{SongZhidaGaplessTopology,FangChenGaplessTopology}, from the finite-singularity-strength regime of $\delta\ne0$ ($d_{\rm max}\ne0$), where the band topology is ill-defined. In the following, the SFB refers to the flat band at $\delta\ne0$.

Compared to a Chern band, this SFB has fundamentally different topological and quantum geometric properties. The quantum metric tensor $\CalG(\bk)$ diverges near the touching point when $d_{\rm max}\ne0$. 
The Berry phase (in units of 2$\pi$) accumulated on an infinitesimal circle in the clockwise direction around the touching point is $\Phi_{\tinyCW}=\sqrt{1-d^2_{\rm max}}$~\cite{SinguFlatWaveFunctPRB2021}, which also equals the Berry flux over the BZ excluding the $\Gamma$ point $\frac{1}{2\pi}\int_{\text{BZ}\backslash\{\Gamma\}}\Omega(\bk)d\bk$. $\Phi_{\tinyCW}$ is non-integer when $d_{\rm max}\ne0$; it decreases continuously from 1 as $\delta$ ($d_{\rm max}$) is enlarged [Fig.~\ref{Fig:DiceSingleParticle}(c) yellow curve]. Thus, in addition to signifying a divergent quantum metric, $d_{\rm max}$ also quantifies the deviation of the SFB from a Chern band in terms of the Berry flux. The distributions of $\Omega(\bk)$ and the trace of the quantum metric $\text{tr}\,\CalG(\bk)$ for $\bk\in\text{BZ}\backslash\{\Gamma\}$ are presented in Fig.~\ref{Fig:DiceSingleParticle}(d) and Fig.~\ref{Fig:DiceSingleParticle}(e)\footnote{Since $\text{tr}\,\CalG(\bk)$ diverges at the $\Gamma$ point, the presented upper limits of quantities involving $\text{tr}\,\CalG(\bk)$ depend on the size of the chosen $\bk$ mesh, but the qualitative behaviors with respect to $\delta$ (or $\alpha$ in the kagome model) remain invariant. We choose a $500\times500$ $\bk$ mesh per BZ in this work.}. When $\delta=0$, both quantities are rather uniform, free of singularity ($d_{\rm max}=0$), and well respect the trace condition, rendering the flat band at $\delta=0$ a nearly ideal Chern band similar to the lowest LL while being gapless\footnote{Details of the $\delta=0$ scenario with lowest LL-like topology and idealness are given in the SI. Such a gapless flat band with $\Phi_{\tinyCW}=1$ (corresponding to a Chern number of 1) was first explored in Ref.~\cite{WenqiPRL2025}. Recent studies show that it saturates the fundamental constraints on topological flat bands in finite-range hopping lattices (hence termed a ``critical topological flat band"~\cite{SongZhidaGaplessTopology}) and that it satisfies two independent topological conditions (hence also termed a ``topological$^2$ flat band"~\cite{FangChenGaplessTopology}).}---a context in which FQAH effects were numerically observed~\cite{WenqiPRL2025}.
However, as $\delta$ ($d_{\rm max}$) increases, $\Omega(\bk)$ becomes sharply concentrated near the $\Gamma$ point with alternating signs, accompanied by strong divergence and fluctuation of $\text{tr}\,\CalG(\bk)$. $T(\bk)=\text{tr}\,\CalG(\bk)-|\Omega(\bk)|$ is also divergent near the $\Gamma$ point when $\delta\ne0$ ($d_{\rm max}\ne0$) and remains large in various regions of the BZ, especially for large $\delta$ [Fig.~\ref{Fig:DiceSingleParticle}(f)]. To quantify the fluctuation, we evaluate the standard deviations $\sigma_{\Omega}$ and $\sigma_{\rm tr\,\CalG}$\footnote{$\sigma_{F}=\frac{1}{2\pi}\sqrt{A_{\rm BZ}\int_{\rm BZ\backslash\{\Gamma\}} \left[F(\bk)-\braket{F}\right]^2 d\bk}$, where $F(\bk)$ denotes $\Omega(\bk)$ or $\text{tr}\,\CalG(\bk)$, $\braket{F}$ is its average in the BZ, and $A_{\rm BZ}$ is the area of the BZ.}, and the trace condition violation $\text{TCV}=\frac{1}{2\pi}\int_{\rm BZ\backslash\{\Gamma\}}T(\bk)d\bk$. All three quantities increase rapidly with $\delta$ [Fig.~\ref{Fig:DiceSingleParticle}(g)]. Notably, $\text{TCV}>2$ when $\delta\gtrsim0.25$, implying a violation of idealness in the SFB stronger than that in the first (excited) LL~\cite{KahlerBandsPRB2021c,FCIgradient2025}. The exploration of FQAH effects in this work focuses on the SFBs at $\delta\ne0$, which have ill-defined band topology and divergent/fluctuating quantum geometry that strongly violates the trace condition---placing them well outside the ideal Chern band paradigm in both respects.

\begin{figure}[t]
	\centering
	\includegraphics[width=3.4in]{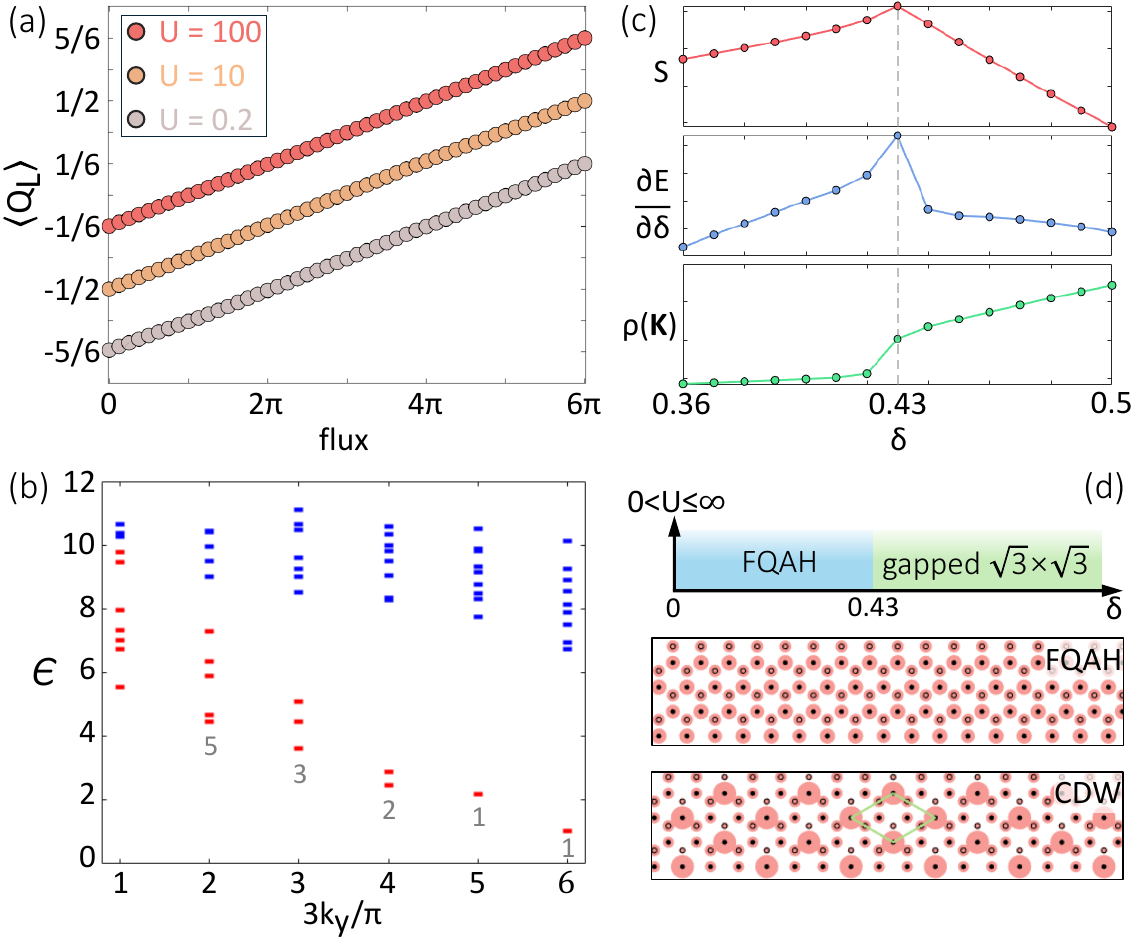}
	\caption{\textbf{DMRG results of the honeycomb model.} (a) Charge pumping under flux insertion in the FQAH phase with $\delta=0.2$ and various $U$ from the weak-interaction to the strong-interaction limit. (b) Momentum-resolved entanglement spectrum $\epsilon$ of the charge sector $Q_L=0$ at $U= 10$ and $\delta=0.2$. (c) Variation of entanglement entropy $S$, the first derivative of the ground state energy $\partial E/\partial \delta$, and the charge distribution at the BZ corner $\rho(\bs{K})$ with $\delta$. (d) Phase diagram in the $\delta$--$U$ parameter space and representative charge patterns in the FQAH and $\sqrt{3}\times\sqrt{3}$ CDW phases.}
	\label{Fig:DiceDMRG}
\end{figure}


\textit{\textcolor{blue}{FQAH and topologically trivial CDW in the honeycomb model---}}We perform DMRG calculations at $\nu=1/3$ filling for spinless fermions in the SFB subject to the NN repulsion $\hat{H}_{\rm int}=U\sum_{\braket{i,j}}\hat{n}_i\hat{n}_j$ (see SI for details). 
Remarkably, an FQAH phase is found for $0\le\delta\lesssim0.43$ ($0\le d_{\rm max}\lesssim0.8$) persisting at arbitrary interaction strengths $0<U\le\infty$ [Fig.~\ref{Fig:DiceDMRG}(a) and Fig.~\ref{Fig:DiceDMRG}(d) top panel]. The FQAH phase features a fractionally quantized Hall conductivity of $\sigma_H=e^2/(3h)$ as revealed by the charge pumping simulation [Fig.~\ref{Fig:DiceDMRG}(a)]. Fig.~\ref{Fig:DiceDMRG}(b) presents the momentum-resolved entanglement spectrum $\epsilon$ of the charge sector $Q_L=0$ at $U=10$ and $\delta=0.2$ within the FQAH regime. It exhibits the edge excitation counting sequence \{1, 1, 2, 3, 5$\cdots$\} of Laughlin states [see red markers in Fig.~\ref{Fig:DiceDMRG}(b)]. In the lowest LL or bands with Chern number $C$, one expects $\sigma_H=\nu C e^2/h$, which clearly breaks down here: The band touching prevents a well-defined Chern number for the SFB, and the Berry flux in $\text{BZ}\backslash\{\Gamma\}$ is non-quantized when $d_{\rm max}\ne0$ ($\delta\ne0$).

\begin{table*}[t]
	\caption{Quantum geometric properties of the honeycomb vs kagome models. ``Profile'' refers to the distribution in the $\bk$ space. ``Trend''---denoted by arrows---refers to the variation of the quantities with respect to the tuning parameter, i.e., $\delta$ ($\alpha$) for the honeycomb (kagome) model.}
	\label{table:system_comparison} 
	\centering
	\renewcommand{\arraystretch}{1.8}  
	\begin{tabularx}{\textwidth}{|>{\centering\arraybackslash}X|>{\centering\arraybackslash}X|>{\centering\arraybackslash}X|}
		\hline
		& \textbf{2-orbital honeycomb model} $\hat{H}_{\tinyhex}$ & \textbf{3-orbital kagome model} $\hat{H}_{\tinykag}$ \\ \hline
		Range of $d_{\text{max}}$ and trend & \makecell{$0 \leq d_{\text{max}} \leq 1$,~ $\nearrow$ \\ $0\le d_{\rm max}\lesssim0.8$ for FQAH} & \makecell{$0.6\lesssim d_{\text{max}} \leq 1$,~ $\searrow$\\ $0.61\lesssim d_{\rm max}\lesssim0.93$ for FQAH} \\ \hline
		Range of $\Phi_{\tinyCW}$ and trend & \makecell{$0\le \Phi_{\tinyCW} \le 1$,~~ $\searrow$ \\ $0.6\lesssim \Phi_{\tinyCW} \leq 1$ for FQAH}  & \makecell{$0\le \Phi_{\tinyCW} \lesssim 0.82$,~ $\nearrow$ \\ $0.36\lesssim \Phi_{\tinyCW} \lesssim 0.79$ for FQAH} \\ \hline
		Profile of $\Omega(\boldsymbol{k})$ & \makecell{concentrated near touching point,\\ exhibit alternating signs} & \makecell{concentrated on two BZ edges,\\ exhibit a single sign} \\ \hline
		Profile of $\text{tr}\,\mathcal{G}(\boldsymbol{k})$ and $T(\boldsymbol{k})$ & 
		\makecell{divergent around touching point,\\ 2D triangular elsewhere} & \makecell{divergent around touching point,\\ 1D stripe elsewhere} \\ \hline
		Trend of $\sigma_{\Omega}$, $\sigma_{\text{tr}\,\CalG}$ and TCV & 
		\makecell{in sync: all $\nearrow$} & 
		\makecell{$\sigma_{\Omega}$ $\nearrow$,~ $\sigma_{\text{tr}\,\CalG}$ $\searrow\nearrow$,~ TCV $\searrow\nearrow$} \\ \hline
	\end{tabularx}
\end{table*}

\begin{figure}[h!]
	\centering
	\includegraphics[width=3.4in]{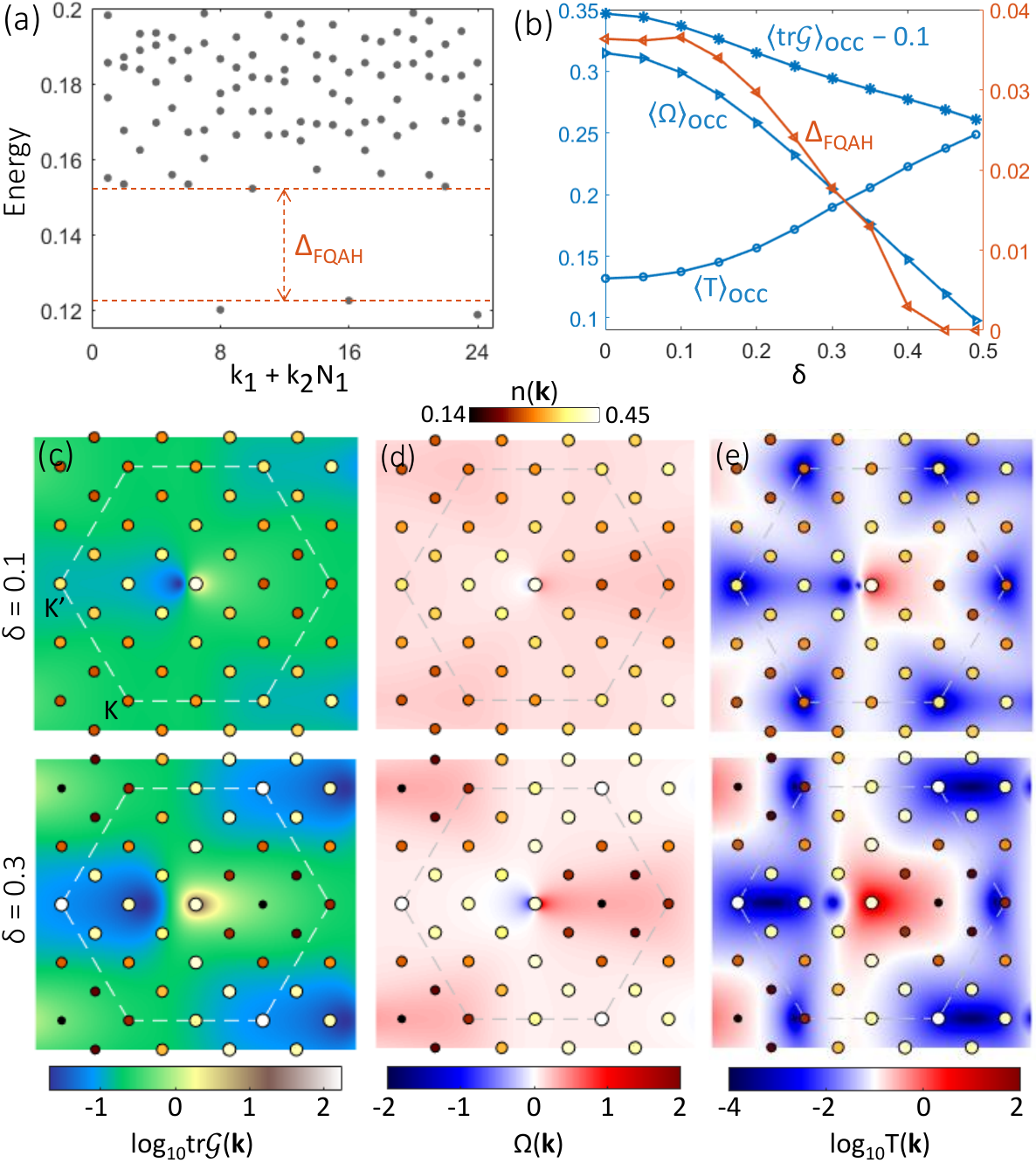}
	\caption{\textbf{ED results of the honeycomb model.} (a) The many-body energy spectrum with $n_{\rm up}=2$ at $U=1$ and $\delta =0.2$ in a rectangular system. (b) The orange curve denotes the variation of the many-body gap of the FQAH phase $\Delta_{\rm FQAH}$ with $\delta$, whereas a vanishing $\Delta_{\rm FQAH}$ means a trivial phase. 
		The blue curves with different symbols represent occupation-weighted $\braket{\text{tr}\,\CalG}_{\rm occ}$, $\braket{\Omega}_{\rm occ}$ and $\braket{T}_{\rm occ}$ averaged over the states with FQAH momenta. Diverging/undefined contributions at the $\Gamma$ point are excluded. $\braket{\text{tr}\,\CalG}_{\rm occ}$ is shifted for clarity. (c--e) Carrier occupation at $\delta=0.1$ (1st row) and $0.3$ (2nd row) represented by dots, whose color and size denote the occupation averaged over the three states with FQAH momenta in a tilted system. The continuous background color displays log$_{10}$tr$\,\CalG(\bk)$, $\Omega(\bk)$ and log$_{10}$$\,T(\bk)$, respectively.}
	\label{Fig:DiceED}
\end{figure}

The existence of FQAH effects is also supported by ED calculations that properly incorporate the effects of band mixing. To reduce numerical costs, we adopt the ``band maximum'' approach~\cite{MultiBandEDPRB2025} on two system configurations (rectangular and tilted, see SI) with 24 unit cells, where the number of particles in the upper band is capped at $n_{\rm up}$, while it is unrestricted in the SFB. Fig.~\ref{Fig:DiceED}(a) presents the many-body spectrum at $\delta=0.2$ and $U=1$ on a rectangular system. Three nearly degenerate ground states can be clearly identified at the expected momenta of FQAH states (details in SI)~\cite{regnault2011fractional}, which remain gapped from the excited states under flux insertion [Fig.~S10(b) of SI]. 
The system is in the FQAH phase when the three ground states lie at the expected momenta and carry nonzero many-body Chern numbers; its many-body gap $\Delta_{\rm FQAH}$ is defined as the energy separation between the fourth-lowest state and the three ground states [Fig.~\ref{Fig:DiceED}(a)]. Otherwise, the system is in the trivial phase and $\Delta_{\rm FQAH}=0$.
The orange curve in Fig.~\ref{Fig:DiceED}(b) shows the evolution of $\Delta_{\rm FQAH}$ with $\delta$ at $U=1$. It decreases and closes at $\delta\sim0.42$, consistent with the FQAH phase boundary from DMRG. This descending behavior of $\Delta_{\rm FQAH}$ anti-correlates with the ascending profile of $\sigma_{\Omega}$, $\sigma_{\rm tr\,\CalG}$ and TCV [Fig.~\ref{Fig:DiceSingleParticle}(g)], implying that improved uniformity of quantum geometry is generally favorable for stabilizing the FQAH phase, not only in Chern bands~\cite{JainCurvatureGap,RoyPRB2014,RoyNatCommun2015} but also in SFBs. However, unlike in Chern bands, $\Omega(\bk)$ and $\text{tr}\,\CalG(\bk)$ could be largely uncorrelated in SFBs with non-quantized Berry flux and strong violation of idealness.

The dots in Figs.~\ref{Fig:DiceED}(c--e) present the $\bk$-space carrier occupation $n(\bk)$ of the FQAH states in a tilted system. $n(\bk)$ is nonuniform and stays quantitatively similar for different $U$ [Figs.~S3(a, b) of SI]. Intriguingly, contrasting behaviors of $n(\bk)$ are observed around the band touching vs elsewhere. (i) Away from the $\Gamma$ point, the neighborhoods of the three $\bs{K}$ ($\bs{K}'$) corners with large (small) $\text{tr}\,\CalG(\bk)$ repel (attract) occupation [Fig.~\ref{Fig:DiceED}(c)]. This tendency becomes more pronounced as $\text{tr}\,\CalG(\bk)$ varies more strongly, quantitatively manifested in the drop of the occupation-weighted $\braket{\text{tr}\,\CalG}_{\rm occ}=\frac{1}{2\pi}\int_{\rm BZ\backslash\{\Gamma\}} \text{tr}\,\CalG(\bk)n(\bk)d\bk$ with $\delta$ [Fig.~\ref{Fig:DiceED}(b)]. This anti-correlation between $n(\bk)$ and $\text{tr}\,\CalG(\bk)$ was also noted in isolated Chern bands~\cite{FCIQuantumMetricPRR2023,FCIQuantumMetricLiuZhaoPRL2020,FCIQuantumMetricYangBo2024}. 
(ii) At and around the $\Gamma$ point, where $\text{tr}\,\CalG(\bk)$ is divergent, accompanied by strongly varying $\Omega(\bk)$ and diverging $T(\bk)$, unexpectedly pronounced $n(\bk)$ is observed. The large $n(\bk\approx\Gamma)$ should not be ascribed to a trivial doubling of states at the band touching.  
The upper-band occupation concentrates around the $\Gamma$ point and decays exponentially with $|\bk|$ [Figs.~S3(c) and Fig.~S7 of SI]. The upper-band occupation right at the $\Gamma$ point is ill-defined due to the degeneracy; nevertheless, its value in the immediate vicinity of $\Gamma$ can be estimated under twisted boundary conditions and is significantly smaller than that in the SFB [Figs.~S7(b--d) of SI].

The occupation pattern can be intuitively understood as electrons preferring smaller overlaps with each other to lower the repulsive interaction. Importantly, $\text{tr}\,\CalG(\bk)$ characterizes the real-space spread of Bloch electrons~\cite{Wannier1997,QuantumGeometryNatRevPhys2026}. Away from the $\Gamma$ point, upper-band occupation is unfavorable due to the kinetic energy cost; electrons thus tend to occupy only the SFB and prefer regions where its $\text{tr}\,\CalG(\bk)$ is small. In contrast, in the vicinity of the $\Gamma$ point, the two bands are nearly degenerate and their Bloch states can be mixed at low kinetic energy cost to form less extended orbitals, for which a small two-band quantum metric trace ($\text{tr}\,\CalG^{(\rm 2band)}$) is a good indicator~\cite{HeavyFermionGapless}. The vanishing $\text{tr}\,\CalG^{(\rm 2band)}$ in the two-orbital honeycomb model is consistent with the observed large carrier occupation in the vicinity of the band touching. Additionally, comparing the two-band ED with the projected one-band ED in the SFB (i.e., enforcing a constraint of $n_{\rm up}=0$), the latter with band mixing artificially suppressed indeed yields higher ground state energies, along with a smaller carrier occupation in the vicinity of $\Gamma$ (Fig.~S8 of SI). 
This illustrates an intriguing scenario in which interacting carriers in a flat band exploit the singular band touching with vanishing $\text{tr}\,\CalG^{(\rm 2band)}$ to lower the energy.
The momentum-space region within which the $\text{tr}\,\CalG^{(\rm 2band)}$ is relevant---determined by the competition between repulsive interaction and kinetic energy---can be estimated from the upper-band occupation (Fig.~S11 in SI).

Properties of the many-body states depend on the interplay of the quantum metric-modulated $n(\bk)$ with its underlying quantum geometry. The FQAH phase exhibits a larger $\Delta_{\rm FQAH}$ at smaller $\delta$, which can be understood from the increase of occupation-weighted $\braket{\Omega}_{\rm occ}$ and the decrease of $\braket{T}_{\rm occ}$ as $\delta$ is reduced [Fig.~\ref{Fig:DiceED}(b)]. The transition to a trivial charge density wave (CDW) phase at large $\delta$ [Fig.~\ref{Fig:DiceDMRG}(d) top panel] can be ascribed to the strong electron localization that is accompanied by greatly reduced $\braket{\Omega}_{\rm occ}$ and enlarged $\braket{T}_{\rm occ}$. In this model, we find $\braket{\Omega}_{\rm occ}\gtrsim0.12$ and $\braket{T}_{\rm occ}\lesssim0.23$ for the FQAH phase. The upper bound on $\braket{T}_{\rm occ}$ is likely underestimated since the suppression of the FQAH phase at large $\delta$ is more likely due to strong electron localization and low $\braket{\Omega}_{\rm occ}$ (cf. discussions for the kagome model).  We remark that the characterizations of SFB quantum geometry in Fig.~\ref{Fig:DiceED}(b) have excluded the diverging contributions at the $\Gamma$ point. Under twisted boundary conditions where contributions close to the singular point can be accounted for (Fig.~S10 of SI), one finds $\braket{T}_{\rm occ}$ in the SFB can well exceed the corresponding value in the first LL\footnote{This counting under small twist likely overestimates the violation of idealness, as two-band quantum geometry may become the more relevant quantity in the vicinity of the $\Gamma$ point, while the characterizations simply excluding the $\Gamma$ point in Fig.~\ref{Fig:DiceED}(b) are more conservative.}. From the perspective of projected one-band quantum geometry, it is somewhat unexpected that the FQAH states can survive such a singular environment. The resolution is that, in the vicinity of the singular band touching, two-band quantum geometry comes into play because of the band (near) degeneracy.

We briefly comment on the CDW phase in DMRG simulations. The ground state transitions to the CDW when $\delta\gtrsim0.43$, which exhibits a $\sqrt{3}\times\sqrt{3}$ periodic charge pattern in contrast to the rather uniform distribution in the FQAH phase [Fig.~\ref{Fig:DiceDMRG}(d)]. In the $\bk$ space, the charge distribution at the BZ corner $\rho(\bs{K})$ rapidly rises when $\delta\gtrsim0.43$ [Fig.~\ref{Fig:DiceDMRG}(c) bottom panel], also pointing to the formation of a $\sqrt{3}\times\sqrt{3}$ charge order. The CDW phase is gapped and topologically trivial, exhibiting no quantized charge pumping in DMRG calculations (Fig.~S12 of SI). To characterize the nature of the phase transition, we examine the variation of various physical quantities with $\delta$. Fig.~\ref{Fig:DiceDMRG}(c) top and middle panels illustrate this by fixing $U = 10$: a sharp discontinuity exists in the derivative of the ground state energy $\partial E/\partial \delta$ at $\delta\approx0.43$, which is accompanied by a peak in the entanglement entropy $S$, pointing to a first-order phase transition.

\begin{figure}[b]
	\centering
	\includegraphics[width=3.4in]{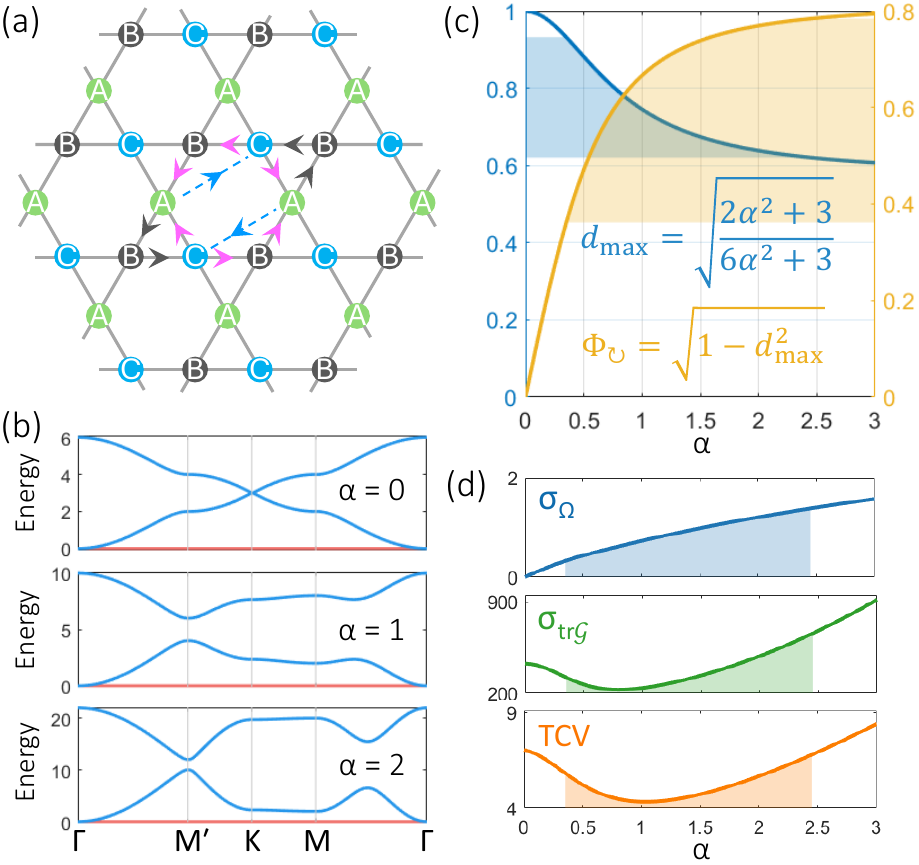}
	\caption{\textbf{Single-particle results of the kagome model}: (a) Schematic illustration of the kagome lattice and the hopping processes in $\hat{H}_{\tinykag}(\bk)$. Black/pink arrows denote NN hopping with complex amplitude $1\pm i\alpha$, blue arrows represent next-NN hopping with complex amplitude $i\alpha-\alpha^2$. (b) Band structures of $\hat{H}_{\tinykag}(\bk)$ for various $\alpha$. (c) Plots of $d_{\rm max}$ (blue) and Berry phase $\Phi_{\tinyCW}$ (yellow) around the touching point as functions of $\alpha$.
		(d) Quantum geometry fluctuations as a function of $\alpha$. The shaded areas within $0.35\lesssim\alpha\lesssim2.46$ in panels (c) and (d) host FQAH effects.}
	\label{Fig:KagomeSingleParticle}
\end{figure}

\begin{figure}[t]
	\centering
	\includegraphics[width=3.4in]{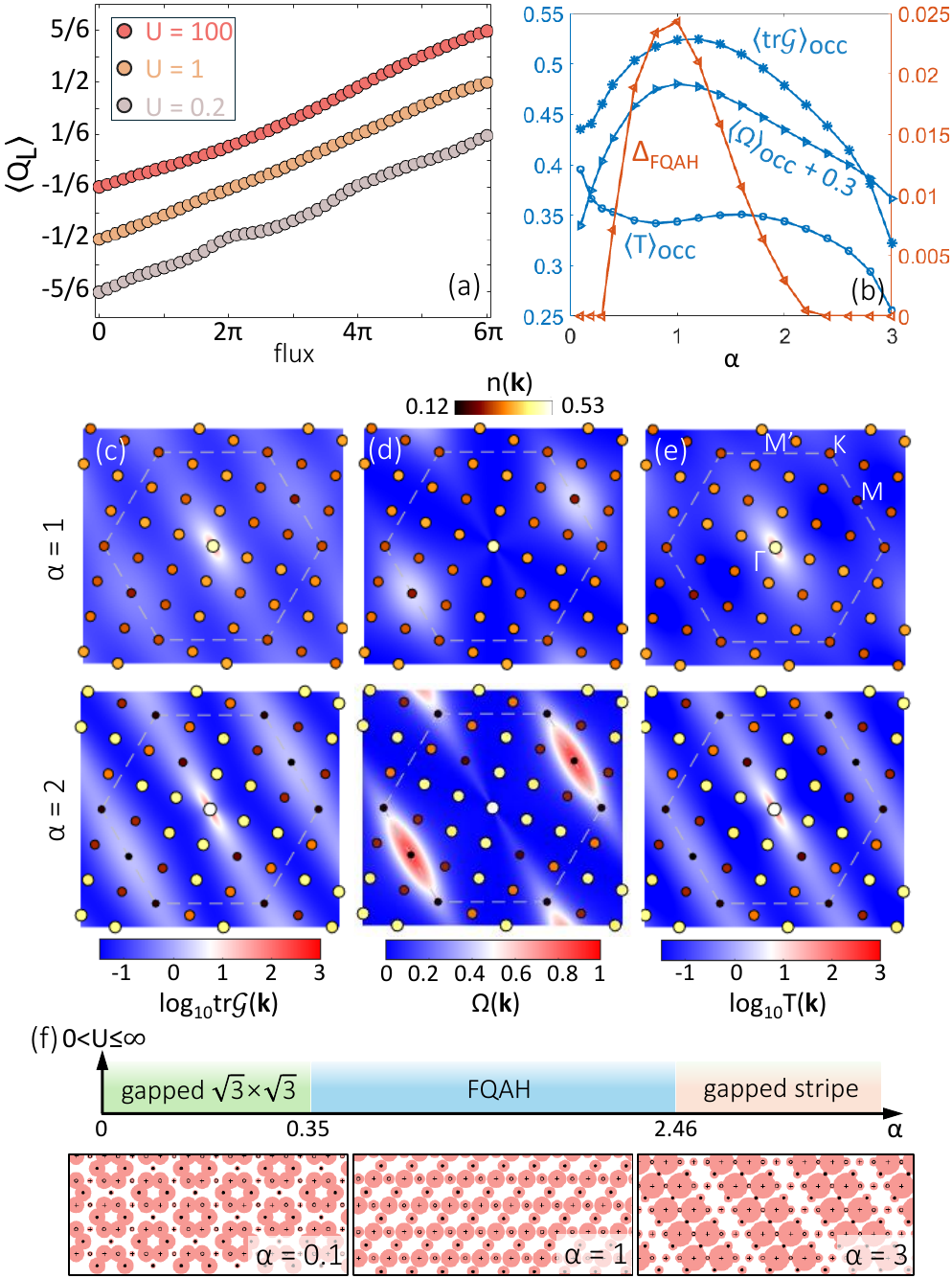}
	\caption{\textbf{Many-body results of the kagome model}: (a) Charge pumping simulation results obtained via DMRG at $\alpha=0.8$ and three different interaction strengths. (b) The orange curve denotes the variation of the many-body gap of the FQAH phase $\Delta_{\rm FQAH}$ with $\alpha$. The blue curves with different symbols represent occupation-weighted $\braket{\text{tr}\,\CalG}_{\rm occ}$, $\braket{\Omega}_{\rm occ}$ and $\braket{T}_{\rm occ}$ averaged over the states with FQAH momenta. Diverging/undefined contributions at the $\Gamma$ point are excluded. $\braket{\Omega}_{\rm occ}$ is shifted for clarity. (c--e) Carrier occupation at $\alpha=1$ and 2 represented by dots, whose color and size denote the occupation averaged over the three states with FQAH momenta in a tilted system. The continuous background color displays log$_{10}$tr$\,\CalG(\bk)$, $\Omega(\bk)$ and log$_{10}$$\,T(\bk)$, respectively. (f) Phase diagram obtained by DMRG and typical charge distributions in the three different phases. $U=1$ in (b--f).}
	\label{Fig:KagomeManyBody}
\end{figure}

\textit{\textcolor{blue}{Kagome model and its quantum geometry---}}We also consider a three-orbital kagome lattice model [Fig.~\ref{Fig:KagomeSingleParticle}(a)]~\cite{SinguFlatCommunPhys2023}. Details of its Hamiltonian $\hat{H}_{\tinykag}(\bk)$ are provided in SI, with a tuning parameter denoted by $\alpha$. It also has a zero-energy SFB with a band touching at the $\Gamma$ point [Fig.~\ref{Fig:KagomeSingleParticle}(b)]. The two touching bands have a total Chern number of $1$ when $\alpha\neq0$, unlike the honeycomb case. The two models also feature qualitatively different quantum geometric properties, summarized in Table~\ref{table:system_comparison}. 
Here the singularity is quantified by $d_{\rm max}=\sqrt{\frac{2\alpha^2+3}{6\alpha^2+3}}$. As $\alpha$ increases from 0, $d_{\rm max}$ drops from 1 and saturates at $\sim0.6$ [Fig.~\ref{Fig:KagomeSingleParticle}(c) blue curve]. At $\alpha=0$, the pristine kagome lattice model is recovered with $\Omega(\bk)\equiv0$ on the SFB. When $\alpha$ is enlarged, $\Omega(\bk)$ spreads out and concentrates around two edges of the BZ [Fig.~\ref{Fig:KagomeManyBody}(d) background color]. The Berry flux excluding the touching point $\Phi_{\tinyCW}=\sqrt{1-d_{\rm max}^2}$ increases from 0 and saturates at $\sim0.8$ [Fig.~\ref{Fig:KagomeSingleParticle}(c) yellow curve]. The quantum metric $\CalG(\bk)$ of the SFB is divergent around the $\Gamma$ point for any $\alpha\ge0$ [Fig.~\ref{Fig:KagomeManyBody}(c) background color]; its divergence becomes milder for larger $\alpha$ and stripe features of high magnitude emerge.  Fig.~\ref{Fig:KagomeSingleParticle}(d) quantifies the quantum geometry fluctuations on the SFB: $\sigma_\Omega$ grows monotonically with $\alpha$, while $\sigma_{\text{tr}\,\CalG}$ and TCV initially decrease, followed by an increase. In the parameter window of interest, the quantum geometry fluctuates more strongly in the kagome model than in the honeycomb model. In particular, the TCV is above 4 in Fig.~\ref{Fig:KagomeSingleParticle}(d), beyond that of the second LL~\cite{KahlerBandsPRB2021c}. The strong stripe anisotropy at large $\alpha$ is expected to lead to anisotropic competing many-body phases in contrast to the $\sqrt{3}\times\sqrt{3}$ CDW in the honeycomb model with 2D-triangular fluctuation.

\textit{\textcolor{blue}{FQAH and topologically trivial phases in the kagome model---}}An FQAH phase is identified at $\nu=1/3$ filling of the SFB via both DMRG [Fig.~\ref{Fig:KagomeManyBody}(a)] and ED [Fig.~S17(b) and Fig.~S21(b) of SI] calculations. The FQAH phase also persists for any $0<U\le\infty$ from the weak to the strong interaction regimes. The orange curve in Fig.~\ref{Fig:KagomeManyBody}(b) shows the many-body gap $\Delta_{\rm FQAH}$ from ED calculations, which emerges at $\alpha\sim0.4$, rises to the maximum at $\alpha\sim1$ and subsequently decreases to zero at $\alpha\sim2.4$. The FQAH phase boundaries extracted from $\Delta_{\rm FQAH}$ (i.e., $\alpha\approx0.4$ and $2.4$) are consistent with the DMRG results [Fig.~\ref{Fig:KagomeManyBody}(f) upper panel]. The Berry flux at $\alpha\sim0.4$ is as low as $\Phi_{\tinyCW}\sim0.4$ [Fig.~\ref{Fig:KagomeSingleParticle}(c)], much smaller than the lower bound in the honeycomb model ($\sim0.6$). The profile of $\Delta_{\rm FQAH}$ anti-correlates with those of $\sigma_{\text{tr}\,\CalG}$ and TCV but not with $\sigma_{\Omega}$ [Fig.~\ref{Fig:KagomeSingleParticle}(d)], implying again that the occupation-weighted quantum geometry should be holistically considered with regard to the stability of FQAH effects in SFBs, as detailed below.

The dots in Figs.~\ref{Fig:KagomeManyBody}(c--e) show the carrier occupation $n(\bk)$ of the FQAH states. Analogously to the honeycomb model, the largest occupation is observed at the $\Gamma$ point, where the SFB's $\text{tr}\,\CalG(\bk)$ and $T(\bk)$ are divergent; elsewhere in the BZ, electrons tend to occupy (avoid) regions with small (large) $\text{tr}\,\CalG(\bk)$. 
Notably, as in the honeycomb model, we also find the large occupation is accompanied by a moderate band mixing in the vicinity of the singular touching point, where the SFB's $\text{tr}\,\CalG(\bk)$ diverges but $\text{tr}\,\CalG^{(\rm 2band)}$ of the two touching bands is vanishing (Fig.~S16 of SI). In comparison, the projected one-band ED in the SFB (i.e., enforcing $n_{\rm up}=0$) yields higher ground state energies (Fig.~S19 of SI).
Unlike in the honeycomb model, here $\braket{T}_{\rm occ}$ stays nearly constant at $\sim0.35$ within a broad window $0.4\lesssim\alpha\lesssim2.2$, suggesting that $\braket{\Omega}_{\rm occ}$ dominates in stabilizing/destabilizing the FQAH phase. Indeed, the profile of $\Delta_{\rm FQAH}$ correlates well with that of $\braket{\Omega}_{\rm occ}$. The suppression of the FQAH phase for $\alpha\lesssim0.4$ and $\alpha\gtrsim2.4$ can be attributed to low $\braket{\Omega}_{\rm occ}$, but the two regimes have different origins. For $\alpha\lesssim0.4$, the small $\braket{\Omega}_{\rm occ}$ is due to the vanishing $\Omega(\bk)$ and low Berry flux $\Phi_{\tinyCW}$ [Fig.~\ref{Fig:KagomeSingleParticle}(c) yellow curve]; when $\alpha\gtrsim2.4$, even though hot spots of $\Omega(\bk)$ and large Berry flux $\Phi_{\tinyCW}$ exist, electrons are localized in regions of vanishing $\Omega(\bk)$, leading to low $\braket{\Omega}_{\rm occ}$ [Fig.~\ref{Fig:KagomeManyBody}(d) bottom panel]. For the FQAH phase, $\braket{\Omega}_{\rm occ}\gtrsim0.11$ is required here, close to the value in the previous model ($\gtrsim0.12$); while here $\braket{T}_{\rm occ}\approx0.35$ is much larger, representing a strong violation of idealness without counting the contribution from the $\Gamma$ point.

We now briefly comment on the two trivial phases competing with the FQAH phase, in the parameter ranges $\alpha\lesssim0.4$ and $\alpha\gtrsim2.4$ respectively [Fig.~\ref{Fig:KagomeManyBody}(f) upper panel]. Both phases are gapped and topologically trivial without quantized charge pumping in DMRG calculations (Fig.~S23 of SI). The two trivial phases exhibit distinct real-space charge orders: a `donut' order with $\sqrt{3}\times\sqrt{3}$ periodicity for $\alpha\lesssim0.4$, and a stripe order with tripled periodicity for $\alpha\gtrsim2.4$ [Fig.~\ref{Fig:KagomeManyBody}(f) lower panel]. We note that stripe phases have also been observed in LLs with an anisotropic effective mass tensor or an anisotropic dielectric constant tensor, caused by softened magnetoroton modes at large anisotropies~\cite{AnisotropicLLYangBo2012,AnisotropicLLZhangFC2012}. In our case, the stripe phase is instead driven by anisotropic quantum geometry at large $\alpha$.


\textit{\textcolor{blue}{Discussions---}}The SFBs here represent a scenario that contrasts with existing venues for FQAH effects in a number of aspects, including the absence of a band gap, the presence of singularities in Bloch functions and quantum geometry, and the strong quantum geometric divergence, fluctuation and non-idealness beyond the first or even second LL. The findings offer new clues toward the question of which conditions are truly indispensable for FQAH effects to emerge and which are the good-to-have ones. The tunable quantum geometric landscape enables a systematic exploration in such unconventional settings with neither band topology nor ideal flat-band quantum geometry. In two SFB models of distinct characteristics, we observe common trends in how quantum geometry influences the FQAH states. While similar qualitative trends of FQAH stability against variations of TCV have also been observed in Chern bands, the emergence of FQAH states in the highly singular contexts of SFBs is striking in the first place. Remarkably, at dilute fillings of the SFBs, FQAH phases are found for the NN interaction $0<U\le\infty$ from the weak to the strong interaction regimes. Our work advances the exploration of FQAH effects toward extreme conditions and implies possibilities for discovering novel topological many-body states.

Our findings highlight the robustness of the many-body topological order in sufficiently flat bands with short-range repulsion. When kinetic energy is negligible, this robustness arises from an intricate interplay between many-body interaction and quantum geometric properties, where interacting carriers spontaneously develop inhomogeneous distributions to occupy regions with favorable local quantum geometry. 
Importantly, band (near) degeneracy may render the multi-band quantum metric the relevant quantity in its vicinity. A sufficiently large occupation-weighted Berry flux $\braket{\Omega}_{\rm occ}$ appears to be necessary to stabilize FQAH states. 
The distributions of these quantum geometric quantities, instead of their simple BZ-averaged values and global band topology, can be more relevant indicators for assessing whether FQAH states can emerge in various flat-band platforms, including Chern bands and non-Chern band systems~\cite{FCIC=0Lin2025,FCIC=0Lu2025,FCIC=0Bergholtz2025}.
The crossover between one-band and multi-band quantum geometry for the interplay with many-body interactions could also be relevant in narrow-gap systems where flat bands are not well separated from dispersive ones in certain regions of the BZ, e.g., in rhombohedral multilayer graphene moir\'e systems~\cite{lu2024fractional,JuLongFCI2025,LuXiaoboGrapheneFCI2025,GrapheneFCIPRX2025}. With such a crossover, the single-band quantities conventionally adopted for characterizing the overall idealness of isolated flat bands may no longer be sufficient for a complete understanding of the properties of many-body states.


\emph{{\color{blue} Acknowledgments---}}The work is supported by the National Natural Science Foundation of China (No. 12425406), Research Grants Council of Hong Kong (AoE/P-701/20, HKU SRFS21227S05), and New Cornerstone Science Foundation. The authors thank Beijing PARATERA Tech Co., Ltd. (https://cloud.paratera.com) for providing HPC resources that supported the research results reported in this paper.

\section{Methods}

\subsection{Density matrix renormalization group (DMRG)}

The charge pumping simulation and the phase diagrams presented in Fig.~\ref{Fig:DiceDMRG}(a) and Fig.~\ref{Fig:KagomeManyBody}(a) are obtained via real-space DMRG simulations, where the maximal bond dimension is set to $D = 400$. The lattice is configured on a finite cylinder with open (periodic) boundary conditions along the $x$ ($y$) direction. The total number of lattice sites is $N_x\times N_y\times2\,(3)$ for the honeycomb (kagome) model, where $N_x$ ($N_y$) denotes the number of unit cells along the $x$ ($y$) direction. For the honeycomb model, we set $N_x=24$ and $N_y=6$; while for the kagome model, we use $N_x = 18$ and $N_y=4$. 

The entanglement entropy, the first-order energy derivative and the CDW order parameter across the phase transition point, as shown in Fig.~\ref{Fig:DiceDMRG}(c), are calculated using iDMRG simulations~\cite{iDMRG}. In the iDMRG simulations, the cylinder is infinite along the $x$ direction and consists of 4 cells in the $y$ direction. The bond dimension is increased to $D=800$ to ensure convergence across all parameter values.

The momentum-resolved entanglement spectra shown in Fig.~\ref{Fig:DiceDMRG}(b) and Fig.~S17(a) of SI are obtained through DMRG simulations performed in a mixed real ($x$ direction) and momentum ($y$ direction) space, where each site is labeled by the pair $\{x,\,k_y\}$~\cite{HybridizeDMRG}. The momentum vector, $k_y$, along the cylinder circumference is used as a conserved quantity, allowing the entanglement spectrum to be categorized into distinct momentum sectors. For both the honeycomb and kagome models, the lattices we use consist of 6 cells along the $y$ direction with $k_y\in\{\frac{\pi}{6},\frac{\pi}{3},\cdots,2\pi\}$, and 60 cells along the $x$ direction. The maximal bond dimension is set to $D = 800$.

All the DMRG simulations in this paper are performed using the ITensor library with U(1) symmetry~\cite{ITensor}.

\subsection{Exact diagonalization (ED)}

Details of the ED calculations are provided in the SI.

\bibliography{Refs}

@article{tang2011high,
  title={High-temperature fractional quantum Hall states},
  author={Tang, Evelyn and Mei, Jia-Wei and Wen, Xiao-Gang},
  journal={Phys. Rev. lett.},
  volume={106},
  number={23},
  pages={236802},
  year={2011},
  publisher={APS},
  url={https://journals.aps.org/prl/abstract/10.1103/PhysRevLett.106.236802}
}

@article{sun2011nearly,
  title={Nearly flatbands with nontrivial topology},
  author={Sun, Kai and Gu, Zhengcheng and Katsura, Hosho and Sarma, S Das},
  journal = {Phys. Rev. Lett.},
  volume={106},
  number={23},
  pages={236803},
  year={2011},
  publisher={APS},
  url={https://journals.aps.org/prl/abstract/10.1103/PhysRevLett.106.236803}
}

@article{neupert2011fractional,
  title={Fractional quantum Hall states at zero magnetic field},
  author={Neupert, Titus and Santos, Luiz and Chamon, Claudio and Mudry, Christopher},
  journal = {Phys. Rev. Lett.},
  volume={106},
  number={23},
  pages={236804},
  year={2011},
  publisher={APS},
  url={https://journals.aps.org/prl/abstract/10.1103/PhysRevLett.106.236804}
}

@article{sheng2011fractional,
  title={Fractional quantum Hall effect in the absence of Landau levels},
  author={Sheng, DN and Gu, Zheng-Cheng and Sun, Kai and Sheng, L},
  journal={Nat. Commun.},
  volume={2},
  number={1},
  pages={389},
  year={2011},
  publisher={Nature Publishing Group UK London},
  url={https://www.nature.com/articles/ncomms1380}
}

@article{wang2011fractional,
  title={Fractional quantum Hall effect of hard-core bosons in topological flat bands},
  author={Wang, Yi-Fei and Gu, Zheng-Cheng and Gong, Chang-De and Sheng, DN},
  journal = {Phys. Rev. Lett.},
  volume={107},
  number={14},
  pages={146803},
  year={2011},
  publisher={APS},
  url={https://journals.aps.org/prl/abstract/10.1103/PhysRevLett.107.146803}
}

@article{regnault2011fractional,
  title={Fractional chern insulator},
  author={Regnault, Nicolas and Bernevig, B Andrei},
  journal={Phys. Rev. X},
  volume={1},
  number={2},
  pages={021014},
  year={2011},
  publisher={APS},
  url={https://journals.aps.org/prx/abstract/10.1103/PhysRevX.1.021014}
}

@article{xiao2011interface,
  title={Interface engineering of quantum Hall effects in digital transition metal oxide heterostructures},
  author={Xiao, Di and Zhu, Wenguang and Ran, Ying and Nagaosa, Naoto and Okamoto, Satoshi},
  journal={Nat. Commun.},
  volume={2},
  number={1},
  pages={596},
  year={2011},
  publisher={Nature Publishing Group UK London},
  url={https://www.nature.com/articles/ncomms1602}
}

@article{InteractionExceedGapPRL2014,
  title = {Fractional Chern Insulators with Strong Interactions that Far Exceed Band Gaps},
  author = {Kourtis, Stefanos and Neupert, Titus and Chamon, Claudio and Mudry, Christopher},
  journal = {Phys. Rev. Lett.},
  volume = {112},
  issue = {12},
  pages = {126806},
  numpages = {5},
  year = {2014},
  month = {Mar},
  publisher = {American Physical Society},
  doi = {10.1103/PhysRevLett.112.126806},
  url = {https://link.aps.org/doi/10.1103/PhysRevLett.112.126806}
}

@article{InteractionExceedGapPRB2015,
  title = {Characterization and stability of a fermionic $\ensuremath{\nu}=1/3$ fractional Chern insulator},
  author = {Grushin, Adolfo G. and Motruk, Johannes and Zaletel, Michael P. and Pollmann, Frank},
  journal = {Phys. Rev. B},
  volume = {91},
  issue = {3},
  pages = {035136},
  numpages = {6},
  year = {2015},
  month = {Jan},
  publisher = {American Physical Society},
  doi = {10.1103/PhysRevB.91.035136},
  url = {https://link.aps.org/doi/10.1103/PhysRevB.91.035136}
}

@article{lu2024fractional,
  title={Fractional quantum anomalous Hall effect in multilayer graphene},
  author={Lu, Zhengguang and Han, Tonghang and Yao, Yuxuan and Reddy, Aidan P and Yang, Jixiang and Seo, Junseok and Watanabe, Kenji and Taniguchi, Takashi and Fu, Liang and Ju, Long},
  journal={Nature},
  volume={626},
  number={8000},
  pages={759--764},
  year={2024},
  publisher={Nature Publishing Group UK London},
  url={https://www.nature.com/articles/s41586-023-07010-7}
}

@Article{JuLongFCI2025,
author={Lu, Zhengguang
and Han, Tonghang
and Yao, Yuxuan
and Hadjri, Zach
and Yang, Jixiang
and Seo, Junseok
and Shi, Lihan
and Ye, Shenyong
and Watanabe, Kenji
and Taniguchi, Takashi
and Ju, Long},
title={Extended quantum anomalous Hall states in graphene/hBN moir{\'e} superlattices},
journal={Nature},
year={2025},
month={Jan},
day={01},
volume={637},
number={8048},
pages={1090-1095},
issn={1476-4687},
doi={10.1038/s41586-024-08470-1},
url={https://doi.org/10.1038/s41586-024-08470-1}
}

@Article{LuXiaoboGrapheneFCI2025,
author={Xie, Jian
and Huo, Zihao
and Lu, Xin
and Feng, Zuo
and Zhang, Zaizhe
and Wang, Wenxuan
and Yang, Qiu
and Watanabe, Kenji
and Taniguchi, Takashi
and Liu, Kaihui
and Song, Zhida
and Xie, X. C.
and Liu, Jianpeng
and Lu, Xiaobo},
title={Tunable fractional Chern insulators in rhombohedral graphene superlattices},
journal={Nat. Mater.},
year={2025},
month={Jul},
day={01},
volume={24},
number={7},
pages={1042-1048},
issn={1476-4660},
doi={10.1038/s41563-025-02225-7},
url={https://doi.org/10.1038/s41563-025-02225-7}
}

@article{GrapheneFCIPRX2025,
  title = {Displacement Field-Controlled Fractional Chern Insulators and Charge Density Waves in a Graphene/hBN Moir\'e Superlattice},
  author = {Aronson, Samuel H. and Han, Tonghang and Lu, Zhengguang and Yao, Yuxuan and Butler, Jackson P. and Watanabe, Kenji and Taniguchi, Takashi and Ju, Long and Ashoori, Raymond C.},
  journal = {Phys. Rev. X},
  volume = {15},
  issue = {3},
  pages = {031026},
  numpages = {9},
  year = {2025},
  month = {Jul},
  publisher = {American Physical Society},
  doi = {10.1103/75gl-jzl6},
  url = {https://link.aps.org/doi/10.1103/75gl-jzl6}
}

@article{BM,
author = {Rafi Bistritzer  and Allan H. MacDonald },
title = {Moiré bands in twisted double-layer graphene},
journal = {Proc. Natl. Acad. Sci. U.S.A.},
volume = {108},
number = {30},
pages = {12233-12237},
year = {2011},
doi = {10.1073/pnas.1108174108},
URL = {https://www.pnas.org/doi/abs/10.1073/pnas.1108174108}
}

@Article{FCIMoTe2Park2023,
author={Park, Heonjoon
and Cai, Jiaqi
and Anderson, Eric
and Zhang, Yinong
and Zhu, Jiayi
and Liu, Xiaoyu
and Wang, Chong
and Holtzmann, William
and Hu, Chaowei
and Liu, Zhaoyu
and Taniguchi, Takashi
and Watanabe, Kenji
and Chu, Jiun-Haw
and Cao, Ting
and Fu, Liang
and Yao, Wang
and Chang, Cui-Zu
and Cobden, David
and Xiao, Di
and Xu, Xiaodong},
title={Observation of fractionally quantized anomalous Hall effect},
journal={Nature},
year={2023},
month={Oct},
day={01},
volume={622},
number={7981},
pages={74-79},
issn={1476-4687},
doi={10.1038/s41586-023-06536-0},
url={https://doi.org/10.1038/s41586-023-06536-0}
}

@Article{FCIMoTe2Jiaqi2023,
author={Cai, Jiaqi
and Anderson, Eric
and Wang, Chong
and Zhang, Xiaowei
and Liu, Xiaoyu
and Holtzmann, William
and Zhang, Yinong
and Fan, Fengren
and Taniguchi, Takashi
and Watanabe, Kenji
and Ran, Ying
and Cao, Ting
and Fu, Liang
and Xiao, Di
and Yao, Wang
and Xu, Xiaodong},
title={Signatures of fractional quantum anomalous Hall states in twisted MoTe2},
journal={Nature},
year={2023},
month={Oct},
day={01},
volume={622},
number={7981},
pages={63-68},
issn={1476-4687},
doi={10.1038/s41586-023-06289-w},
url={https://doi.org/10.1038/s41586-023-06289-w}
}

@Article{FCIMoTe2ShanJie2023,
author={Zeng, Yihang
and Xia, Zhengchao
and Kang, Kaifei
and Zhu, Jiacheng
and Kn{\"u}ppel, Patrick
and Vaswani, Chirag
and Watanabe, Kenji
and Taniguchi, Takashi
and Mak, Kin Fai
and Shan, Jie},
title={Thermodynamic evidence of fractional Chern insulator in moir{\'e} MoTe2},
journal={Nature},
year={2023},
month={Oct},
day={01},
volume={622},
number={7981},
pages={69-73},
issn={1476-4687},
doi={10.1038/s41586-023-06452-3},
url={https://doi.org/10.1038/s41586-023-06452-3}
}

@article{FCIMoTe2PRX2023,
  title = {Observation of Integer and Fractional Quantum Anomalous Hall Effects in Twisted Bilayer ${\mathrm{MoTe}}_{2}$},
  author = {Xu, Fan and Sun, Zheng and Jia, Tongtong and Liu, Chang and Xu, Cheng and Li, Chushan and Gu, Yu and Watanabe, Kenji and Taniguchi, Takashi and Tong, Bingbing and Jia, Jinfeng and Shi, Zhiwen and Jiang, Shengwei and Zhang, Yang and Liu, Xiaoxue and Li, Tingxin},
  journal = {Phys. Rev. X},
  volume = {13},
  issue = {3},
  pages = {031037},
  numpages = {12},
  year = {2023},
  month = {Sep},
  publisher = {American Physical Society},
  doi = {10.1103/PhysRevX.13.031037},
  url = {https://link.aps.org/doi/10.1103/PhysRevX.13.031037}
}

@article{HongyiNSR2020,
    author = {Yu, Hongyi and Chen, Mingxing and Yao, Wang},
    title = "{Giant magnetic field from moiré induced Berry phase in homobilayer semiconductors}",
    journal = {Natl. Sci. Rev.},
    volume = {7},
    number = {1},
    pages = {12-20},
    year = {2020},
    month = {08},
    issn = {2095-5138},
    doi = {10.1093/nsr/nwz117},
    url = {https://doi.org/10.1093/nsr/nwz117}
}

@article{RoyPRB2012,
  title = {Fractional Chern insulators and the ${W}_{\ensuremath{\infty}}$ algebra},
  author = {Parameswaran, S. A. and Roy, R. and Sondhi, S. L.},
  journal = {Phys. Rev. B},
  volume = {85},
  issue = {24},
  pages = {241308},
  numpages = {4},
  year = {2012},
  month = {Jun},
  publisher = {American Physical Society},
  doi = {10.1103/PhysRevB.85.241308},
  url = {https://link.aps.org/doi/10.1103/PhysRevB.85.241308}
}

@article{RoyPRB2014,
  title = {Band geometry of fractional topological insulators},
  author = {Roy, Rahul},
  journal = {Phys. Rev. B},
  volume = {90},
  issue = {16},
  pages = {165139},
  numpages = {7},
  year = {2014},
  month = {Oct},
  publisher = {American Physical Society},
  doi = {10.1103/PhysRevB.90.165139},
  url = {https://link.aps.org/doi/10.1103/PhysRevB.90.165139}
}

@Article{RoyNatCommun2015,
author={Jackson, T. S.
and M{\"o}ller, Gunnar
and Roy, Rahul},
title={Geometric stability of topological lattice phases},
journal={Nat. Commun.},
year={2015},
month={Nov},
day={04},
volume={6},
number={1},
pages={8629},
issn={2041-1723},
doi={10.1038/ncomms9629},
url={https://doi.org/10.1038/ncomms9629}
}

@article{FCIPositionMomentumDualityPRL2015,
  title = {Position-Momentum Duality and Fractional Quantum Hall Effect in Chern Insulators},
  author = {Claassen, Martin and Lee, Ching Hua and Thomale, Ronny and Qi, Xiao-Liang and Devereaux, Thomas P.},
  journal = {Phys. Rev. Lett.},
  volume = {114},
  issue = {23},
  pages = {236802},
  numpages = {6},
  year = {2015},
  month = {Jun},
  publisher = {American Physical Society},
  doi = {10.1103/PhysRevLett.114.236802},
  url = {https://link.aps.org/doi/10.1103/PhysRevLett.114.236802}
}

@article{IdealFCIPRB2017,
  title = {Band structure engineering of ideal fractional Chern insulators},
  author = {Lee, Ching Hua and Claassen, Martin and Thomale, Ronny},
  journal = {Phys. Rev. B},
  volume = {96},
  issue = {16},
  pages = {165150},
  numpages = {16},
  year = {2017},
  month = {Oct},
  publisher = {American Physical Society},
  doi = {10.1103/PhysRevB.96.165150},
  url = {https://link.aps.org/doi/10.1103/PhysRevB.96.165150}
}

@article{AshvinFCIPRR2020,
  title = {Fractional Chern insulator states in twisted bilayer graphene: An analytical approach},
  author = {Ledwith, Patrick J. and Tarnopolsky, Grigory and Khalaf, Eslam and Vishwanath, Ashvin},
  journal = {Phys. Rev. Res.},
  volume = {2},
  issue = {2},
  pages = {023237},
  numpages = {12},
  year = {2020},
  month = {May},
  publisher = {American Physical Society},
  doi = {10.1103/PhysRevResearch.2.023237},
  url = {https://link.aps.org/doi/10.1103/PhysRevResearch.2.023237}
}

@article{WangjieIdealBandPRL2021,
  title = {Exact Landau Level Description of Geometry and Interaction in a Flatband},
  author = {Wang, Jie and Cano, Jennifer and Millis, Andrew J. and Liu, Zhao and Yang, Bo},
  journal = {Phys. Rev. Lett.},
  volume = {127},
  issue = {24},
  pages = {246403},
  numpages = {6},
  year = {2021},
  month = {Dec},
  publisher = {American Physical Society},
  doi = {10.1103/PhysRevLett.127.246403},
  url = {https://link.aps.org/doi/10.1103/PhysRevLett.127.246403}
}

@article{WangjieIdealBandPRR2023,
  title = {Origin of model fractional Chern insulators in all topological ideal flatbands: Explicit color-entangled wave function and exact density algebra},
  author = {Wang, Jie and Klevtsov, Semyon and Liu, Zhao},
  journal = {Phys. Rev. Res.},
  volume = {5},
  issue = {2},
  pages = {023167},
  numpages = {20},
  year = {2023},
  month = {Jun},
  publisher = {American Physical Society},
  doi = {10.1103/PhysRevResearch.5.023167},
  url = {https://link.aps.org/doi/10.1103/PhysRevResearch.5.023167}
}

@article{VortexablePRB2023,
  title = {Vortexability: A unifying criterion for ideal fractional Chern insulators},
  author = {Ledwith, Patrick J. and Vishwanath, Ashvin and Parker, Daniel E.},
  journal = {Phys. Rev. B},
  volume = {108},
  issue = {20},
  pages = {205144},
  numpages = {20},
  year = {2023},
  month = {Nov},
  publisher = {American Physical Society},
  doi = {10.1103/PhysRevB.108.205144},
  url = {https://link.aps.org/doi/10.1103/PhysRevB.108.205144}
}

@article{VortexablePRL2025,
  title = {Higher Vortexability: Zero-Field Realization of Higher Landau Levels},
  author = {Fujimoto, Manato and Parker, Daniel E. and Dong, Junkai and Khalaf, Eslam and Vishwanath, Ashvin and Ledwith, Patrick},
  journal = {Phys. Rev. Lett.},
  volume = {134},
  issue = {10},
  pages = {106502},
  numpages = {8},
  year = {2025},
  month = {Mar},
  publisher = {American Physical Society},
  doi = {10.1103/PhysRevLett.134.106502},
  url = {https://link.aps.org/doi/10.1103/PhysRevLett.134.106502}
}

@article{WangjieHigherIdealBandPRX2025,
  title = {Theory of Generalized Landau Levels and Its Implications for Non-Abelian States},
  author = {Liu, Zhao and Mera, Bruno and Fujimoto, Manato and Ozawa, Tomoki and Wang, Jie},
  journal = {Phys. Rev. X},
  volume = {15},
  issue = {3},
  pages = {031019},
  numpages = {39},
  year = {2025},
  month = {Jul},
  publisher = {American Physical Society},
  doi = {10.1103/1zg9-qbd6},
  url = {https://link.aps.org/doi/10.1103/1zg9-qbd6}
}

@article{KahlerBandsPRB2021a,
  title = {K\"ahler geometry and Chern insulators: Relations between topology and the quantum metric},
  author = {Mera, Bruno and Ozawa, Tomoki},
  journal = {Phys. Rev. B},
  volume = {104},
  issue = {4},
  pages = {045104},
  numpages = {13},
  year = {2021},
  month = {Jul},
  publisher = {American Physical Society},
  doi = {10.1103/PhysRevB.104.045104},
  url = {https://link.aps.org/doi/10.1103/PhysRevB.104.045104}
}

@article{KahlerBandsPRB2021b,
  title = {Engineering geometrically flat Chern bands with Fubini-Study K\"ahler structure},
  author = {Mera, Bruno and Ozawa, Tomoki},
  journal = {Phys. Rev. B},
  volume = {104},
  issue = {11},
  pages = {115160},
  numpages = {16},
  year = {2021},
  month = {Sep},
  publisher = {American Physical Society},
  doi = {10.1103/PhysRevB.104.115160},
  url = {https://link.aps.org/doi/10.1103/PhysRevB.104.115160}
}

@article{KahlerBandsPRB2021c,
  title = {Relations between topology and the quantum metric for Chern insulators},
  author = {Ozawa, Tomoki and Mera, Bruno},
  journal = {Phys. Rev. B},
  volume = {104},
  issue = {4},
  pages = {045103},
  numpages = {13},
  year = {2021},
  month = {Jul},
  publisher = {American Physical Society},
  doi = {10.1103/PhysRevB.104.045103},
  url = {https://link.aps.org/doi/10.1103/PhysRevB.104.045103}
}

@article{IdealChernCurveSpaceLL2023,
  title = {Ideal Chern bands as Landau levels in curved space},
  author = {Estienne, Benoit and Regnault, Nicolas and Cr\'epel, Valentin},
  journal = {Phys. Rev. Res.},
  volume = {5},
  issue = {3},
  pages = {L032048},
  numpages = {6},
  year = {2023},
  month = {Sep},
  publisher = {American Physical Society},
  doi = {10.1103/PhysRevResearch.5.L032048},
  url = {https://link.aps.org/doi/10.1103/PhysRevResearch.5.L032048}
}

@article{ChiralMultiGrapheneLedwith2022,
  title = {Family of Ideal Chern Flatbands with Arbitrary Chern Number in Chiral Twisted Graphene Multilayers},
  author = {Ledwith, Patrick J. and Vishwanath, Ashvin and Khalaf, Eslam},
  journal = {Phys. Rev. Lett.},
  volume = {128},
  issue = {17},
  pages = {176404},
  numpages = {7},
  year = {2022},
  month = {Apr},
  publisher = {American Physical Society},
  doi = {10.1103/PhysRevLett.128.176404},
  url = {https://link.aps.org/doi/10.1103/PhysRevLett.128.176404}
}

@article{ChiralMultiGrapheneWangJie2022,
  title = {Hierarchy of Ideal Flatbands in Chiral Twisted Multilayer Graphene Models},
  author = {Wang, Jie and Liu, Zhao},
  journal = {Phys. Rev. Lett.},
  volume = {128},
  issue = {17},
  pages = {176403},
  numpages = {6},
  year = {2022},
  month = {Apr},
  publisher = {American Physical Society},
  doi = {10.1103/PhysRevLett.128.176403},
  url = {https://link.aps.org/doi/10.1103/PhysRevLett.128.176403}
}

@article{ChiralMultiGrapheneDongJunkai2023,
  title = {Many-body ground states from decomposition of ideal higher Chern bands: Applications to chirally twisted graphene multilayers},
  author = {Dong, Junkai and Ledwith, Patrick J. and Khalaf, Eslam and Lee, Jong Yeon and Vishwanath, Ashvin},
  journal = {Phys. Rev. Res.},
  volume = {5},
  issue = {2},
  pages = {023166},
  numpages = {17},
  year = {2023},
  month = {Jun},
  publisher = {American Physical Society},
  doi = {10.1103/PhysRevResearch.5.023166},
  url = {https://link.aps.org/doi/10.1103/PhysRevResearch.5.023166}
}

@article{IdealFlatStrainSunKaiPRL2023,
  title = {Topological Exact Flat Bands in Two-Dimensional Materials under Periodic Strain},
  author = {Wan, Xiaohan and Sarkar, Siddhartha and Lin, Shi-Zeng and Sun, Kai},
  journal = {Phys. Rev. Lett.},
  volume = {130},
  issue = {21},
  pages = {216401},
  numpages = {7},
  year = {2023},
  month = {May},
  publisher = {American Physical Society},
  doi = {10.1103/PhysRevLett.130.216401},
  url = {https://link.aps.org/doi/10.1103/PhysRevLett.130.216401}
}

@article{ITensor,
	title={{The ITensor Software Library for Tensor Network Calculations}},
	author={Matthew Fishman and Steven R. White and E. Miles Stoudenmire},
	journal={SciPost Phys. Codebases},
	pages={4},
	year={2022},
	publisher={SciPost},
	doi={10.21468/SciPostPhysCodeb.4},
	url={https://scipost.org/10.21468/SciPostPhysCodeb.4}
}

@article{FCIQuantumMetricPRR2023,
  title = {Quantum metric induced phases in Moir\'e materials},
  author = {Abouelkomsan, Ahmed and Yang, Kang and Bergholtz, Emil J.},
  journal = {Phys. Rev. Res.},
  volume = {5},
  issue = {1},
  pages = {L012015},
  numpages = {8},
  year = {2023},
  month = {Feb},
  publisher = {American Physical Society},
  doi = {10.1103/PhysRevResearch.5.L012015},
  url = {https://link.aps.org/doi/10.1103/PhysRevResearch.5.L012015}
}

@article{FCIQuantumMetricLiuZhaoPRL2020,
  title = {Particle-Hole Duality, Emergent Fermi Liquids, and Fractional Chern Insulators in Moir\'e Flatbands},
  author = {Abouelkomsan, Ahmed and Liu, Zhao and Bergholtz, Emil J.},
  journal = {Phys. Rev. Lett.},
  volume = {124},
  issue = {10},
  pages = {106803},
  numpages = {6},
  year = {2020},
  month = {Mar},
  publisher = {American Physical Society},
  doi = {10.1103/PhysRevLett.124.106803},
  url = {https://link.aps.org/doi/10.1103/PhysRevLett.124.106803}
}

@article{FCIQuantumMetricYangBo2024,
  title={Quantum metric induced hole dispersion and emergent particle-hole symmetry in topological flat bands},
  author={Ji, Guangyue and Yang, Bo},
  journal={arXiv:2409.08324},
  url={https://arxiv.org/abs/2409.08324},
  year={2024}
}

@article{FlatChernTheorem,
doi = {10.1088/1751-8113/47/15/152001},
url = {https://dx.doi.org/10.1088/1751-8113/47/15/152001},
year = {2014},
month = {mar},
publisher = {IOP Publishing},
volume = {47},
number = {15},
pages = {152001},
author = {Li Chen and Tahereh Mazaheri and Alexander Seidel and Xiang Tang},
title = {The impossibility of exactly flat non-trivial Chern bands in strictly local periodic tight binding models},
journal = {J. Phys. A: Math. Theor.}
}

@article{BandTouchingBalents2008,
  title = {Band touching from real-space topology in frustrated hopping models},
  author = {Bergman, Doron L. and Wu, Congjun and Balents, Leon},
  journal = {Phys. Rev. B},
  volume = {78},
  issue = {12},
  pages = {125104},
  numpages = {9},
  year = {2008},
  month = {Sep},
  publisher = {American Physical Society},
  doi = {10.1103/PhysRevB.78.125104},
  url = {https://link.aps.org/doi/10.1103/PhysRevB.78.125104}
}

@article{SinguFlatWaveFunctPRB2021,
  title = {Wave-function geometry of band crossing points in two dimensions},
  author = {Hwang, Yoonseok and Jung, Junseo and Rhim, Jun-Won and Yang, Bohm-Jung},
  journal = {Phys. Rev. B},
  volume = {103},
  issue = {24},
  pages = {L241102},
  numpages = {6},
  year = {2021},
  month = {Jun},
  publisher = {American Physical Society},
  doi = {10.1103/PhysRevB.103.L241102},
  url = {https://link.aps.org/doi/10.1103/PhysRevB.103.L241102}
}

@article{SinguFlatClassificationPRB2019,
  title = {Classification of flat bands according to the band-crossing singularity of Bloch wave functions},
  author = {Rhim, Jun-Won and Yang, Bohm-Jung},
  journal = {Phys. Rev. B},
  volume = {99},
  issue = {4},
  pages = {045107},
  numpages = {21},
  year = {2019},
  month = {Jan},
  publisher = {American Physical Society},
  doi = {10.1103/PhysRevB.99.045107},
  url = {https://link.aps.org/doi/10.1103/PhysRevB.99.045107}
}

@Article{SinguFlatCommunPhys2023,
author={Kim, Hyeongseop
and Oh, Chang-geun
and Rhim, Jun-Won},
title={General construction scheme for geometrically nontrivial flat band models},
journal={Commun. Phys.},
year={2023},
month={Oct},
day={18},
volume={6},
number={1},
pages={305},
issn={2399-3650},
doi={10.1038/s42005-023-01407-6},
url={https://doi.org/10.1038/s42005-023-01407-6}
}

@article{SinguFlatAdvPhysX2021,
author = {Jun-Won Rhim and Bohm-Jung Yang},
title = {Singular flat bands},
journal = {Adv. Phys.: X},
volume = {6},
number = {1},
pages = {1901606},
year = {2021},
publisher = {Taylor \& Francis},
doi = {10.1080/23746149.2021.1901606},
URL = {https://doi.org/10.1080/23746149.2021.1901606}
}

@article{JainCurvatureGap,
  title = {Adiabatic continuity between Hofstadter and Chern insulator states},
  author = {Wu, Ying-Hai and Jain, J. K. and Sun, Kai},
  journal = {Phys. Rev. B},
  volume = {86},
  issue = {16},
  pages = {165129},
  numpages = {13},
  year = {2012},
  month = {Oct},
  publisher = {American Physical Society},
  doi = {10.1103/PhysRevB.86.165129},
  url = {https://link.aps.org/doi/10.1103/PhysRevB.86.165129}
}

@article{WenqiPRL2025,
  title = {Fractional Quantum Anomalous Hall Effect in a Singular Flat Band},
  author = {Yang, Wenqi and Zhai, Dawei and Tan, Tixuan and Fan, Feng-Ren and Lin, Zuzhang and Yao, Wang},
  journal = {Phys. Rev. Lett.},
  volume = {134},
  issue = {19},
  pages = {196501},
  numpages = {7},
  year = {2025},
  month = {May},
  publisher = {American Physical Society},
  doi = {10.1103/PhysRevLett.134.196501},
  url = {https://link.aps.org/doi/10.1103/PhysRevLett.134.196501}
}

@article{FCIFarFromIdealPRL2024,
  title = {Quantum Geometry and Stabilization of Fractional Chern Insulators Far from the Ideal Limit},
  author = {Shavit, Gal and Oreg, Yuval},
  journal = {Phys. Rev. Lett.},
  volume = {133},
  issue = {15},
  pages = {156504},
  numpages = {6},
  year = {2024},
  month = {Oct},
  publisher = {American Physical Society},
  doi = {10.1103/PhysRevLett.133.156504},
  url = {https://link.aps.org/doi/10.1103/PhysRevLett.133.156504}
}

@article{FCIgradient2025,
  title={Gradient-based search of quantum phases: discovering unconventional fractional Chern insulators},
  author={Fonseca, Andr{\'e} Grossi and Wang, Eric and Vaidya, Sachin and Ledwith, Patrick J and Vishwanath, Ashvin and Solja{\v{c}}i{\'c}, Marin},
  journal={arXiv:2509.10438},
  url={https://arxiv.org/abs/2509.10438v1},
  year={2025}
}

@Article{FCIC=0Lin2025,
author={Lin, Zuzhang
and Lu, Hongyu
and Yang, Wenqi
and Zhai, Dawei
and Yao, Wang},
title={Fractional Chern insulator states in an isolated flat band of zero Chern number},
journal={Newton},
year={2026},
publisher={Elsevier},
volume = {2},
issue = {3},
pages = {100339},
issn={2950-6360},
doi={10.1016/j.newton.2025.100339},
url={https://doi.org/10.1016/j.newton.2025.100339}
}

@article{FCIC=0Lu2025,
  title={Bosonic Laughlin and Moore-Read states from non-Chern flat bands},
  author={Lu, Hongyu and Yao, Wang},
  journal={arXiv:2510.14685},
  url={https://arxiv.org/abs/2510.14685},
  year={2025}
}

@article{FCIC=0Bergholtz2025,
  title = {Topological order without band topology in moir\'e graphene},
  author = {Liu, Hui and Perea-Causin, Raul and Liu, Zhao and Bergholtz, Emil J.},
  journal = {Phys. Rev. Res.},
  volume = {8},
  issue = {3},
  pages = {L032039},
  numpages = {8},
  year = {2026},
  month = {Sep},
  publisher = {American Physical Society},
  doi = {10.1103/jx5w-35z9},
  url = {https://link.aps.org/doi/10.1103/jx5w-35z9}
}

@article{MultiBandEDPRB2025,
  title = {Moir\'e fractional Chern insulators. IV. Fluctuation-driven collapse in multiband exact diagonalization calculations on rhombohedral graphene},
  author = {Yu, Jiabin and Herzog-Arbeitman, Jonah and Kwan, Yves H. and Regnault, Nicolas and Bernevig, B. Andrei},
  journal = {Phys. Rev. B},
  volume = {112},
  issue = {7},
  pages = {075110},
  numpages = {73},
  year = {2025},
  month = {Aug},
  publisher = {American Physical Society},
  doi = {10.1103/PhysRevB.112.075110},
  url = {https://link.aps.org/doi/10.1103/PhysRevB.112.075110}
}

@article{TiltedRegnaultPRB2014,
  title = {${\mathbb{Z}}_{2}$ fractional topological insulators in two dimensions},
  author = {Repellin, C. and Bernevig, B. Andrei and Regnault, N.},
  journal = {Phys. Rev. B},
  volume = {90},
  issue = {24},
  pages = {245401},
  numpages = {14},
  year = {2014},
  month = {Dec},
  publisher = {American Physical Society},
  doi = {10.1103/PhysRevB.90.245401},
  url = {https://link.aps.org/doi/10.1103/PhysRevB.90.245401}
}

@article{TiltedBergholtzPRL2013,
  title = {Hierarchy of Fractional Chern Insulators and Competing Compressible States},
  author = {L\"auchli, A. M. and Liu, Zhao and Bergholtz, E. J. and Moessner, R.},
  journal = {Phys. Rev. Lett.},
  volume = {111},
  issue = {12},
  pages = {126802},
  numpages = {5},
  year = {2013},
  month = {Sep},
  publisher = {American Physical Society},
  doi = {10.1103/PhysRevLett.111.126802},
  url = {https://link.aps.org/doi/10.1103/PhysRevLett.111.126802}
}

@article{iDMRG,
  title={Infinite size density matrix renormalization group, revisited},
  author={McCulloch, Ian P},
  journal={arXiv:0804.2509},
  url={https://arxiv.org/abs/0804.2509},
  year={2008}
}

@article{HybridizeDMRG,
  title = {Density matrix renormalization group on a cylinder in mixed real and momentum space},
  author = {Motruk, Johannes and Zaletel, Michael P. and Mong, Roger S. K. and Pollmann, Frank},
  journal = {Phys. Rev. B},
  volume = {93},
  issue = {15},
  pages = {155139},
  numpages = {9},
  year = {2016},
  month = {Apr},
  publisher = {American Physical Society},
  doi = {10.1103/PhysRevB.93.155139},
  url = {https://link.aps.org/doi/10.1103/PhysRevB.93.155139}
}

@article{GMPalgebra1986,
  title = {Magneto-roton theory of collective excitations in the fractional quantum Hall effect},
  author = {Girvin, S. M. and MacDonald, A. H. and Platzman, P. M.},
  journal = {Phys. Rev. B},
  volume = {33},
  issue = {4},
  pages = {2481--2494},
  numpages = {0},
  year = {1986},
  month = {Feb},
  publisher = {American Physical Society},
  doi = {10.1103/PhysRevB.33.2481},
  url = {https://link.aps.org/doi/10.1103/PhysRevB.33.2481}
}

@article{SongZhidaGaplessTopology,
  title={Stable Topology in Exactly Flat Bands},
  author={Li, Yan-Qi and Wang, Yi-Jie and Lin, Pei-Han and Wang, Bin and Song, Zhi-Da},
  journal={arXiv:2603.12258},
  url={https://arxiv.org/abs/2603.12258v2},
  year={2026}
}

@article{FangChenGaplessTopology,
  title = {General Construction from Topological Loop States to Topological Invariants in Exactly Flat Bands},
  author = {Liu, Rui-Heng and Hu, Jiangping and Fang, Chen},
  journal = {Phys. Rev. Lett.},
  volume = {137},
  issue = {14},
  pages = {146601},
  numpages = {9},
  year = {2026},
  month = {Sep},
  publisher = {American Physical Society},
  doi = {10.1103/mgd4-jnts},
  url = {https://link.aps.org/doi/10.1103/mgd4-jnts}
}

@article{HeavyFermionGapless,
  title={Topological heavy fermion principle for flat (narrow) bands with concentrated quantum geometry},
  author={Herzog-Arbeitman, Jonah and Yu, Jiabin and C{\u{a}}lug{\u{a}}ru, Dumitru and Hu, Haoyu and Regnault, Nicolas and Liu, Chaoxing and Vafek, Oskar and Coleman, Piers and Tsvelik, Alexei and Song, Zhi-da and others},
  journal={arXiv:2404.07253},
  url={https://arxiv.org/abs/2404.07253},
  year={2024}
}

@article{AnisotropicLLYangBo2012,
  title = {Band mass anisotropy and the intrinsic metric of fractional quantum Hall systems},
  author = {Yang, Bo and Papi\ifmmode \acute{c}\else \'{c}\fi{}, Z. and Rezayi, E. H. and Bhatt, R. N. and Haldane, F. D. M.},
  journal = {Phys. Rev. B},
  volume = {85},
  issue = {16},
  pages = {165318},
  numpages = {9},
  year = {2012},
  month = {Apr},
  publisher = {American Physical Society},
  doi = {10.1103/PhysRevB.85.165318},
  url = {https://link.aps.org/doi/10.1103/PhysRevB.85.165318}
}

@article{AnisotropicLLZhangFC2012,
  title = {Fractional quantum Hall states in two-dimensional electron systems with anisotropic interactions},
  author = {Wang, Hao and Narayanan, Rajesh and Wan, Xin and Zhang, Fuchun},
  journal = {Phys. Rev. B},
  volume = {86},
  issue = {3},
  pages = {035122},
  numpages = {7},
  year = {2012},
  month = {Jul},
  publisher = {American Physical Society},
  doi = {10.1103/PhysRevB.86.035122},
  url = {https://link.aps.org/doi/10.1103/PhysRevB.86.035122}
}

@article{Wannier1997,
  title = {Maximally localized generalized Wannier functions for composite energy bands},
  author = {Marzari, Nicola and Vanderbilt, David},
  journal = {Phys. Rev. B},
  volume = {56},
  issue = {20},
  pages = {12847--12865},
  numpages = {0},
  year = {1997},
  month = {Nov},
  publisher = {American Physical Society},
  doi = {10.1103/PhysRevB.56.12847},
  url = {https://link.aps.org/doi/10.1103/PhysRevB.56.12847}
}

@Article{QuantumGeometryNatRevPhys2026,
author={Verma, Nishchhal
and Moll, Philip J. W.
and Holder, Tobias
and Queiroz, Raquel},
title={Quantum geometry and the hidden scales in materials},
journal={Nat. Rev. Phys.},
year={2026},
month={Apr},
day={01},
volume={8},
number={4},
pages={226-239},
issn={2522-5820},
doi={10.1038/s42254-026-00923-y},
url={https://doi.org/10.1038/s42254-026-00923-y}
}



\clearpage 

\renewcommand{\thefigure}{S\arabic{figure}}       
\renewcommand{\thesection}{Supplementary Sec.~\arabic{section}}
\renewcommand{\thepage}{S\arabic{page}}
\renewcommand{\theequation}{S\arabic{equation}}
\setcounter{figure}{0}
\setcounter{section}{0}
\setcounter{page}{1}
\setcounter{equation}{0}

\title{Supplemental Materials for ``Fractional quantization in gapless flat bands with ill-defined topology and divergent quantum geometry''}
\maketitle
\onecolumngrid
\tableofcontents


\section{Detailed methods for many-body numerical simulations}

\subsection{Density matrix renormalization group (DMRG)}
The charge pumping simulation and the phase diagrams presented in Fig.~2(a) and Fig.~5(a) of the main text are obtained via real-space DMRG simulations, where the maximal bond dimension is set to $D = 400$. The lattice  is configured on a finite cylinder with open (periodic) boundary conditions along the $x$ ($y$) direction. The total number of lattice sites is given by $N_x\times N_y\times2\,(3)$ for the honeycomb (kagome) model, where $N_x$ ($N_y$) denotes the number of unit cells along the $x$ ($y$) direction. For the honeycomb model, we set $N_x=24$ and $N_y=6$, while for the kagome model, we use $N_x = 18$ and $N_y=4$. 

The entanglement entropy, the first-order energy derivative and the CDW order parameter across the phase transition point, as shown in Fig.~2(c) of the main text, are calculated using iDMRG simulations~\cite{iDMRG}. In the iDMRG simulations, the cylinder is infinite along the $x$ direction and consists of 4 cells in the $y$ direction. The bond dimension is increased to $D=800$ to ensure convergence across all parameter values.

The momentum-resolved entanglement spectra shown in Fig.~2(b) of the main text and Fig.~\ref{FigSupp:Kagome}(a) are obtained through DMRG simulations performed in a mixed real ($x$ direction) and momentum ($y$ direction) space, where each site is labeled by the pair $\{x,\,k_y\}$~\cite{HybridizeDMRG}. The momentum vector, $k_y$, around the cylinder is used as a conserved quantity, allowing the entanglement spectrum to be categorized into distinct momentum sectors. For both the honeycomb and kagome models, the lattices we use consist of 6 cells along the $y$ direction, with $k_y$ values belonging to $\{\frac{\pi}{6},\frac{\pi}{3},\cdots,2\pi\}$, and 60 cells along the $x$ direction. The maximal bond dimension is set to $D = 800$.

All the DMRG simulations in this paper are performed using the ITensor library with U(1) symmetry~\cite{ITensor}.


\subsection{Exact diagonalization (ED)}

We consider nearest-neighbor interactions in the tight-binding models, with the interaction strength characterized by the constant $U$. In momentum space, the interaction term takes the following form:
\begin{equation}\label{HI_orbital}
	\hat{H}_I=\sum_{\bk_1-\bk_4\in \text{BZ},\,\boldsymbol{\delta}_m}\frac{U}{N}B_{\bk_1}^{\dagger}C_{\bk_2}^{\dagger}C_{\bk_4}B_{\bk_3}\delta_{\lceil\bk_1+\bk_2-\bk_3-\bk_4\rceil,0}e^{i(\bk_4-\bk_2)\cdot\boldsymbol{\delta}_m}+\cdots.
\end{equation}
For the honeycomb model, ~\eqref{HI_orbital} only includes the interaction term between B-C orbitals, while for the kagome model, ~\eqref{HI_orbital} also contains interaction terms between A-B, A-C orbitals [denoted by $\cdots$ in ~\eqref{HI_orbital}]. $N$ denotes the number of unit cells and the operator $\lceil\cdots\rceil$ is employed to project its argument back into the Brillouin Zone (BZ). $\boldsymbol{\delta}_m$ is the vector connecting the two unit cells hosing neighboring orbitals, where the subscript $m$ indicates the three different orientations.

After diagonalizing the single-particle Hamiltonian, we obtain a transformation matrix $\Cal{U}$ that maps the atomic orbital basis to the single-particle Bloch band basis. The relation between the creation operators in the two bases is given by:
\begin{equation}
	\gamma_{\bk,n}^{\dagger} = \sum_\alpha \Cal{U}_{\alpha, n}(\bk)\alpha_{\bk} ^\dagger,
\end{equation}
where $\alpha$ is the orbital index and $n$ is the band index. $n=1,\,2$ for the honeycomb model and $n=1,\,2,\,3$ for the kagome model. 

In the single-particle band basis, the interaction Hamiltonian reads
\begin{equation}\label{ED-HI}
	\hat{H}_{I}=\sum_{\substack{\bm{k}_1-\bm{k}_4 \in \text{BZ},\,\bm{\delta}_m,\\ n_1,n_2,n_3,n_4}}\frac{U_1}{N}\gamma^{\dagger}_{\bm{k}_1,n_1}\gamma^{\dagger}_{\bm{k}_2,n_2}\gamma_{\bm{k}_4,n_4}\gamma_{\bm{k}_3,n_3}\delta_{\lceil \bm{k}_1+\bm{k}_2-\bm{k}_3-\bm{k}_4\rceil ,0}e^{i(\bm{k}_4-\bm{k}_2)\cdot \bm{\delta}_m}\Cal{U}^*_{B,n_1}\Cal{U}^*_{C,n_2} \Cal{U}_{C,n_4}\Cal{U}_{B,n_3}+\cdots,
\end{equation}
and the single-particle Hamiltonian reads
\begin{equation}\label{ED-H0}
	\hat{H}_0=\sum_{\bk\in \text{BZ},\,n} \varepsilon_n(\bk)\gamma_{\bk,n}^{\dagger}\gamma_{\bk,n}.
\end{equation}
In the honeycomb model, both energy bands are considered; whereas for the kagome model, due to the high energy of the topmost band [Fig.~4(b) of the main text], the Hilbert space is projected onto the two lower touching bands. Consequently, the summation over the band index $n$ in ~\eqref{ED-H0} and ~\eqref{ED-HI} is restricted to $n=1,\,2$ for both models. 

We consider different system configurations. The first type of $\bk$ grid is the $N_1 \times N_2=4\times6$ rectangular grid, where each $\bk$ point is labeled by the integer $k_1+N_1k_2$. $k_1$ and $k_2$ are determined by the following expression
\begin{equation}
	\bk = \frac{k_1}{N_1}\boldsymbol{g}_1+\frac{k_2}{N_2}\boldsymbol{g}_2, ~~~k_1=1,\cdots,N_1~\text{and}~k_2=1,\cdots,N_2,
\end{equation}
where $\boldsymbol{g}_i$ are primitive reciprocal lattice vectors.

\begin{figure}[t]
	\centering
	\includegraphics[width=4.4in]{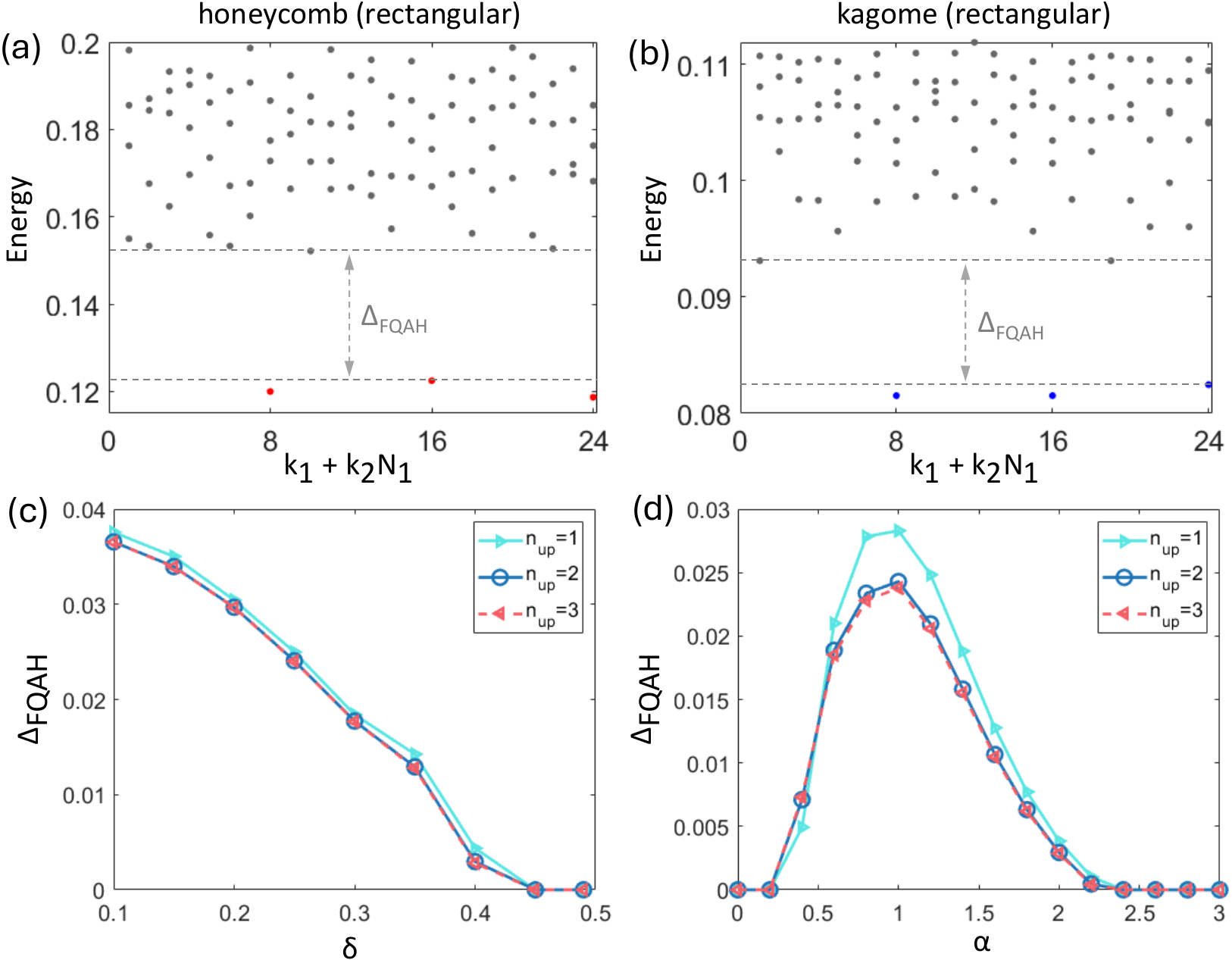}
	\caption{(a, b) Many-body energy spectra computed on a $4\times6$ rectangular $\bk$ grid for the honeycomb model at $\delta=0.2$ and the kagome model at $\alpha=1.6$, respectively. The three nearly degenerate ground states are highlighted in red and blue, respectively. (c, d) Plots of the many-body gap of the FQAH phase, $\Delta_{\rm FQAH}$, as a function of $\delta$ (honeycomb model) or $\alpha$ (kagome model) with respect to $n_{\rm up}$ ranging from 1 to 3. The interaction strength $U$ is set to 1 for both models.}~\label{FigSupp:GapCoverge}
\end{figure}

To reduce computational costs, we adopt the "band maximum" approximation introduced in Ref.~\cite{MultiBandEDPRB2025}, where the number of particles in the upper dispersive band is capped at $n_{\rm up}$, while it is unrestricted in the singular flat band (SFB). After diagonalizing the many-body Hamiltonian $\hat{H}_0+\hat{H}_I$ at $\nu=1/3$ filling of the SFB, we observe three nearly degenerate ground states at momenta $\bs{\Cal{K}}_{\rm rect}=\{8,\,16,\,24\}$ within the FQAH parameter region [see Fig.~3(a) of the main text and Figs.~\ref{FigSupp:GapCoverge}(a, b)], consistent with the degeneracy pattern of the Laughlin states and the generalized Pauli exclusion principle~\cite{regnault2011fractional}. 
In the rectangular configuration, $\bs{\Cal{K}}_{\rm rect}$ are the FQAH momenta mentioned in the main text.

If the three lowest-energy states in the $\bs{\Cal{K}}_{\rm rect}$ momentum sectors are the ground states, and if they also have nonzero many-body Chern numbers, we define the many-body gap of the FQAH phase $\Delta_{\rm FQAH}$ as their energy separation to the fourth-lowest energy [Figs.~\ref{FigSupp:GapCoverge}(a, b)]. If the lowest-energy states in the $\bs{\Cal{K}}_{\rm rect}$ momentum sectors are not the ground states and/or if their many-body Chern numbers vanish, the system is in a non-FQAH trivial phase, for which $\Delta_{\rm FQAH}=0$.

To assess the convergence of the results with respect to $n_{\rm up}$, we set $U=1$ and plot the many-body gap $\Delta_{\rm FQAH}$ as a function of $\delta$ in the honeycomb model (or as a function of $\alpha$ in the kagome model) for various values of $n_{\rm up}$. As illustrated in Figs.~\ref{FigSupp:GapCoverge}(c, d), for both the honeycomb and kagome models, the differences between the results at $n_{\rm up}=2$ and $n_{\rm up}=3$ are negligible, indicating satisfactory convergence is achieved at $n_{\rm up}=2$. The ED results presented in the main text are all obtained with $n_{\rm up}=2$.
The many-body gap $\Delta_{\rm FQAH}$ and occupation-weighted geometric quantities $\braket{\Omega}_\text{occ}$, $\braket{\text{tr}\CalG}_\text{occ}$ and $\braket{\text{T}}_\text{occ}$ [see Fig.~3(b) and Fig.~5(b) of the main text] were computed using the $4\times6$ rectangular $\bk$ grid.

\begin{figure}[h!]
	\centering
	\includegraphics[width=4.4in]{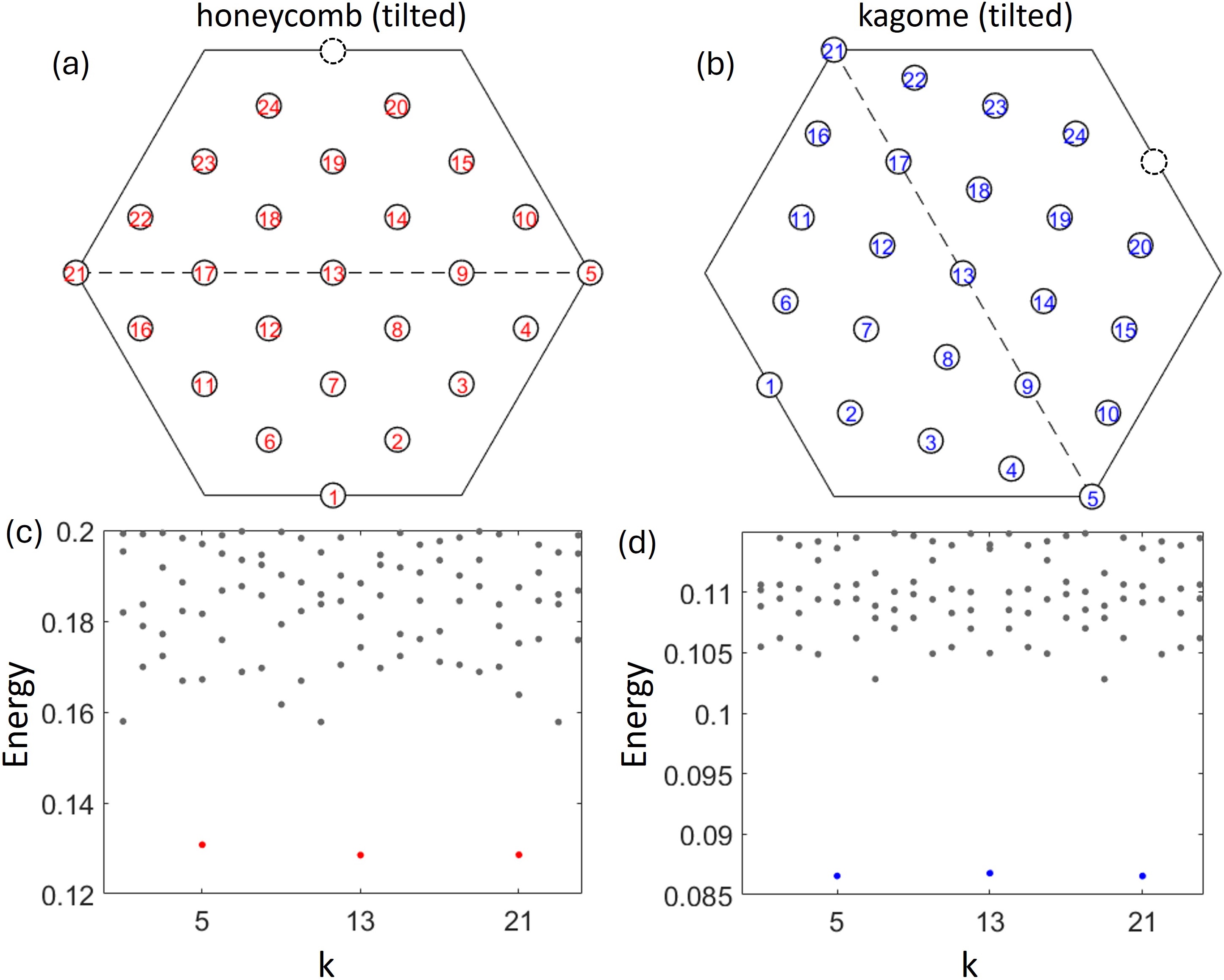}
	\caption{(a, b) Illustration of the $1\times24$ tilted $\bk$ grids for the honeycomb lattice and the kagome lattice, respectively, with all $\bk$ points indexed. The $\bk$ grids are mirror symmetric with respect to the dashed lines. The $\bk$ points marked by the dashed circles are equivalent to the $\bk$ points indexed by 1 under tilted boundary conditions. (c, d) Many-body energy spectra computed on the $1\times24$ tilted $\bk$ grids at $\delta=0.2$ for the honeycomb model and at $\alpha =1.6$ for the kagome model, respectively. $U$ is fixed at 1, and $n_{\rm up}$ is set to 2.}~\label{FigSupp:kmesh24}
\end{figure}

We also consider an alternative $1\times24$ tilted $\bk$ grid~\cite{TiltedBergholtzPRL2013,TiltedRegnaultPRB2014} to examine the dependence of the results on the system configuration [Figs.~\ref{FigSupp:kmesh24}(a, b)]. The corresponding many-body energy spectra are displayed in Figs.~\ref{FigSupp:kmesh24}(c, d) for the honeycomb and kagome model, respectively. We observe FQAH states with topological many-body gaps similar to those in the $4\times6$ rectangular case [Figs.~\ref{FigSupp:GapCoverge}(a, b)], where the three nearly degenerate ground states appear in the momentum sectors $\bs{\Cal{K}}_{\rm tilt}=\{5,\,13,\,21\}$, consistent with the generalized Pauli exclusion principle. 
The $1\times24$ tilted $\bk$ grids are mirror symmetric with respect to the dashed lines [Figs.~\ref{FigSupp:kmesh24}(a, b)], sharing the same symmetry as that of the quantum geometry of the models. Therefore, we adopt the $1\times24$ tilted $\bk$ grids for presenting the carrier occupation $n(\bk)$ in the main text [Figs.~3(c--e) and Figs.~5(c--e)]. The results of $n(\bk)$ shown in Figs.~3(c--e) and Figs.~5(c--e) are averaged over the three lowest states with momenta $\bs{\Cal{K}}_{\rm tilt}$.

\begin{figure}[h!]
	\centering
	\includegraphics[width=7in]{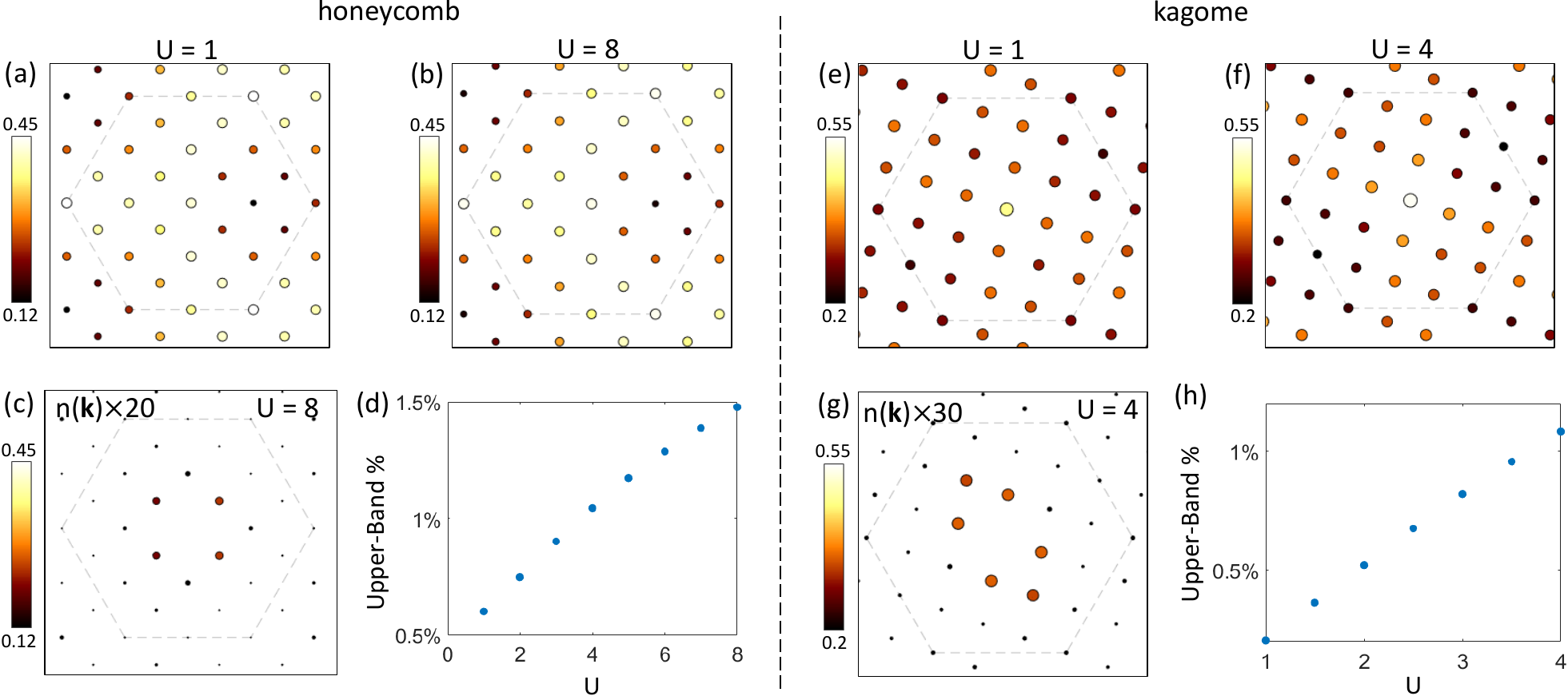}
	\caption{(a, b) Charge distribution $n(\bk)$ of the ground states in $\bk$ space at $U=1$ and $U=8$ respectively for the honeycomb model with $\delta=0.3$. (c) Occupation of the upper band at $U=8$ excluding the $\Gamma$ point. The values are amplified by a factor of 20. (d) Percentage of the states from the upper band contributing to the FQAH states at various $U$ values. (e, f) Charge distribution $n(\bk)$ of the ground states in $\bk$ space at $U=1$ and $U=4$ respectively for the kagome model with $\alpha=1$. (g) Occupation of the upper band at $U=4$ excluding the $\Gamma$ point. The values are amplified by a factor of 30. (h) Percentage of the states from the upper band contributing to the FQAH states at various $U$ values. $n_{\rm up}$ is set to 3 here.}~\label{FigSupp:BandMixHoneyKagome}
\end{figure}

We also examine the effect of the interaction strength $U$ on the charge distribution $n(\bk)$ of the FQAH states. Taking the honeycomb model as an example, the $n(\bk)$ for $U=8$ (comparable to the width of the upper dispersive band) is nearly identical to that at $U=1$ [Figs.~\ref{FigSupp:BandMixHoneyKagome}(a, b)], indicating that the interaction strength has weak influence on the charge distribution. Due to the degeneracy at the $\Gamma$ point, occupation in the SFB and the upper band cannot be unambiguously determined. The upper-band occupation---excluding the $\Gamma$ point---at $U=8$ is shown in Fig.~\ref{FigSupp:BandMixHoneyKagome}(c), with the values amplified by a factor of 20. The total contribution of the upper band (excluding the $\Gamma$ point) to the FQAH state at various $U$ values is displayed in Fig.~\ref{FigSupp:BandMixHoneyKagome}(d). We observe that the contribution from the upper band (excluding the $\Gamma$ point) remains quite small, $\Cal{O}(1\%)$, even at large $U$ values.
Qualitatively similar behaviors are observed for the Kagome model [Figs.~\ref{FigSupp:BandMixHoneyKagome}(e--h)].
Although the weight of the upper band away from the $\Gamma$ point is small, band mixing is non-negligible and important near $\Gamma$: Setting $n_{\rm up}=0$ results in a many-body spectrum that is completely different from the $n_{\rm up}\ne0$ case (see e.g., $n_{\rm up}=2$ in Fig.~\ref{FigSupp:1Bandvs2BandEDHoney} and Fig.~\ref{FigSupp:1Bandvs2BandEDKagome}), the latter can well reproduce the two-band ED results without constraining the number of particles in the upper band ($n_{\rm up}>2$ is needed when $U$ becomes too large).

In the main text, we also present results that involve the charge distribution $n(\bk)$ of the trivial phases [Fig.~3(b) when $\delta\gtrsim0.43$, Figs.~5(b) when $\alpha\lesssim0.4$ or $\alpha\gtrsim2.4$]. In these cases, we also consider the three lowest states with momenta $\bs{\Cal{K}}_{\rm rect}$ in the rectangular geometry or $\bs{\Cal{K}}_{\rm tilt}$ in the tilted geometry. Using states with other momenta yields quantitatively similar results.


\section{Extra details on the flat band in the honeycomb model: from critical topology to maximal strength singularity}

\subsection{Connection to the fluxed dice lattice model and twisted bilayer MoTe$_2$}

\begin{figure}[h!]
	\centering
	\includegraphics[width=4.25in]{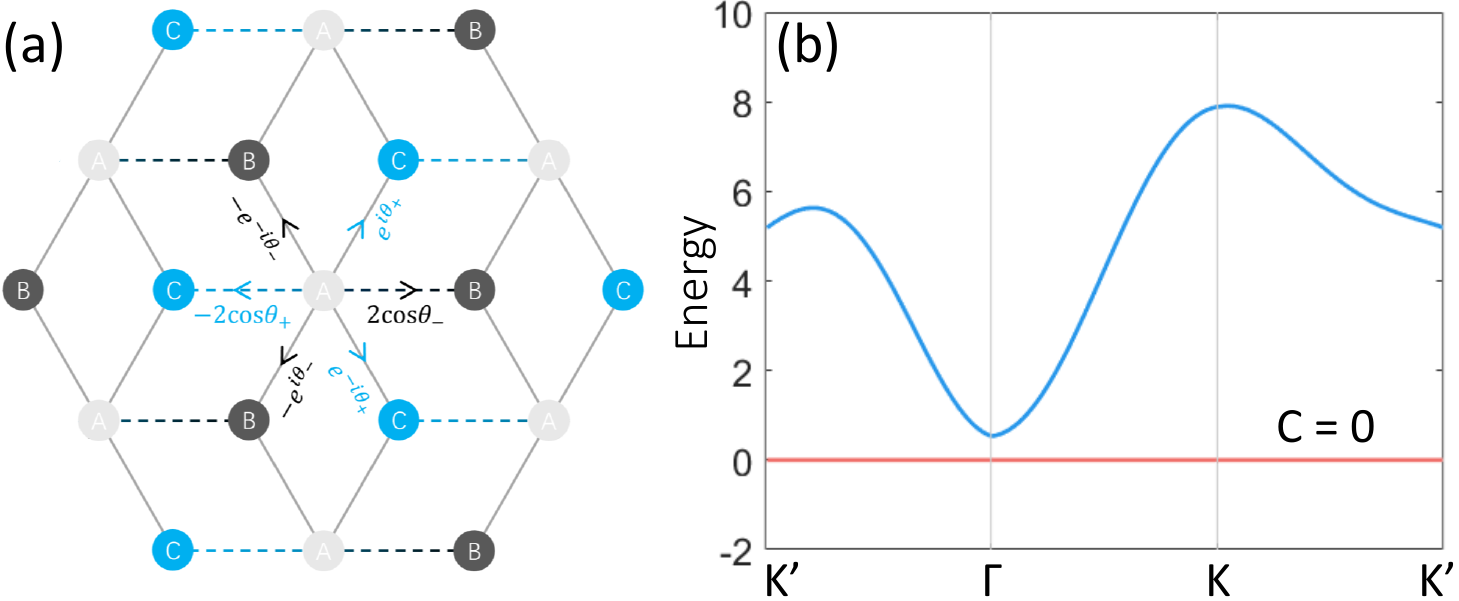}
	\caption{(a) Schematics of the Dice lattice model. (b) A representative example of obtaining an isolated trivial exact flat band in the model of $\hat{H}_{\tinyhex}(\bk)$ with either finite or zero maximal quantum distance $d_{\rm max}$. For the figure, $d_{\rm max}=0.44$ at $\delta=0.2$ is used, and $\eta=0.5$.}~\label{FigSupp:DiceGappedTrivialBand}
\end{figure}

The honeycomb lattice model $\hat{H}_{\tinyhex}(\bk)$ is built upon the $2\pi$-fluxed dice lattice model with three orbitals~\cite{WenqiPRL2025}:
\begin{equation}
	\hat{H}_{\rm dice}=
	\begin{pmatrix}
		\epsilon_A&f(\bk)&g(\bk)\\
		f^{*}(\bk)&0&0\\
		g^{*}(\bk)&0&0
	\end{pmatrix},
\end{equation}
where the orbitals are denoted as A, B and C.
We generalize the model in Ref.~\cite{WenqiPRL2025} by introducing a parameter $\delta$ in $f(\bk)$ and $g(\bk)$ [see Fig.~\ref{FigSupp:DiceGappedTrivialBand}(a)]:
\begin{equation}
	\begin{aligned}
		f(\bk) &=-t\left(2\cos\theta_-e_1 - e^{i\theta_-} e_2 - e^{-i\theta_-}  e_3\right)\\
		g(\bk) &=-t\left(-2\cos\theta_+e_1^{*} + e^{i\theta_+} e_2^{*} + e^{-i\theta_+}  e_3^{*}\right)
	\end{aligned},
\end{equation}
where $\theta_\pm=-\pi/3\pm\delta$, $e_i=e^{-i\bk\cdot\bd_i}$ and $\bd_{1,2,3}$ are the nearest-neighbor vectors. The model in Ref.~\cite{WenqiPRL2025} is recovered when $\delta=0$, while $\delta\ne0$ adjusts the hopping phases and amplitudes, and more importantly, the quantum geometric properties.
We note that with $\delta=0$, the model $\hat{H}_{\rm dice}$ corresponds to the bipartite limit of the three-orbital tight-binding model of twisted bilayer MoTe$_2$~\cite{HongyiNSR2020} (where hopping between B and C orbitals is finite but weak as they are localized in opposite layers). Experimentally, the scenario of $\delta\ne0$ can potentially be realized by applying strain to the system of $\delta=0$. The model can also possibly be realized with artificial lattices of cold atoms with anisotropic potential landscapes~\cite{WenqiPRL2025}.

In the large $\epsilon_A$ limit, one arrives at a two-orbital model~\cite{WenqiPRL2025}:
\begin{equation}\label{H-honey}
	\hat{H}'=
	-\epsilon_A^{-1}\begin{pmatrix} 
		|f|^2&gf^{*}\\
		fg^{*}&|g|^2 
	\end{pmatrix},
\end{equation}
which reproduces the properties of the two lower touching bands of $\hat{H}_{\rm dice}$ contributed by the B and C orbitals. The honeycomb lattice model $\hat{H}_{\tinyhex}(\bk)$ in the main text is equivalent to $\hat{H}'$.

\subsection{Stable band topology at $\delta=0$}

By varying the value of $\delta$, the flat band in the dice or honeycomb models can switch from a gapless Chern band with well-defined Chern number $C=1$ to SFB with ill-defined topology. In the main text, we have focused on the singular aspects of the model at $\delta\ne0$. Here we elaborate on the stable topology at $\delta=0$~\cite{SongZhidaGaplessTopology,FangChenGaplessTopology}. 

First, we consider the momentum-space perspective by repeating the discussions in Ref.~\cite{WenqiPRL2025} and show that the flat band has stable topology~\cite{SongZhidaGaplessTopology}. Around the $\Gamma$ point, $f(\bk\rightarrow0)=g(\bk\rightarrow0)=-ie^{i\phi}$, where $\phi$ is the polar angle of $\bk$. Therefore, the Bloch state on the flat band $\psi_{0}^{\rm dice}=(0,g,-f)^T/\sqrt{|f|^2+|g|^2}$ in dice model and $\psi_{0}^{\tinyhex}=(g,-f)^T/\sqrt{|f|^2+|g|^2}$ in the honeycomb model are continuous (up to the phase $e^{i\phi}$) around the $\Gamma$ point. This leads to a vanishing maximal quantum distance $d_{\rm max}=0$. The continuous Bloch states also renders the projector on the flat band well-defined throughout the BZ, thus the Berry curvature and Chern number. The phase winding of $e^{i\phi}$ leads to $\Phi_{\tinyCW}=1$, which is nothing but the Chern number of the flat band.

\begin{figure}[t]
	\centering
	\includegraphics[width=5in]{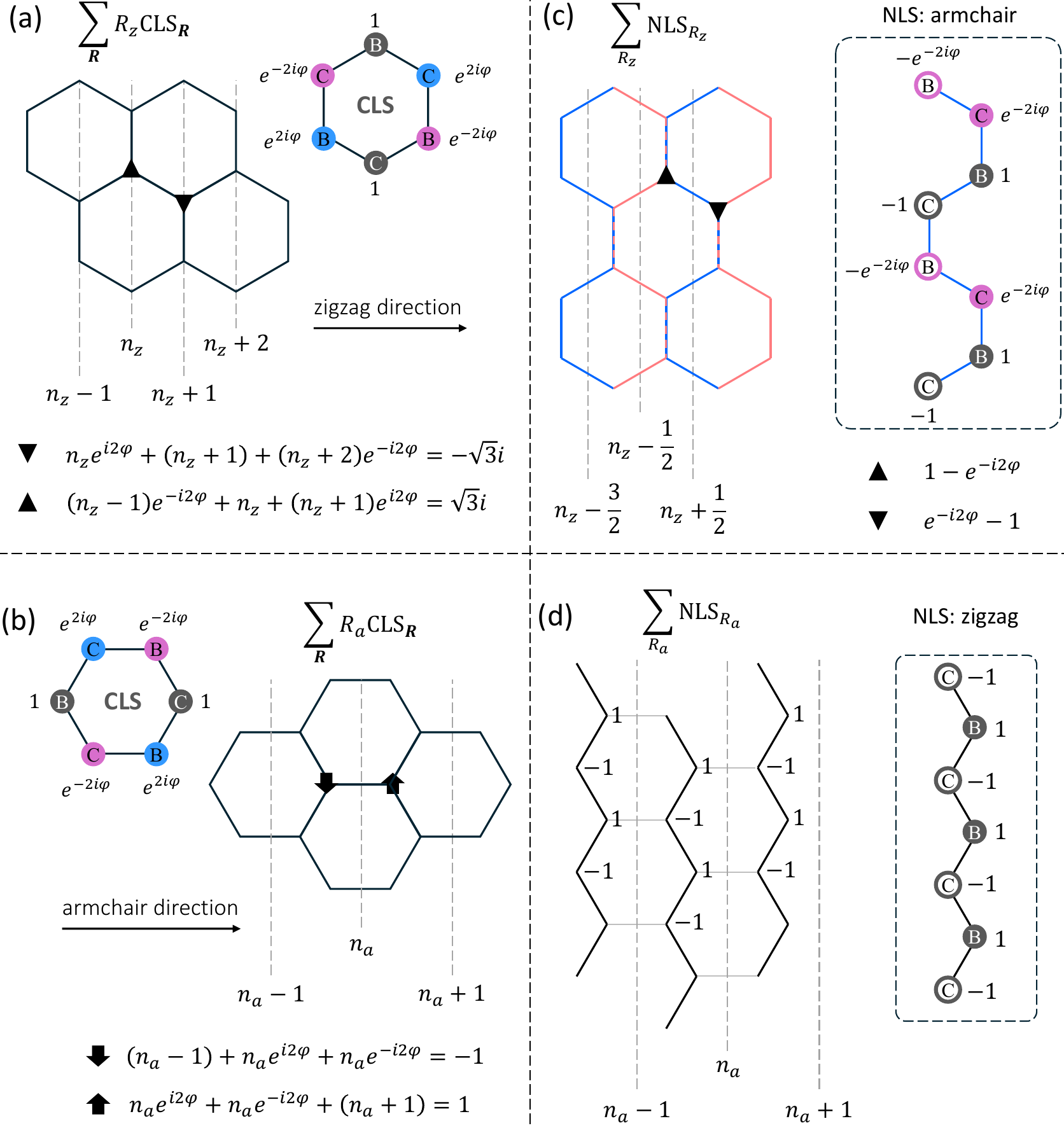}
	\caption{(a) Schematics of CLS in the dice lattice model and $-i\frac{\partial}{\partial k_i}u_{\bk}|_{\bk=0}=\sum_{\bs{R}}\text{CLS}_{\bs{R}}\,R_i$ along the zigzag direction ($i=z$). The vertical dashed lines mark the coordinates of the CLS centers along the zigzag direction. The numbers in the CLS denote the weights on different orbitals. The two black triangles denote the opposite weights on the honeycomb lattice formed by $\sum_{\bs{R}}\text{CLS}_{\bs{R}}\,R_z$. (b) Similar to (a) but along the armchair direction ($i=a$). (c) Schematics of NLS along the armchair direction and $\sum_{R_i}\text{NLS}_{R_i}$ along the zigzag direction ($i=z$). The vertical dashed lines mark the coordinates of the NLS centers along the zigzag direction. The numbers in the NLS denote the weights on different orbitals. The two black triangles denote the opposite weights on the honeycomb lattice formed by $\sum_{R_z}\text{NLS}_{R_z}$. (d) Similar to (c) but along the armchair direction ($i=a$).}~\label{FigSupp:CLSNLS}
\end{figure}

Next we show that the stable topology can also be understood from the real-space perspective~\cite{FangChenGaplessTopology}. The existence of an exact flat band at the zero energy with a band touching can be understood by employing the zero-energy compact localized states (CLS) and non-contractible loop states (NLS)~\cite{BandTouchingBalents2008}. The CLS is strictly bounded within a finite region, while the NLS is extended in one direction but compactly localized in the other. As explicitly discussed in Ref.~\cite{WenqiPRL2025}, in a dice (thus honeycomb) lattice of $N$ unit cells, $N-1$ independent CLS and 2 independent NLS can be constructed---a total number of $N+1$ zero-energy states on the $N$-cell lattice dictates the existence of a touching point. Fig.~\ref{FigSupp:CLSNLS} shows schematics of CLS and two NLS in the armchair and zigzag direction respectively in the dice lattice model. Given the real-space representation of the CLS centered at $\bs{R}$, which will be denoted as CLS$_{\bs{R}}$, the unnormalized Bloch state can be constructed as $u_{\bk}(\br)=\sum_{\bs{R}}\text{CLS}_{\bs{R}}\,e^{i\bk\cdot\bs{R}}$. Its derivative with respect to $k_i$ at $\bk=0$ is $-i\frac{\partial}{\partial k_i}u_{\bk}|_{\bk=0}=\sum_{\bs{R}}\text{CLS}_{\bs{R}}\,R_i$. Figs.~\ref{FigSupp:CLSNLS}(a) and (b) show schematically the $-i\frac{\partial}{\partial k_i}u_{\bk}|_{\bk=0}$ in the dice lattice along the zigzag ($i=z$) and armchair ($i=a$) direction respectively. Both of them form a honeycomb lattice with neighboring sites having opposite weights, which are denoted by black triangles in Fig.~\ref{FigSupp:CLSNLS}(a) or black arrows in Fig.~\ref{FigSupp:CLSNLS}(b). On the other hand, denoting the real-space representation of the NLS centered at $R_i$ as NLS$_{R_i}$, it is straightforward to see that $\sum_{R_i}\text{NLS}_{R_i}$ also form a honeycomb lattice with neighboring sites having opposite weights along the zigzag [Fig.~\ref{FigSupp:CLSNLS}(c)] and armchair [Fig.~\ref{FigSupp:CLSNLS}(d)] direction. Apparently, the honeycomb lattices in Figs.~\ref{FigSupp:CLSNLS}(a)--(d) are all equivalent up to a phase. And $\sum_{R_z}\text{NLS}_{R_z}=\lambda\sum_{R_a}\text{NLS}_{R_a}$ with $\text{Im}\,\lambda\ne0$ is the so-called second topological condition introduced in Ref.~\cite{FangChenGaplessTopology}. This condition also leads to a well-defined Chern number in the current case. In fact, $-i\frac{\partial}{\partial k_z}u_{\bk}|_{\bk=0}=-i\frac{\partial}{\partial k_a}u_{\bk}|_{\bk=0}$ (up to a phase) is equivalent to the continuity of the normalized Bloch state on the flat band.

\subsection{Realization of isolated trivial exact flat bands at either zero or finite maximal quantum distance}

Due to the bipartite structure of $\hat{H}_{\rm dice}$, an exact flat band at the zero energy always exists independent of the details of $f$ and $g$~\cite{WenqiPRL2025}: One can easily verify that $\hat{H}_{\rm dice}\psi_0^{\rm dice}\equiv0$. Similarly, $\hat{H}'\psi_0^{\tinyhex}=\hat{H}_{\tinyhex}\psi_0^{\tinyhex}\equiv0$. Therefore, by perturbing $f$ and/or $g$ such that they do not simultaneously vanish at the same $\bk$ point, the band touching can be broken and the SFB becomes an isolated trivial exact flat band with Chern number $C=0$~\cite{FlatChernTheorem,FCIC=0Lin2025,FCIC=0Lu2025}. Fig.~\ref{FigSupp:DiceGappedTrivialBand}(b) illustrates this by replacing $f(\bk)=-t\left(2\eta\cos\theta_-e_1 - e^{i\theta_-} e_2 - e^{-i\theta_-}  e_3\right)$ and $g(\bk)=-t\left(-2\eta\cos\theta_+e_1^{*} + e^{i\theta_+} e_2^{*} + e^{-i\theta_+}  e_3^{*}\right)$ with $\eta<1$ in the model of $\hat{H}_{\tinyhex}(\bk)$.


\section{Extra many-body results of the honeycomb model}

\subsection{Carrier occupation of FQAH ground states at $\delta=0$}

\begin{figure}[h!]
	\centering
	\includegraphics[width=5.8in]{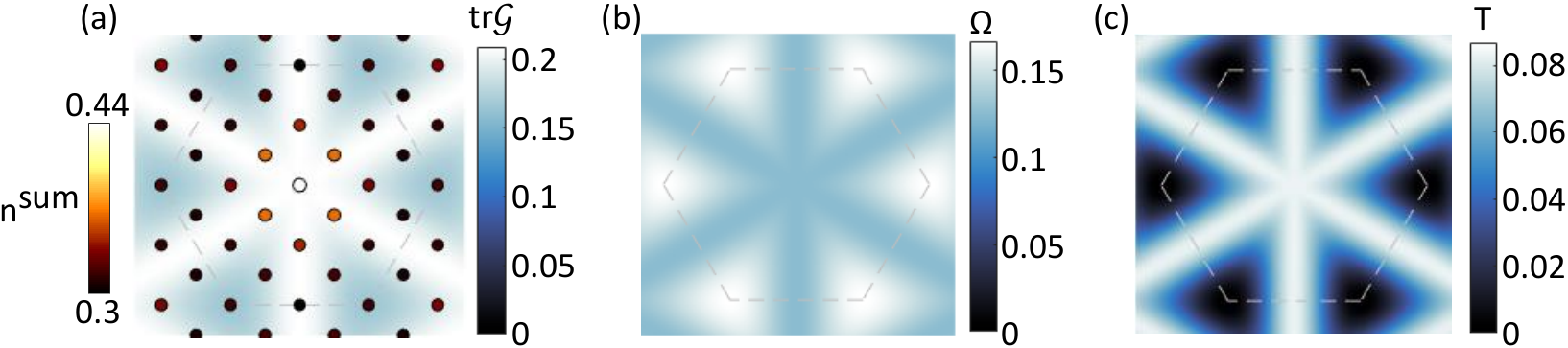}
	\caption{(a) Carrier occupation of the FQAH ground states at $\delta=0$ and $U=1$ in the two bands. (c, d) Distribution of Berry curvature $\Omega(\bk)$ and $T(\bk)=\text{tr}\,\CalG(\bk)-|\Omega(\bk)|$.}~\label{FigSupp:OccHoneyDelta0}
\end{figure}

Fig.~\ref{FigSupp:OccHoneyDelta0} shows the carrier occupation of the FQAH ground states at $\delta=0$. As the quantum geometry of the flat band is very uniform, it is clear that the occupation is nearly constant away from the $\Gamma$ point, while there is a clear enhancement around $\Gamma$.

\subsection{Results under twisted boundary conditions: carrier occupation}

\begin{figure}[h!]
	\centering
	\includegraphics[width=5.5in]{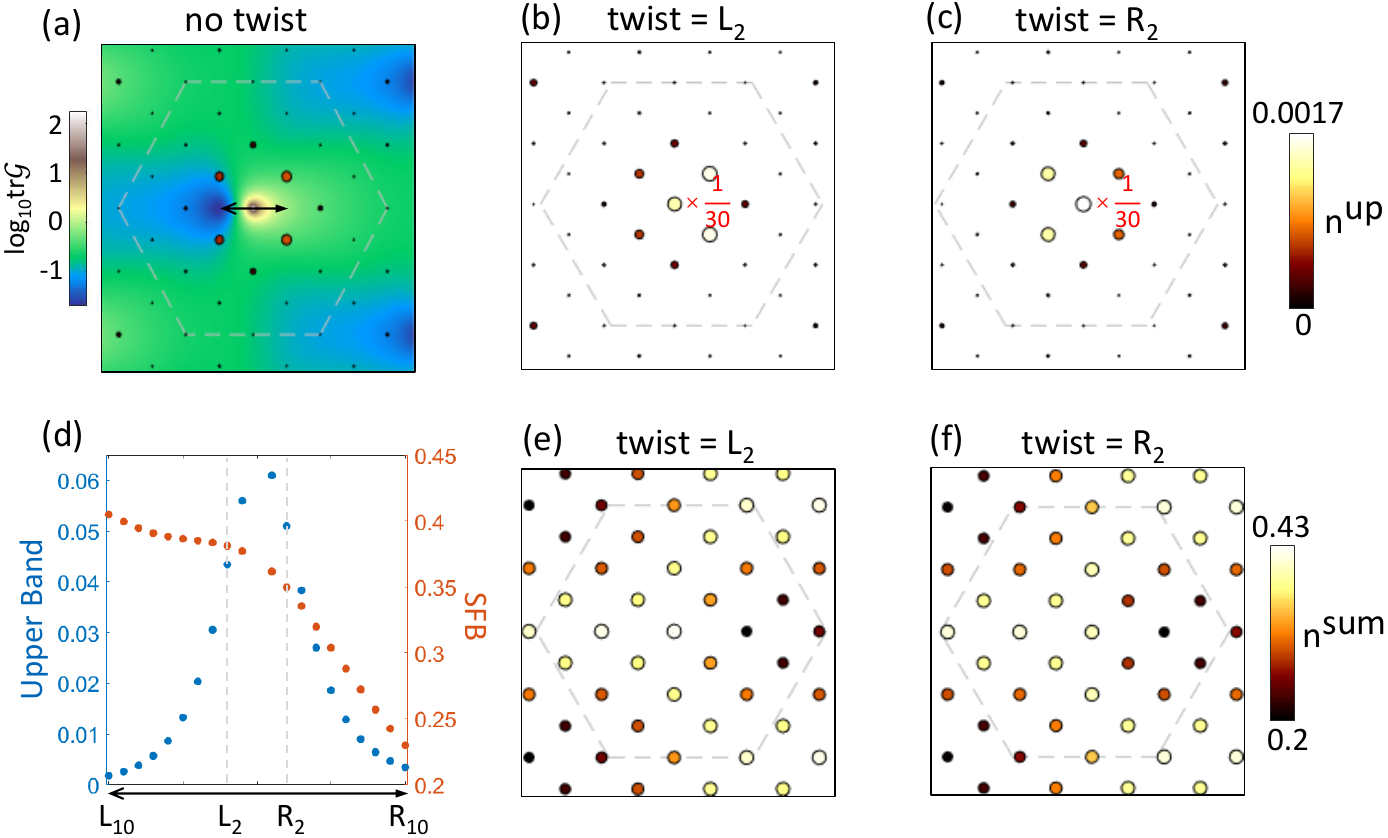}
	\caption{(a-c) Occupation in the upper band under various TBC. The $\Gamma$ point is shifted to the left edge or the right edge of the double-headed arrow in (a) in 10 steps as the boundary condition is changed, which are denoted by $L_1,\cdots,L_{10}$ and $R_{1},\cdots,R_{10}$ [also see panel (d)]. The continuous background color in (a) displays log$_{10}$tr$\mathcal{G}$. The occupation at the central point in (b) and (c) is rescaled by a factor of 1/30 for clarity. (d) Occupation variation in the upper band and the SFB at the $\bk$ point offset from $\Gamma$ by the twist under TBC. (e, f) Total occupation in the two touching bands with the TBC set to $L_2$ and $R_2$, respectively. $U=1$ and $\delta = 0.2$ in all the panels.}~\label{FigSupp:UpSfbOccTwistHoney}
\end{figure}

Figs.~\ref{FigSupp:UpSfbOccTwistHoney}(a-c) shows the carrier occupation in the upper band, $n^{\rm up}(\bk)$ (not to confuse with the particle number in the upper band $n_{\rm up}$), under twisted boundary condition (TBC). The TBC is applied along the direction of the black double-headed arrow in Fig.~\ref{FigSupp:UpSfbOccTwistHoney}(a). Twenty twisted values have been used, which are uniformly distributed on the arrow with the largest corresponding to the arrow heads [see Fig.~\ref{FigSupp:UpSfbOccTwistHoney}(d)]. Distribution of tr$\mathcal{G}$ in the SFB is displayed as the continuous background color in Fig.~\ref{FigSupp:UpSfbOccTwistHoney}(a). The occupation in the upper band is small and decays exponentially with the distance from the touching point [Figs.~\ref{FigSupp:UpSfbOccTwistHoney}(a-c) and blue dots in Fig.~\ref{FigSupp:UpSfbOccTwistHoney}(d)]. 
The occupation in the SFB is large around the touching point. Away from the touching point, carriers tend to occupy regions with small tr$\mathcal{G}$. 
The total occupation in the two bands, $n^{\rm sum}(\bk)$, is shown in Fig.~\ref{FigSupp:UpSfbOccTwistHoney}(e, f) for two different TBC.

\subsection{Results under twisted boundary conditions: 1-band vs 2-band ED}

\begin{figure}[h!]
	\centering
	\includegraphics[width=6.5in]{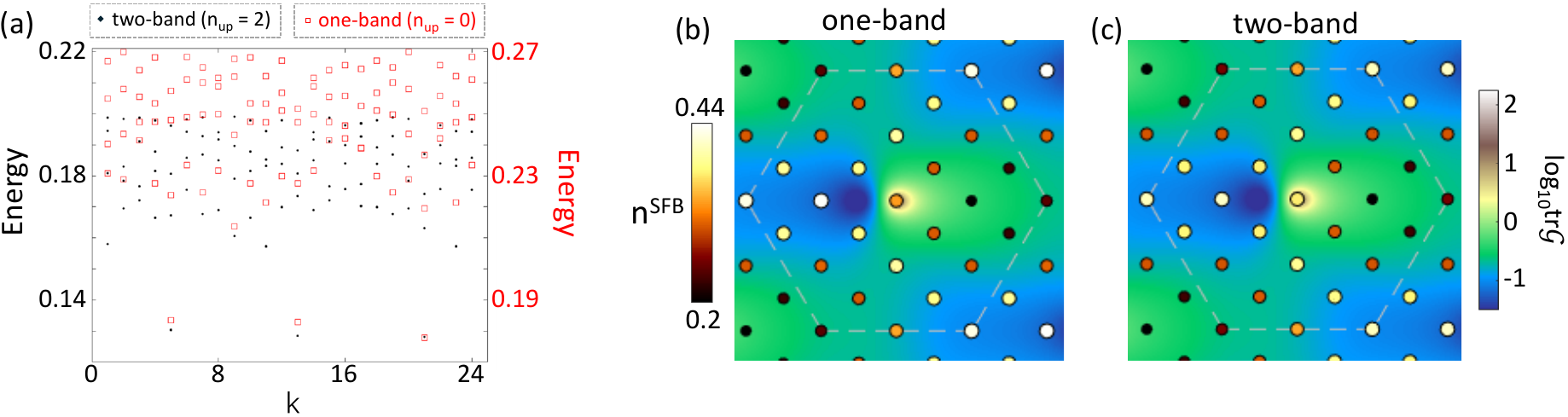}
	\caption{(a) Comparison of many-body energy spectra obtained from one-band ED ($n_{\rm up}=0$) and two-band ED ($n_{\rm up}=2$) in the honeycomb model at $\delta=0.2$. A small twist (1\% of the arrow length in Fig.~\ref{FigSupp:UpSfbOccTwistHoney}(a)) is added. Note that the right energy axis for the one-band ED has larger values. (b, c) Carrier occupation of the ground states on the SFB from one-band ED (b) and two-band ED (c). In two-band ED, the upper band occupation is 0.06 at the central point and negligible elsewhere. The interaction strength is $U=1$.}~\label{FigSupp:1Bandvs2BandEDHoney}
\end{figure}

The red squares in Fig.~\ref{FigSupp:1Bandvs2BandEDHoney}(a) show the ED many-body spectra from one-band ED calculations (setting $n_{\rm up}=0$) under TBC with $U=1$, which have significantly higher energy than its two-band counterpart (black dots, with $n_{\rm up}=2$). The carrier occupation of the ground states in the one-band ED [Fig.~\ref{FigSupp:1Bandvs2BandEDHoney}(b)] is also smaller around the $\Gamma$ point than that of two-band ED [Fig.~\ref{FigSupp:1Bandvs2BandEDHoney}(c)].

\subsection{One-band plus one extra zero mode ED vs two-band ED}

\begin{figure}[h!]
	\centering
	\includegraphics[width=4.5in]{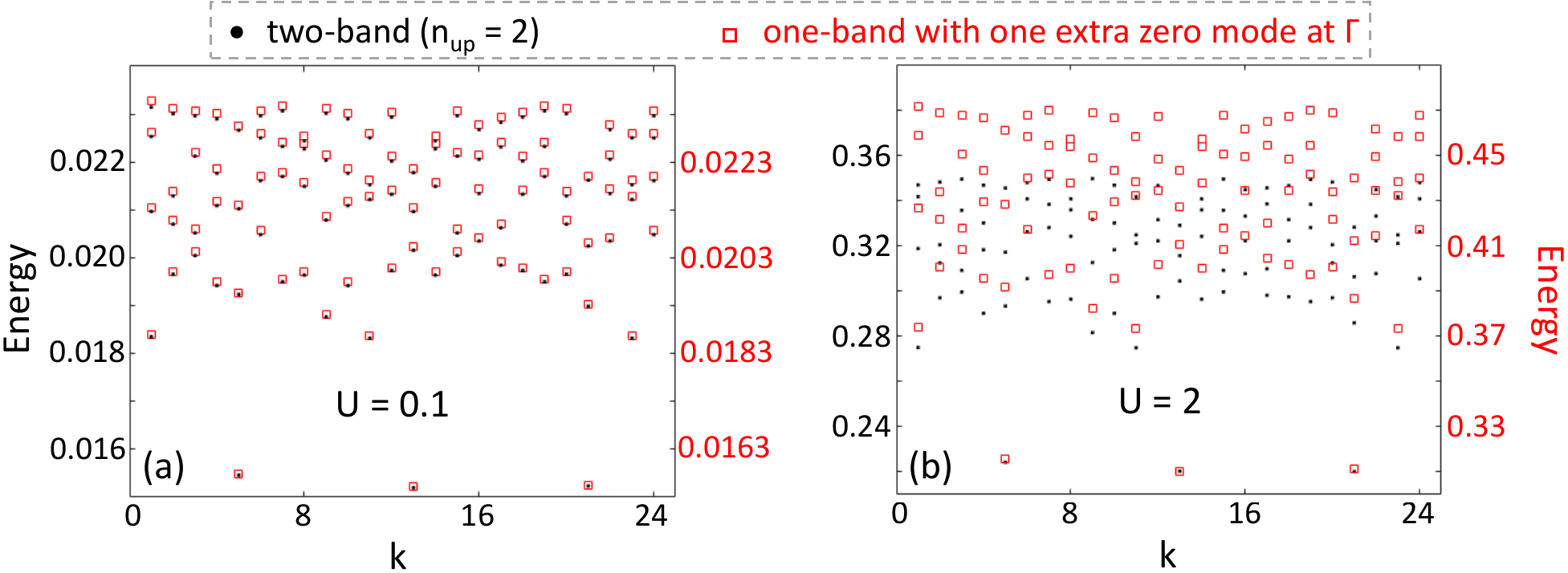}
	\caption{Comparison of many-body spectra from two-band ED (with $n_{\rm up}=2$) and one-band ED (plus an extra zero mode at $\Gamma$) with $U=0.1$ (a) and $U=2$ (b). Model parameter $\delta=0.2$ for both panels and the $1\times24$ tilted $\bk$ grid is used.}~\label{FigSupp:1Bandvs2BandEDNoTwistHoney}
\end{figure}

The red squares in Fig.~\ref{FigSupp:1Bandvs2BandEDNoTwistHoney} show the ED many-body spectra from one-band ED calculations but considering the extra zero-energy mode from upper band at the $\Gamma$ point. The black dots represent the two-band ED results (with $n_{\rm up}=2$). The red squares and black dots overlap well when $U$ is small [Fig.~\ref{FigSupp:1Bandvs2BandEDNoTwistHoney}(a)], while significant discrepancy emerges when $U$ is large [Fig.~\ref{FigSupp:1Bandvs2BandEDNoTwistHoney}(b)]. The latter can be understood as band mixing is expected to be important in a larger area around the $\Gamma$ point with the increase of $U$, which gets less captured in the calculations with one-band plus an extra upper-band state at $\Gamma$ only.

\subsection{Results under twisted boundary conditions: occupation-weighted quantum geometry}
\begin{figure}[h!]
	\centering
	\includegraphics[width=4.25in]{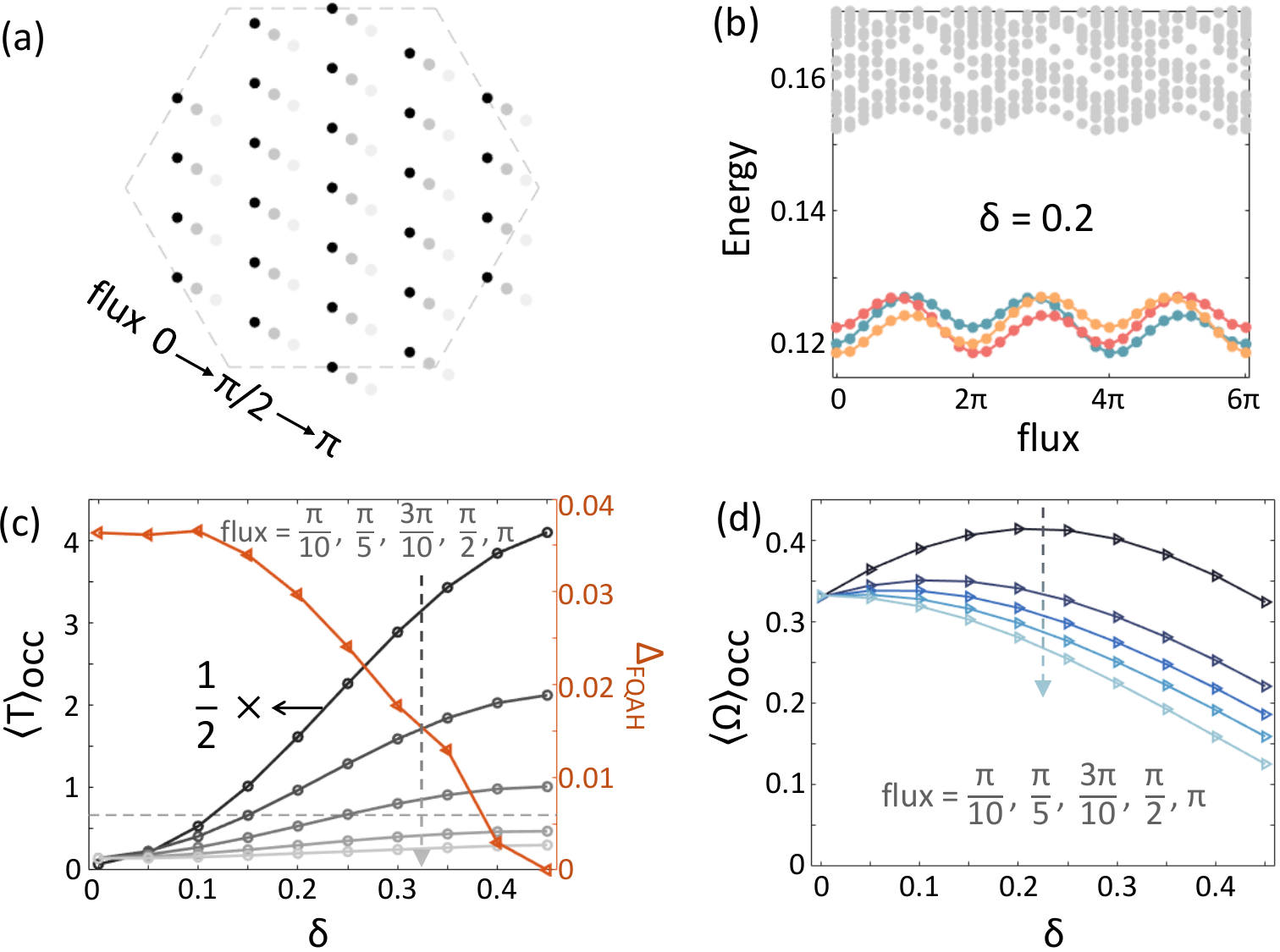}
	\caption{(a) Schematic illustration of the correspondence between flux insertion (equivalently, TBC) and the shift of $\bk$ grid. (b) ED spectral flow under flux insertion at $\delta=0.2$. (c, d) Plots of $\langle T \rangle_{\rm occ}$ and $\langle \Omega \rangle_{\rm occ}$ as functions of $\delta$ for different flux insertions, respectively. The topmost black curve in (c) is rescaled by 1/2 for clarity. The gray horizontal dashed line in (c) indicates the value of $\langle T \rangle_{\rm occ}^{\rm 1LL}=2/3$ in the first Landau level. The variation of many-body gap as a function of $\delta$ is also shown in (c). $U=1$ in panels (b--d).}~\label{FigSupp:OccQuantGeoTwistHoney}
\end{figure}

Fig.~\ref{FigSupp:OccQuantGeoTwistHoney} shows the occupation-weighted quantum geometry for the honeycomb model under TBC. By twisting the $\bk$ grid away from the touching point, contributions around the touching point can be included. As illustrated by the dark black curves in Fig.~\ref{FigSupp:OccQuantGeoTwistHoney}(c), for small twists, $\langle T\rangle _{\rm occ}$ is about one order of magnitude larger than the value of $2/3$ in the first Landau level (1LL), over a wide range of $\delta$. 
This counting under a small twist likely overestimates the violation of idealness, as two-band quantum geometry may become the more relevant quantity in the vicinity of $\Gamma$ point, while the characterizations simply excluding the $\Gamma$ point in Fig.~3(b) of the main text are more conservative.

\subsection{Results under twisted boundary conditions: Energy scale for tr$\CalG^{(\rm 2band)}$ being relevant}

\begin{figure}[h!]
	\centering
	\includegraphics[width=4.1in]{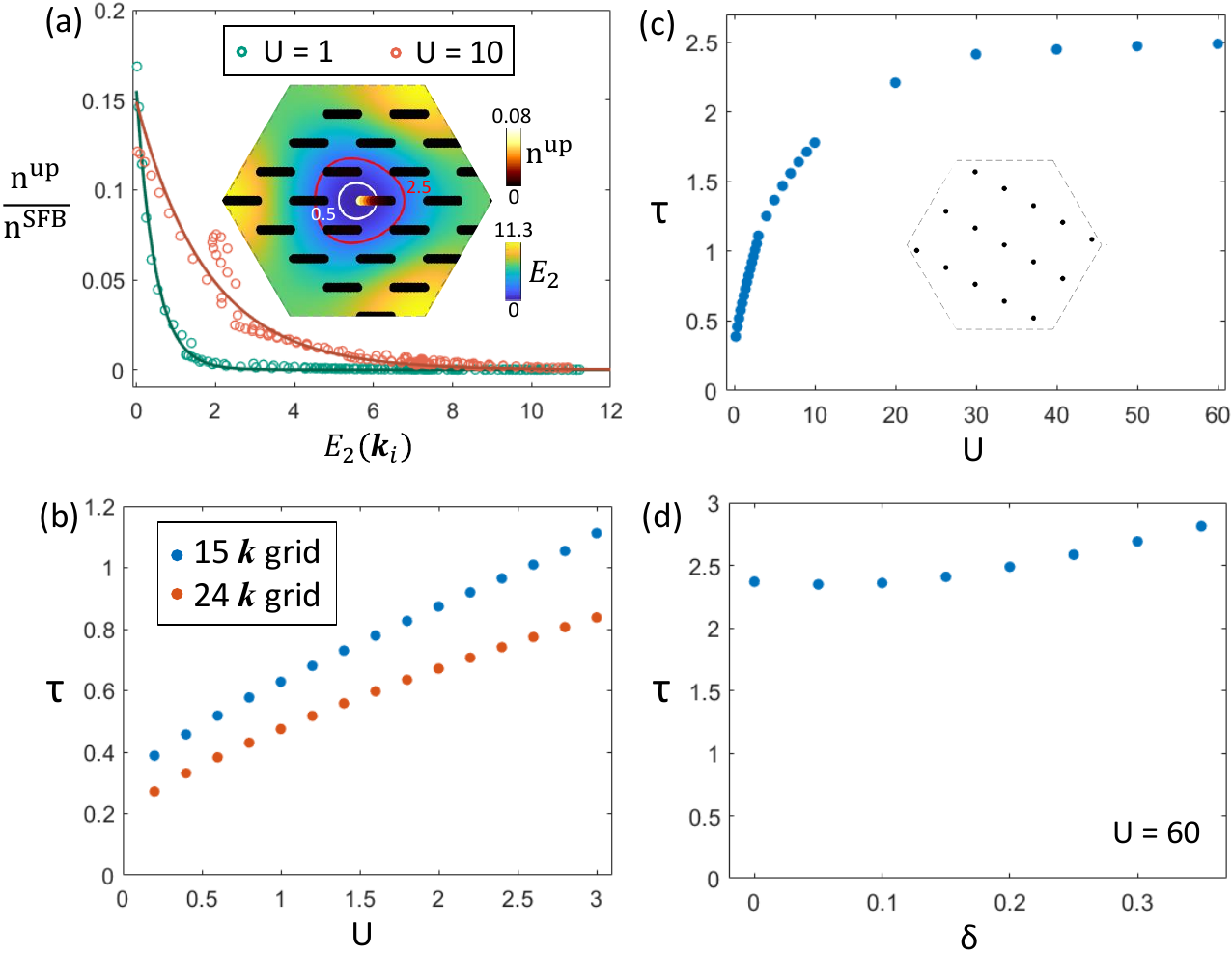}
	\caption{(a) Variation of ratio of occupations in the two touching bands with $E_2(\bk_i)$. The discrete points represent ED results under 10 TBC; the solid curves represent exponential fitting. Colored dots (collapsed discrete points from 10 TBC) in the inset show the upper-band occupations $n^{\rm up}(\bk_i)$ at $U=1$. The BZ is color coded by the upper-band dispersion $E_2$. The white and red circles represent energy contours of $E_2(\bk)=0.5$ and 2.5 respectively. (b) Evolution of $\tau$ with $U$ obtained using two different grid sizes. (c) $\tau$ in a larger $U$ window obtained with the $3\times5$ $\bk$ grid. (d) Variation of $\tau$ with $\delta$, at fixed $U=60$. Model parameter $\delta=0.2$ in (a--c).}~\label{FigSupp:EnergyScaleHoney}
\end{figure}

We consider two different $\bk$ grids: tilted $1\times24$ and rectangular $3\times5$ [Fig.~\ref{FigSupp:EnergyScaleHoney}(c) inset]. Due to computational constraints, the former is limited to small $U$ (for which we set upper-band particle number $n_{\rm up}=2$), while $U$ can take much larger values with the smaller $3\times5$ grid (for which the value of $n_{\rm up}$ is not constrained). In order to estimate the region over which tr$\CalG^{(\rm 2band)}$ is relevant in the BZ, 10 different TBC are considered [e.g., $R_1\cdots R_{10}$ in Figs.~\ref{FigSupp:UpSfbOccTwistHoney}(d)]. We label the discrete $\bk$ points on the grids 
as $\bk_i$ (with $i=1,\,2,\dots,\,15$ or 24) for each TBC and denote the upper-band dispersion as $E_2(\bk)$. Fig.~\ref{FigSupp:EnergyScaleHoney}(a) presents variation of the ratio of occupations in the two bands, $R(\bk_i)=\frac{n^{\rm up}(\bk_i)}{n^{\rm SFB}(\bk_i)}$, obtained from ED calculations under $U=1$ and $U=10$ (full results in the BZ from 10 different TBC are shown). 
As expected, $R(\bk_i)$ decays sharply with $E_2(\bk_i)$ that corresponds to band splitting (the same trend is also observed for $n^{\rm up}(\bk_i)$ vs $E_2(\bk_i)$), indicating that the region for tr$\CalG^{(\rm 2band)}$ to be relevant is well localized around the $\Gamma$ point. The results can be fitted well with $R(\bk_i)\propto e^{-E_2(\bk_i)/\tau}$, as shown by the solid curves in Fig.~\ref{FigSupp:EnergyScaleHoney}(a).
The value $\tau$ quantifies the energy scale, within which tr$\CalG^{(\rm 2band)}$ is relevant, and the contour of $E_2(\bk)=\tau$ denotes the corresponding momentum-space region where tr$\CalG^{(\rm 2band)}$ becomes relevant. In the inset of Fig.~\ref{FigSupp:EnergyScaleHoney}(a), the colored dots (collapsed discrete points from 10 different TBC) show the distribution of upper-band occupation $n^{\rm up}(\bk_i)$ in the BZ. The dominant contributions are well enclosed by the contour of $E_2(\bk)=\tau$.
Figs.~\ref{FigSupp:EnergyScaleHoney}(b, c) present variation of the energy scale $\tau$ with interaction strength $U$.
For large $U$, $\tau$ gradually saturates. The momentum-space region corresponding to the saturated $\tau$ value is shown as the red contour in the inset of Fig.~\ref{FigSupp:EnergyScaleHoney}(a). Fig.~\ref{FigSupp:EnergyScaleHoney}(d) shows the saturated $\tau$ value (obtained at $U=60$) as a function of $\delta$ within the FQAH phase.

\subsection{Many-body results of the trivial phase}

\begin{figure}[h!]
	\centering
	\includegraphics[width=4.25in]{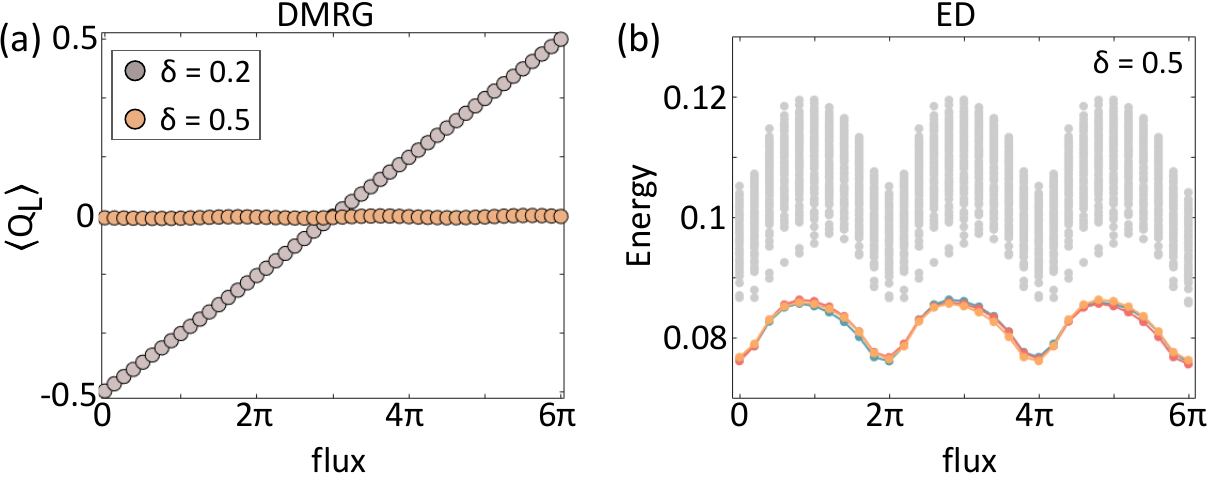}
	\caption{(a) Charge pumping simulation results for $\delta = 0.2$ and $\delta = 0.5$. (b) Evolution of the ED spectrum under flux insertion at $\delta = 0.5$, where the $\bk$ grid corresponds to Fig.~\ref{FigSupp:kmesh24}(a). $U=1$ in both (a) and (b).}~\label{FigSupp:TrivialPhaseHoneycomb}
\end{figure}

Fig.~\ref{FigSupp:TrivialPhaseHoneycomb}(a) shows representative DMRG charge pumping results of the honeycomb model in the FQAH phase ($\delta = 0.2$) and the $\sqrt{3}\times\sqrt{3}$ CDW phase ($\delta = 0.5$). For $\delta=0.5$, the charge pumping does not exhibit quantized nonzero charge transfer, indicating that the CDW phase is topologically trivial. We also employed the tilted $\bk$ grid---commensurate with the real-space charge pattern [lower panel of Fig.~2(d) in the main text]---to examine the ED spectral flow at $\delta=0.5$ [Fig.~\ref{FigSupp:TrivialPhaseHoneycomb}(b)]. A robust gap between the three lowest-energy states and the higher-energy states is observed under flux insertion.
We therefore conclude that the trivial phase in the honeycomb model ($\delta \gtrsim 0.43$) is a gapped, topologically trivial CDW phase.

\subsection{Effects of gap opening perturbation}

Here we examine the effects of single-particle gap-opening perturbations on the quantum geometry and on the stability of the FQAH phase in the honeycomb model. We take a mean-field perspective as guide to choose perturbations. Specifically, the nearest-neighbor interaction $U\hat n_i\hat n_j$ induces two types of effective single-particle contributions within mean-field approximation: a hopping term, $-U\langle\hat C_i^\dagger\hat C_j\rangle\hat C_j^\dagger\hat C_i+h.c.$, and an onsite energy, $U\langle\hat C_i^\dagger\hat C_i\rangle\hat C_j^\dagger\hat C_j+h.c.$. The former arises from inter-orbital coherence, while the latter originates from unequal occupation of different orbitals. By examining the order parameters of the many-body ground states obtained from DMRG, we find that, as $\delta \to 0$, the occupations of the B and C orbitals become nearly identical, and the dominant mean-field contribution arises from a nontrivial correlation value $ \langle\hat{C}_{B_i}^\dagger\hat{C}_{C_j}\rangle<0$, which generates an effective hopping term between the B and C orbitals. In contrast, for larger $\delta$, particles preferentially occupy the B orbitals, and the effective onsite energy offset between the B and C orbitals becomes the dominant mean-field contribution. Motivated by these, here we consider the effects of two types of perturbations in the non-interacting model: (i) an additional hopping term between nearest-neighbor B and C orbitals, $\sum_{\braket{i,j}}t' \hat{C}_{B_i}^\dagger\hat{C}_{C_j}+h.c.$ with $t' > 0$, and (ii) an additional on-site energy term on the B orbitals, $-\delta_B\hat{C}_{B_i}^\dagger\hat{C}_{B_i}$ with $\delta_B>0$.

\begin{figure}[h!]
	\centering
	\includegraphics[width=5.5in]{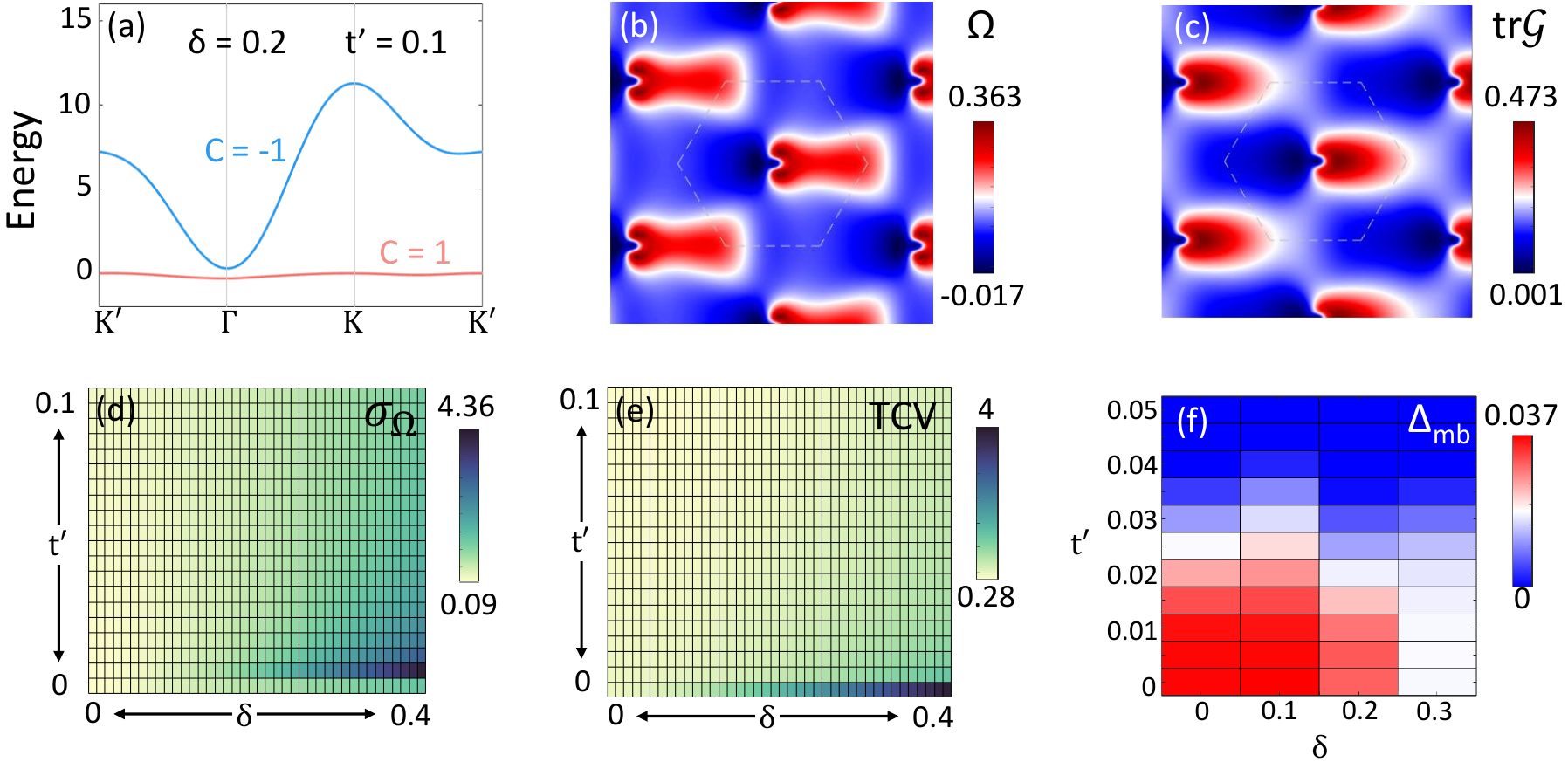}
	\caption{Effects of the additional hopping between nearest-neighbor B and C orbitals, $t'$, on quantum geometry and FQAH stability: (a) Band structures at $\delta=0.2$ and $t'=0.1$. (b, c) Distributions of Berry curvature, $\Omega$, and trace of quantum metric tensor, tr$\CalG$, for the lowest Chern band in (a). (d, e) Distribution of Berry curvature fluctuation, $\sigma_\Omega$, and trace condition violation, TCV, in the $t'-\delta$ parameter space. (f) Variation of the many-body energy gap in the $t'$--$\delta$ parameter space at $U = 1$ and $n_{\rm up}=2$.}~\label{FigSupp:PerturbationHoney}
\end{figure}

\begin{figure}[h!]
	\centering
	\includegraphics[width=5.5in]{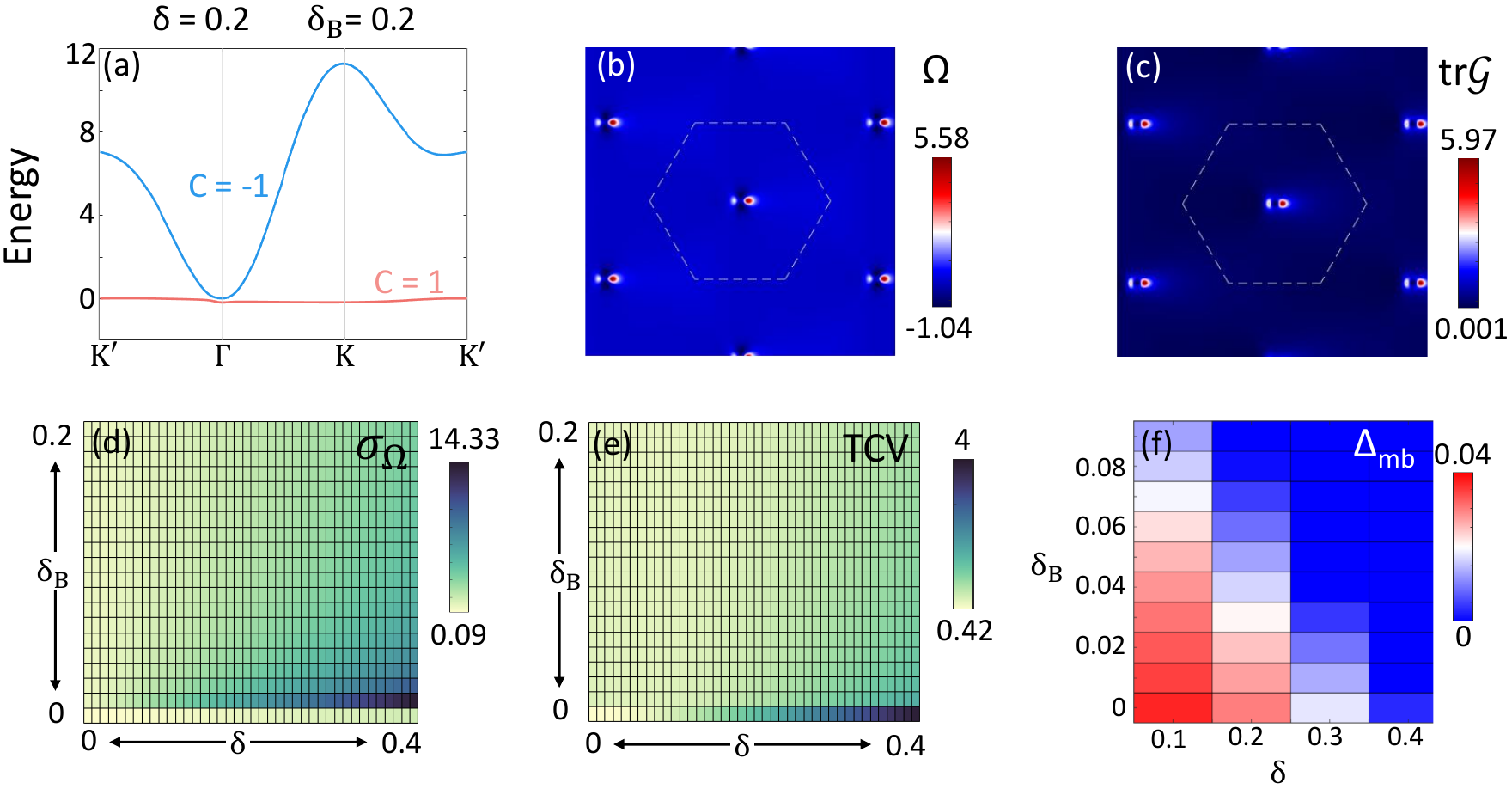}
	\caption{Effects of the additional on-site energy on the B orbitals, $\delta_B$, on quantum geometry and FQAH stability: (a--c) Band structures and the quantum geometry of the lowest Chern band at $\delta=0.2$ and $\delta_B=0.2$. (d--f) Analogous to (d--f) in Fig.~\ref{FigSupp:PerturbationHoney}}~\label{FigSupp:PerturbationHoney2}
\end{figure}

Figs.~\ref{FigSupp:PerturbationHoney}(a--c) and Figs.~\ref{FigSupp:PerturbationHoney2}(a-c) present representative band structures and quantum geometry distribution in the lowest band after introducing the perturbations. The perturbations lead to an isolated Chern band with small dispersion and gap size $\propto t'$ or $\delta_B$, and render the band geometry well-defined throughout the BZ but strongly nonuniform. Figs.~\ref{FigSupp:PerturbationHoney}(d, e) and Figs.~\ref{FigSupp:PerturbationHoney2}(d,e) quantify the `idealness’ of the narrow Chern band. Upon gapping out the band touching, there is an abrupt increase in the Berry curvature fluctuation $\sigma_\Omega$ followed by a decrease as $t'$ or $\delta_B$ increases. Meanwhile, the TCV decreases monotonically with increasing $t'$ for all $\delta$. For $\delta_B$, the TCV decreases monotonically when $\delta > 0.08$, while for $\delta \leq 0.08$ it first increases slightly and then decreases as $\delta_B$ grows. We note that here $\sigma_\Omega$ and especially TCV are much larger than those in twisted bilayer MoTe$_2$.

We performed two-band ($n_{\rm up}=2$) ED calculations to investigate the influence of the perturbations on the stability of the FQAH phase in Fig.~\ref{FigSupp:PerturbationHoney}(f) and Fig.~\ref{FigSupp:PerturbationHoney2}(f). At a fixed $\delta$, the many-body gap gradually closes as the intensity of the perturbation increases. This points to a scenario that is in stark contrast to usual expectations: the FQAH phase is most stable in the SFB exhibiting the largest many-body gap, while gapping out the band touching tends to destabilize it.

\section{Extra single-particle properties of the kagome model}

\subsection{Single-particle lattice model}

The kagome model features three orbitals and its momentum-space Hamiltonian is given by~\cite{SinguFlatCommunPhys2023}:
\begin{equation}\label{H-kagome}
	\begin{aligned}
		\hat{H}_{\tinykag}(\bk) = 
		\begin{pmatrix}
			2\alpha^2+2 & f_3^* & f_2^*+f^*_{4}\\
			f_3 & 2 & f_1^*\\
			f_2+f_{4} & f_1 & 2\alpha^2+2
		\end{pmatrix}
	\end{aligned},
\end{equation}
where $f_{i=1,2,3}=t(e_i+e_i^*)$, $f_{4}=i\alpha t(e_1e_3^*+e_1^*e_3)$ 
and $t=e^{i\theta}\sqrt{1+\alpha^2}=1+i\alpha$ with $\theta=\text{acos}(1/\sqrt{1+\alpha^2})$. Here $\alpha$ is chosen as a tuning parameter. This model has a zero-energy exact flat band with a singular band touching at the $\Gamma$ point. The Bloch state of the SFB reads $\psi_0=(e_3^*-e_2e_1^*,\,-1+i\alpha e_1^{*2}+e_2^2-i\alpha e_3^{*2},\,e_1^*-e_2e_3^*)^{T}$.


\subsection{Single-band quantum geometry}

\begin{figure}[h!]
	\centering
	\includegraphics[width=7in]{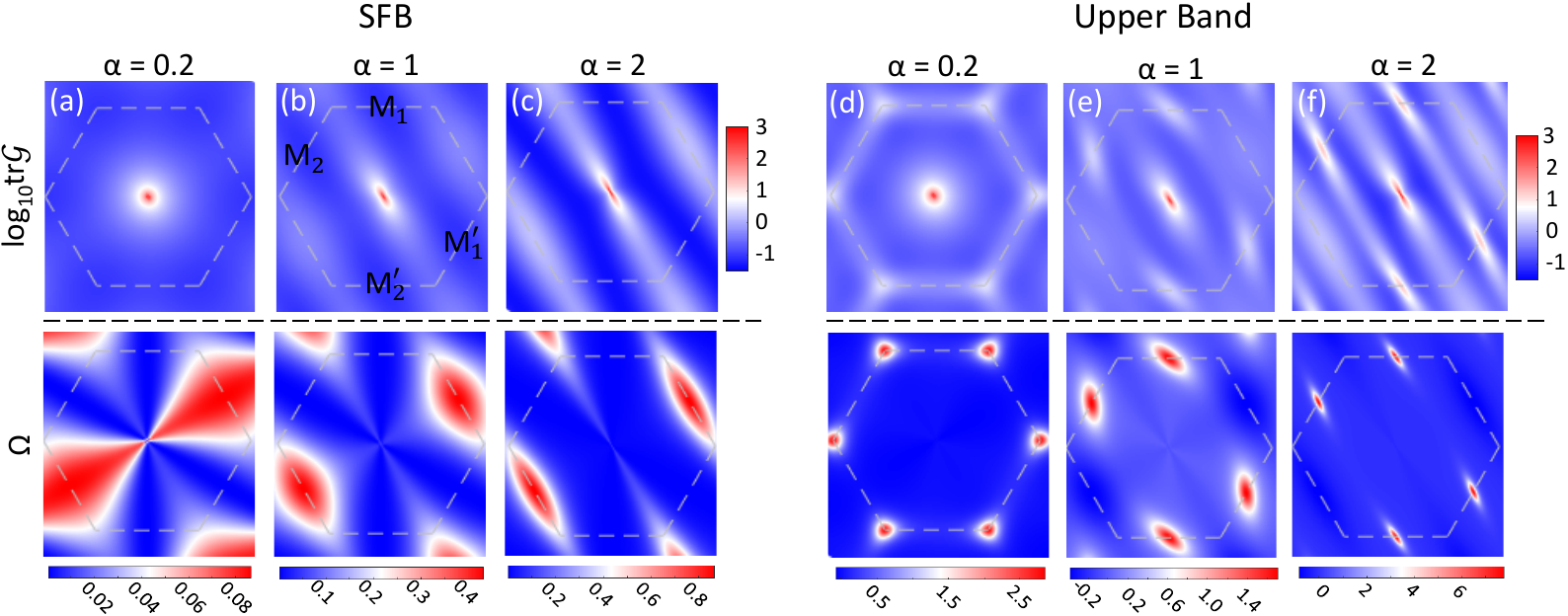}
	\caption{(a--c) Quantum geometry of the SFB in the Kagome model for various $\alpha$. (d--f) Quantum geometry of the dispersive band touching the SFB for various $\alpha$.}~\label{FigSupp:SingleBandQuantGeoSFB_vs_Up}
\end{figure}

Fig.~\ref{FigSupp:SingleBandQuantGeoSFB_vs_Up} presents the distribution of $\Omega(\bk)$ and $\text{tr}\CalG(\bk)$ of the SFB and the neighboring upper dispersive band. The $\Omega(\bk)$ of the SFB differs significantly from that of the upper band, which is expected as the two bands together have total Chern number of 1. On the other hand, these two bands have similar $\text{tr}\CalG(\bk)$ with differences at the BZ boundaries when $\alpha$ is small or along the $M_1$--$M'_1$ and $M_2$--$M'_2$ directions [see Fig.~\ref{FigSupp:SingleBandQuantGeoSFB_vs_Up}(b)] when $\alpha$ is large. Such differences around $M_1$, $M'_1$, $M_2$ and $M'_2$ arise because of the small gaps between the two dispersive bands [cf. Fig.~4(b) of the main text].

\subsection{Two-band quantum geometry}

\begin{figure}[h!]
	\centering
	\includegraphics[width=4.4in]{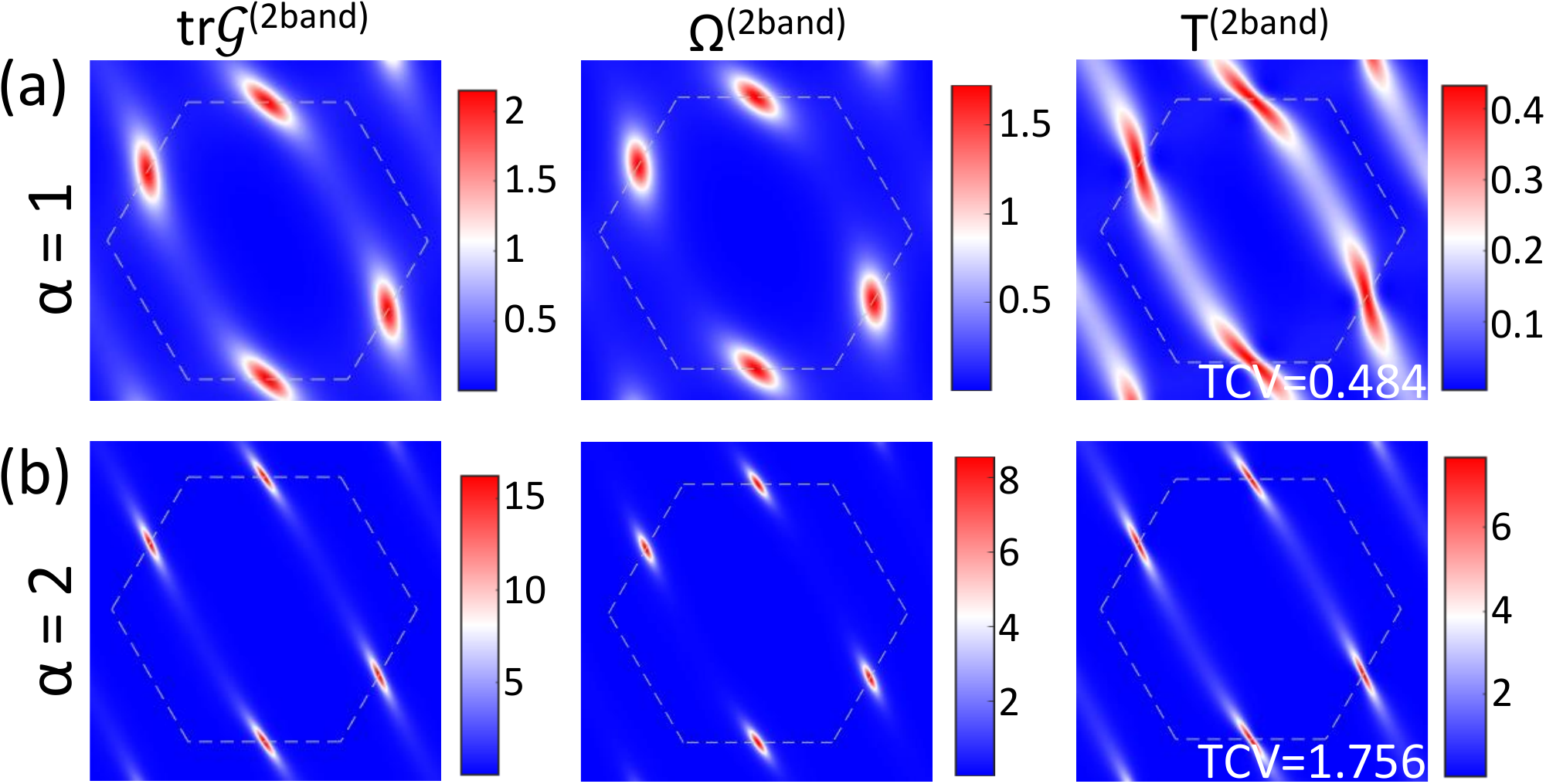}
	\caption{Two-band quantum geometry of the two touching bands in the kagome model at $\alpha=1$ (a) and $\alpha =2$ (b), respectively.}~\label{FigSupp:TwoBandQuantumGeometry}
\end{figure}

The Abelian Berry curvature and quantum metric tensor by projecting onto the two touching bands can be evaluated as $\Omega^{\rm 2band}=i\,\text{Tr}\,P[\partial_xP,\partial_yP]$ and $\CalG_{ij}^{\rm 2band}=\frac{1}{2}\text{Tr}\,\partial_iP\partial_jP$, where $P=\mathbb{U}\mathbb{U}^\dagger$ is the projector onto the two bands, $\mathbb{U}$ is a $3\times2$ matrix whose columns are the Bloch states of the two bands, and `Tr' is over the Hilbert space. For the three-band model, they are related to the corresponding quantities of the third band: $\Omega^{\rm 2band}=-\Omega^{\rm 3rd}$ and $\text{tr}\,\CalG^{\rm 2band}=\text{tr}\,\CalG^{\rm 3rd}$. Fig.~\ref{FigSupp:TwoBandQuantumGeometry} shows representative results of $\Omega^{\rm 2band}$ and $\text{tr}\,\CalG^{\rm 2band}$, where one can see that both quantities are vanishing around the $\Gamma$ point. The hot spots locate around the small gaps between the two dispersive bands [cf. Fig.~4(b) of the main text].

\section{Extra many-body results of the kagome model}

\subsection{DMRG entanglement spectrum and ED spectral flow}

\begin{figure}[h!]
	\centering
	\includegraphics[width=6in]{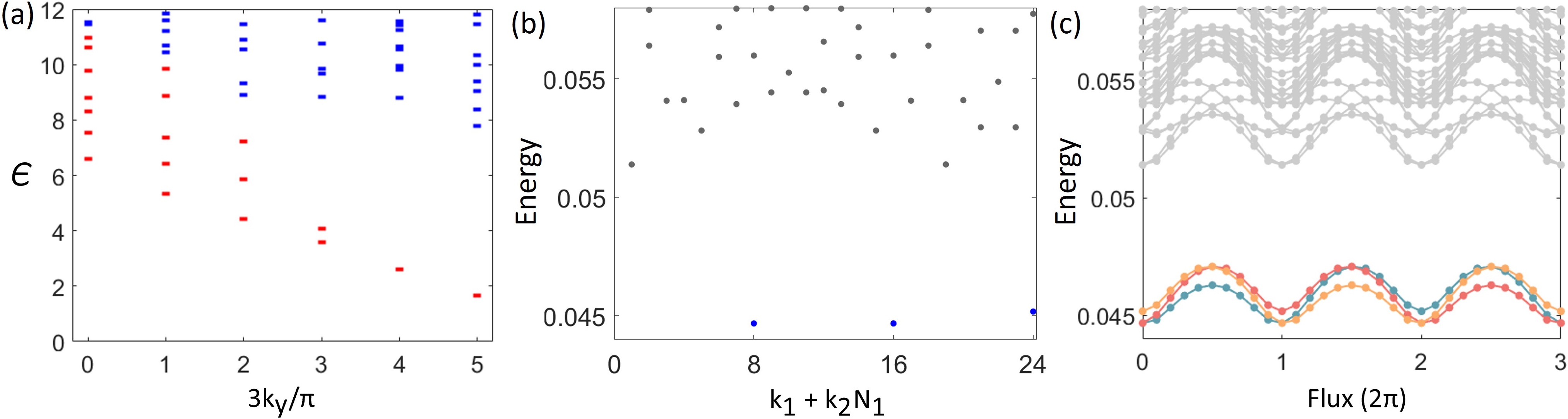}
	\caption{(a) Momentum-resolved entanglement spectrum $\epsilon$ of the kagome model obtained by DMRG simulations in mixed space, at $\alpha = 1$ and $U=3$. (b, c) ED many-body spectrum and spectral flow under flux insertion on a $4\times6$ rectangular $\bk$ grid at $\alpha=1.6$ and $U=0.5$.}~\label{FigSupp:Kagome}
\end{figure}

From DMRG calculations, we observe the FQAH phase characterized by a Hall conductivity of $\sigma_H=e^2/(3h)$ at $\nu=1/3$ filling of the SFB within $0.35\lesssim\alpha\lesssim2.46$. The entanglement spectrum of the FQAH states, exhibiting the sequence $\{1,1,2,3,5,\cdots\}$ is shown in Fig.~\ref{FigSupp:Kagome}(a), indicating that the states are Laughlin-like. The ED spectrum computed on a $4\times 6$ rectangular $\bk$ grid is presented in Fig.~\ref{FigSupp:Kagome}(b), where three nealy degenerate ground states are observed at $\bs{\Cal{K}}_{\rm rect}=\{8,\,16,\,24\}$, consistent with the expected degeneracy for Laughlin states and the generalized Pauli exclusion principle. The spectrum flow under flux insertion, shown in Fig.~\ref{FigSupp:Kagome}(c), demonstrates that the topological gap persists when twisting the boundary conditions.

\subsection{Results under twisted boundary conditions: carrier occupation}

\begin{figure}[h!]
	\centering
	\includegraphics[width=5.5in]{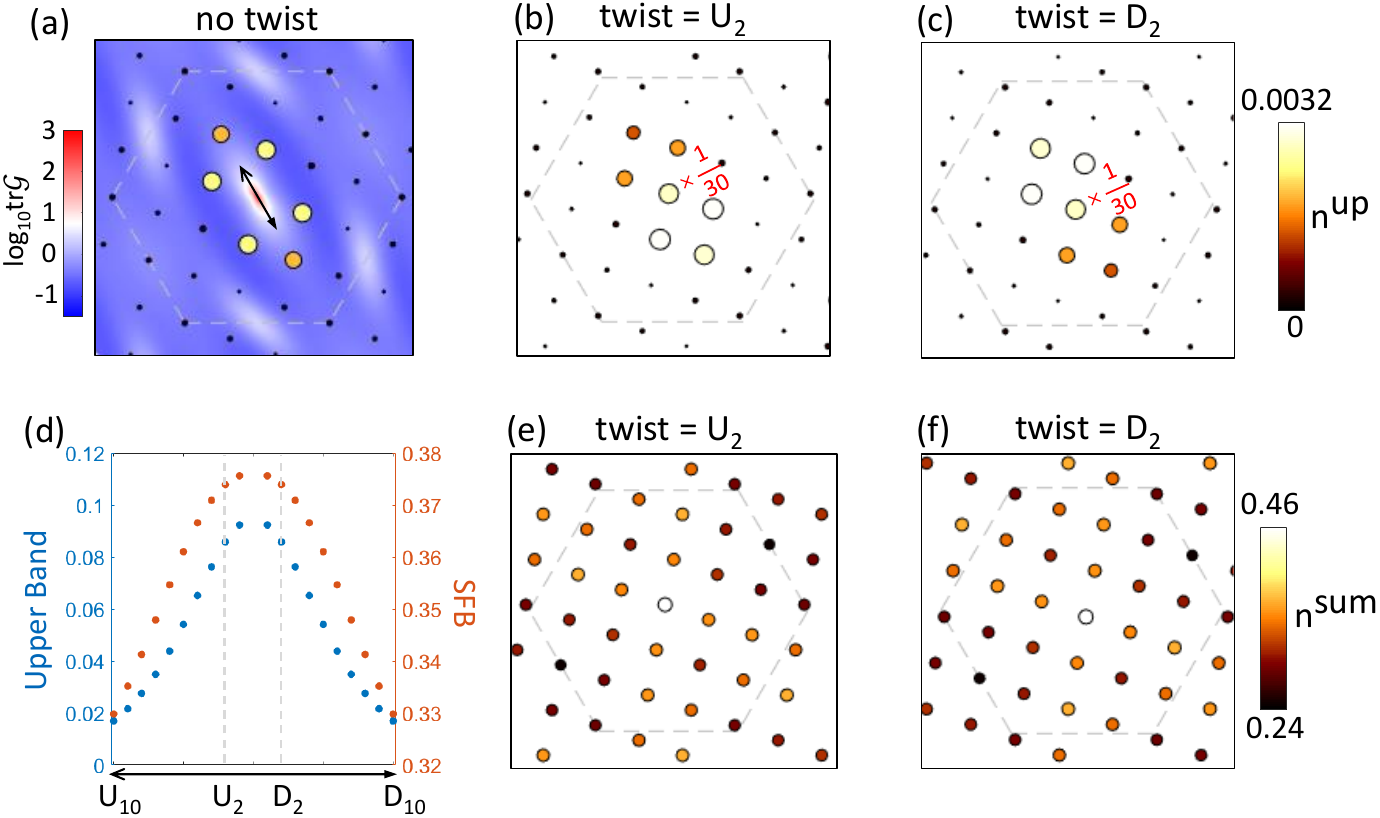}
	\caption{(a--c) Occupation of the upper band (the one touching the SFB) under various TBC. The $\Gamma$ point is shifted to the upper-left end or the lower-right end of the double-headed arrow in (a) in 10 steps as the boundary condition is changed [also see panel (d)]. The continuous background in (a) displays log$_{10}$tr$\mathcal{G}$ of the upper band. The occupation at the central point  in (b) and (c) is scaled by a factor of 1/30. (d) Occupation variation in the upper band and the SFB at the $\bk$ point offset from $\Gamma$ by the twist under different TBC. (e, f) Occupation on the two touching bands with the TBC set to $U_2$ and $D_2$, respectively. $U=1$ and $\alpha = 1$ in all the panels.}~\label{FigSupp:UpSfbOccTwistKagome}
\end{figure}

Fig.~\ref{FigSupp:UpSfbOccTwistKagome} presents the carrier occupation in the $\bk$ space under TBC for the kagome model. The results are presented in a similar manner as those in the honeycomb case (Fig.~\ref{FigSupp:UpSfbOccTwistHoney}). As shown in Fig.~\ref{FigSupp:UpSfbOccTwistKagome}(a--c), the occupation in the upper band (i.e., the dispersive band touching the SFB) concentrates near the touching point and shares a similar elongated anisotropy as tr$\mathcal{G}$. The blue dots in Fig.~\ref{FigSupp:UpSfbOccTwistKagome}(d) highlight the exponential decay of the upper-band occupation with distance from the touching point. The brown dots in Fig.~\ref{FigSupp:UpSfbOccTwistKagome}(d) illustrate that the occupation in the SFB is large and fluctuates weakly around the touching point. Elsewhere on the SFB, the occupation is large (small) in the regions with small (large) tr$\mathcal{G}$ [see Figs.~\ref{FigSupp:UpSfbOccTwistKagome}(e, f)].

\subsection{Results under twisted boundary conditions: 1-band vs 2-band ED}

\begin{figure}[h!]
	\centering
	\includegraphics[width=6.25in]{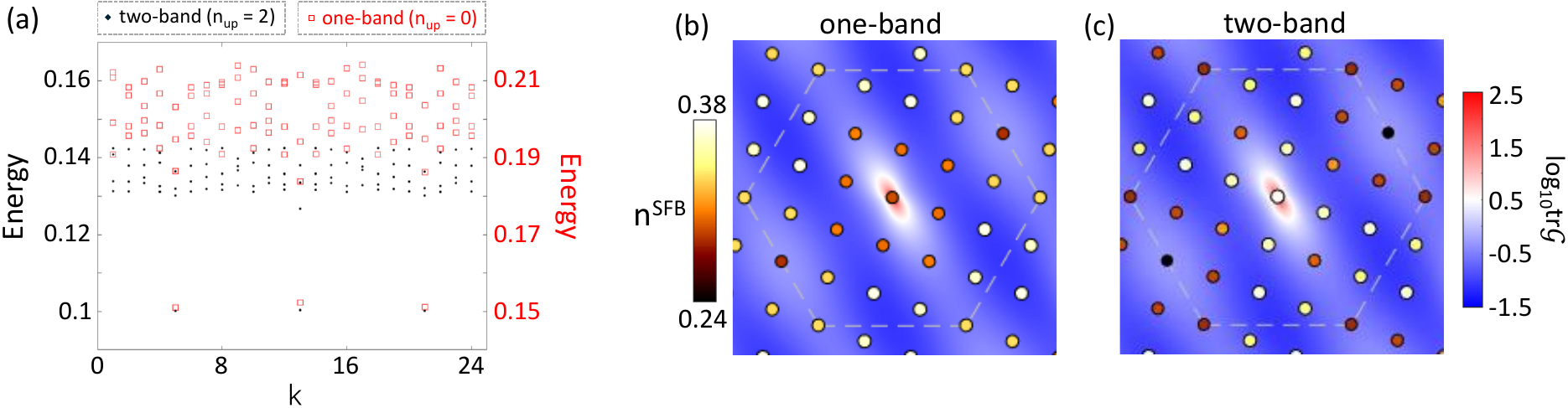}
	\caption{(a) Comparison of many-body energy spectra from one-band ED ($n_{\rm up}=0$) and two-band ED ($n_{\rm up}=2$) in the kagome model at $\alpha=1$. A small twist (1\% of the arrow length in Fig.~\ref{FigSupp:UpSfbOccTwistKagome}(a)) is added. Note that the right energy axis for the one-band ED has larger values. (b, c) Carrier occupation of the ground states on the SFB from one-band ED (b) and two-band ED (c). In two-band ED, the upper band occupation is 0.09 at the central point and negligible elsewhere. The interaction strength is $U=1$.}~\label{FigSupp:1Bandvs2BandEDKagome}
\end{figure}

The red squares in Fig.~\ref{FigSupp:1Bandvs2BandEDKagome}(a) show the ED many-body spectra from one-band ED calculations (setting $n_{\rm up}=0$) under TBC with $U=1$, which have much higher energy than its two-band counterpart (black dots, with $n_{\rm up}=2$). The carrier occupation of the ground states in the one-band ED [Fig.~\ref{FigSupp:1Bandvs2BandEDKagome}(b)] is also smaller around the $\Gamma$ point than that of two-band ED [Fig.~\ref{FigSupp:1Bandvs2BandEDKagome}(c)].

\subsection{One-band plus one extra zero mode ED vs two-band ED}

\begin{figure}[h!]
	\centering
	\includegraphics[width=4.5in]{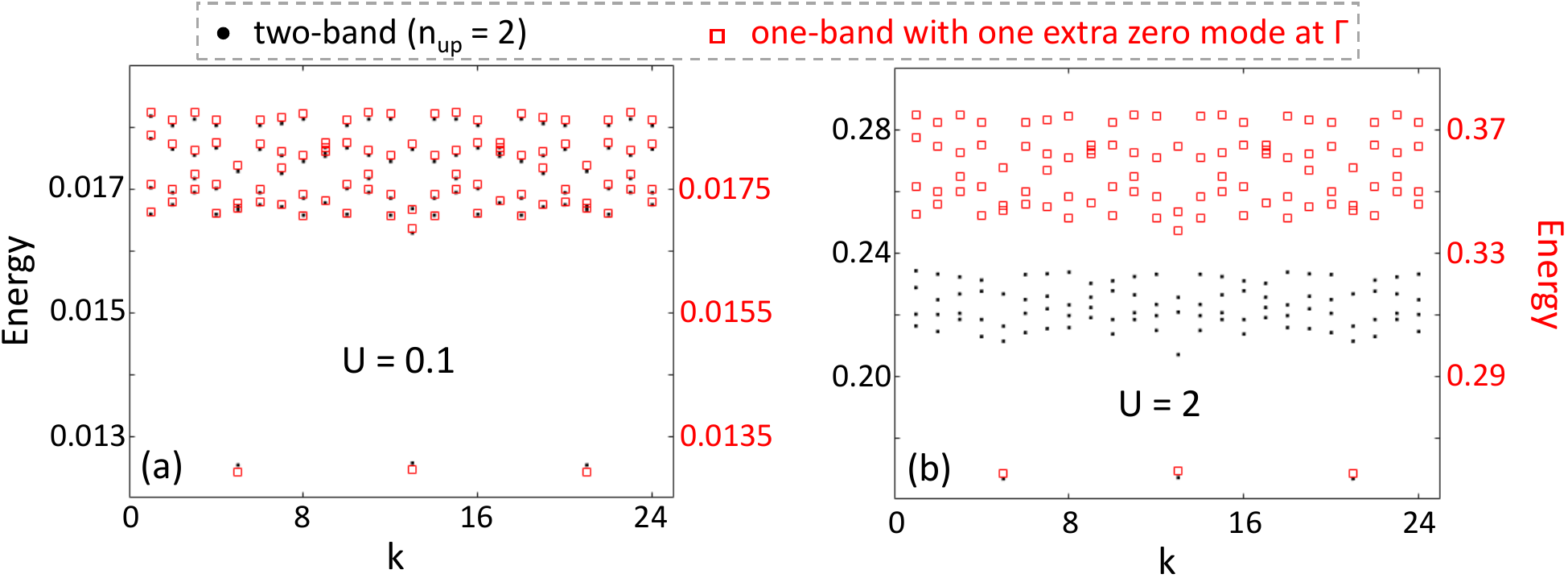}
	\caption{Comparison of many-body spectra from two-band ED (with $n_{\rm up}=2$) and one-band ED (plus the extra zero mode at $\Gamma$) with $U=0.1$ (a) and $U=2$ (b). Model parameter $\alpha=1$ for both panels and the $1\times24$ tilted $\bk$ grid is used.}~\label{FigSupp:1Bandvs2BandEDNoTwistKagome}
\end{figure}

The red squares in Fig.~\ref{FigSupp:1Bandvs2BandEDNoTwistKagome} show the ED many-body spectra from calculations with one band plus the extra zero-energy mode from upper band at $\Gamma$ point. The black dots represent the two-band ED results (with $n_{\rm up}=2$). The red squares and black dots well overlap when $U$ is small [Fig.~\ref{FigSupp:1Bandvs2BandEDNoTwistKagome}(a)], while significant discrepancy emerges when $U$ is large [Figs.~\ref{FigSupp:1Bandvs2BandEDNoTwistKagome}(b)].

\subsection{Results under twisted boundary conditions: occupation-weighted quantum geometry}

\begin{figure}[h!]
	\centering
	\includegraphics[width=4.25in]{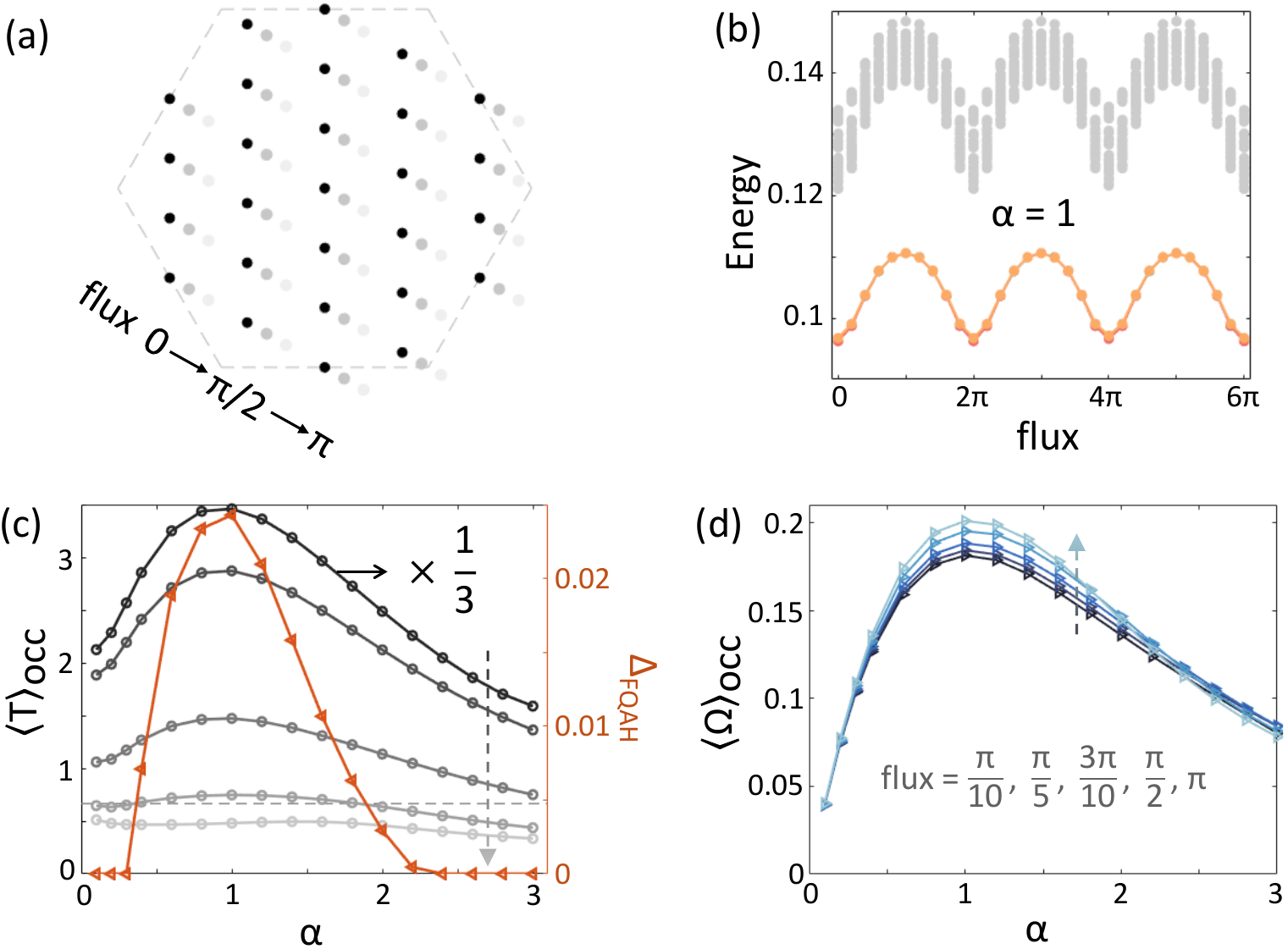}
	\caption{(a) Schematic illustration of the correspondence between flux insertion and the shift of $\bk$ grid. (b) ED spectral flow under flux insertion at $\alpha=1$. (c, d) Plots of $\langle T \rangle_{\rm occ}$ and $\langle \Omega \rangle_{\rm occ}$ as functions of $\alpha$ for different flux insertions, respectively. The topmost black curve in (c) is rescaled by 1/3 for clarity. The gray horizontal dashed line in (c) indicates the value of $\langle T \rangle_{\rm occ}^{\rm 1LL}=2/3$ in the first Landau level. The many-body gap as a function of $\alpha$ is also shown in (c). $U=1$ in (b--d).}~\label{FigSupp:OccQuantGeoTwistKagome}
\end{figure}

Fig.~\ref{FigSupp:OccQuantGeoTwistKagome} shows the occupation-weighted quantum geometry of the kagome model under TBC.  
For small twists, $\langle T\rangle_{\rm occ}$ is about one order of magnitude larger than $\braket{T}_{\rm occ}^{\rm 1LL}=2/3$ in the 1LL over the whole range of $\alpha$ [Fig.~\ref{FigSupp:OccQuantGeoTwistKagome}(c) dark black curves]. 
This counting under a small twist likely overestimates the violation of idealness, as two-band quantum geometry may become the more relevant quantity in the vicinity of $\Gamma$ point, while the characterizations simply excluding the $\Gamma$ point in Fig.~5(b) of the main text are more conservative.
In the region with the largest many-body gap $\Delta_{\rm FQAH}$, one finds a large $\langle T\rangle_{\rm occ}$ [also in Fig.~5(b) of the main text], implying that a large $\langle T\rangle_{\rm occ}$ does not necessarily lead to a small $\Delta_{\rm FQAH}$. These results suggest that the FQAH states can tolerate large violation of idealness and the variation of $\Delta_{\rm FQAH}$ with $\alpha$ is dominantly determined by $\Omega_{\rm occ}$ [see the similar profile of $\Delta_{\rm FQAH}$ in Fig.~\ref{FigSupp:OccQuantGeoTwistKagome}(c) and $\Omega_{\rm occ}$ in Fig.~\ref{FigSupp:OccQuantGeoTwistKagome}(d), also in Fig.~5(b) of the main text].

\subsection{Results under twisted boundary conditions: Energy scale for tr$\CalG^{(\rm 2band)}$ being relevant}

\begin{figure}[h!]
	\centering
	\includegraphics[width=4.25in]{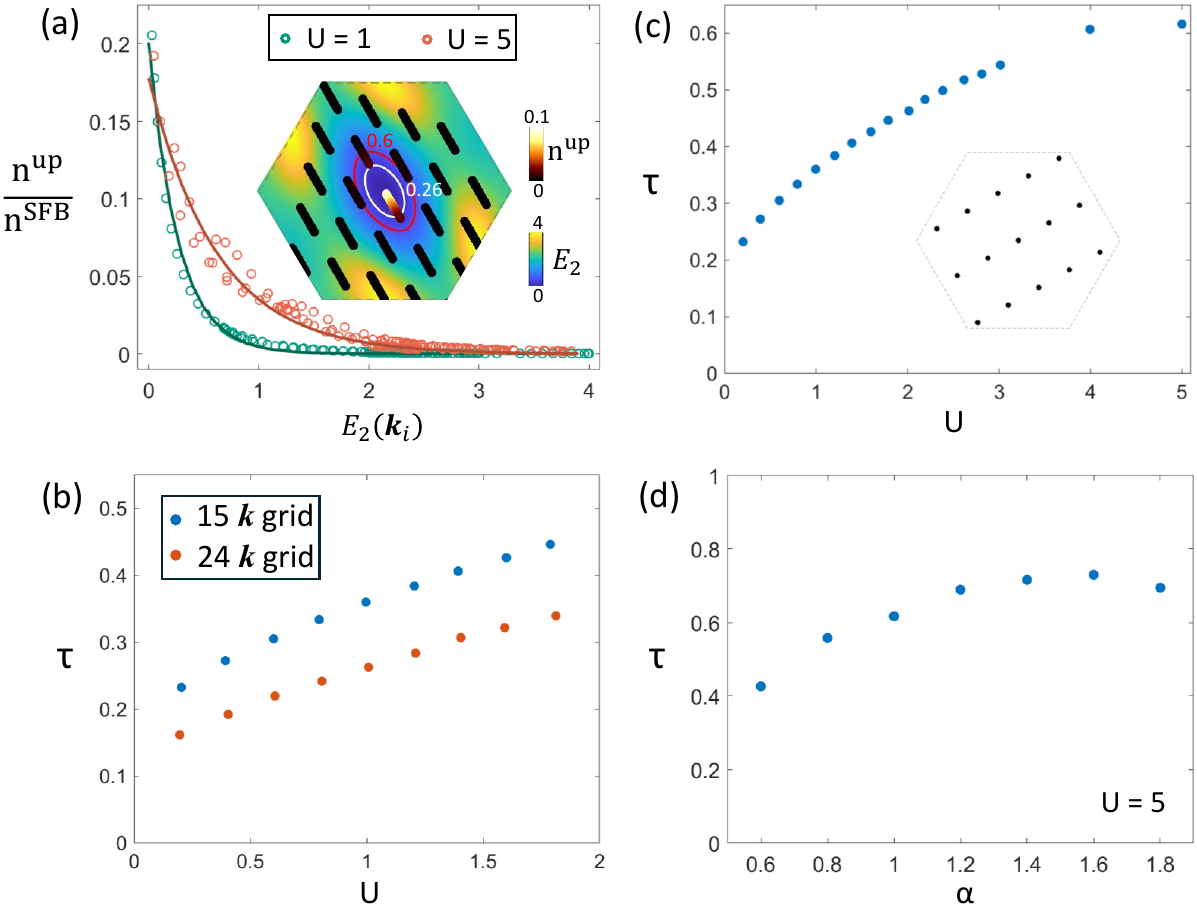}
	\caption{(a) Variation of ratio of occupations in the two touching bands in Kagome model with $E_2(\bk_i)$. The discrete points represent ED results under 10 TBC; the solid curves represent exponential fitting. Colored dots (collapsed discrete points from 10 TBC) in the inset show the upper-band occupations $n^{\rm up}(\bk_i)$ at $U=1$. The BZ is color coded by the upper-band dispersion $E_2$. The white and red circles represent energy contours of $E_2(\bk)=0.26$ and 0.6 respectively. (b) Evolution of $\tau$ with $U$ obtained using two different grid sizes. (c) $\tau$ in a larger $U$ window obtained with the $3\times5$ $\bk$ grid. (d) Variation of $\tau$ with $\alpha$, at fixed $U=5$. Model parameter $\alpha=1$ in (a, b).}~\label{FigSupp:EnergyScaleKagome}
\end{figure}

We consider two different $\bk$ grids: tilted $1\times24$ and rectangular $3\times5$ [Fig.~\ref{FigSupp:EnergyScaleKagome}(c) inset]. Due to computational constraints, the former is limited to small $U$ (for which we set upper-band particle number $n_{\rm up}=2$), while $U$ can take much larger values with the smaller $3\times5$ grid (for which the value of $n_{\rm up}$ is not constrained). The kagome model has three energy bands, but we can only project the interaction onto the two touching bands due to computational constraints, thus we consider $U\le5$ for the $3\times5$ $\bk$ grid. Fig.~\ref{FigSupp:EnergyScaleKagome}(a) presents variation of the ratio of occupations in the two touching bands, $R(\bk_i)=\frac{n^{\rm up}(\bk_i)}{n^{\rm SFB}(\bk_i)}$, obtained from ED calculations (results from 10 different TBC are shown). 
As expected, $R(\bk_i)$ decays sharply with $E_2(\bk_i)$ (the same trend is also observed for $n^{\rm up}(\bk_i)$ vs $E_2(\bk_i)$), indicating that the region for tr$\CalG^{(\rm 2band)}$ to be relevant is well localized around the $\Gamma$ point. The results can be fitted with $R(\bk_i)\propto e^{-E_2(\bk_i)/\tau}$ [solid curves in Fig.~\ref{FigSupp:EnergyScaleKagome}(a)]. The value $\tau$ quantifies the energy scale, within which tr$\CalG^{(\rm 2band)}$ is relevant, and the contour of $E_2(\bk)=\tau$ defines the corresponding momentum-space region. In the inset of Fig.~\ref{FigSupp:EnergyScaleKagome}(a), colored dots (collapsed discrete points from 10 different TBCs) show the distribution of upper-band occupation $n^{\rm up}(\bk_i)$ in the BZ at $U=1$. Points with significant $n^{\rm up}(\bk_i)$ are well enclosed by the contour of $E_2(\bk)=\tau$.
Figs.~\ref{FigSupp:EnergyScaleKagome}(b, c) present variation of the energy scale $\tau$ with interaction strength $U$.
The contour of the largest $\tau$ value (from $U=5$) is shown as the red contour in the inset of Fig.~\ref{FigSupp:EnergyScaleKagome}(a). Notice that, in contrast to the two-band honeycomb model, in the kagome model here the two touching bands have non-zero tr$\CalG^{(\rm 2band)}$, which becomes large when moving away from the $\Gamma$ point (Fig.~\ref{FigSupp:TwoBandQuantumGeometry}). 
Fig.~\ref{FigSupp:EnergyScaleKagome}(d) shows the $\tau$ value obtained at $U=5$ as a function of band parameter $\alpha$ within the FQAH phase.

\subsection{Many-body results of the trivial phase}

\begin{figure}[h!]
	\centering
	\includegraphics[width=4.25in]{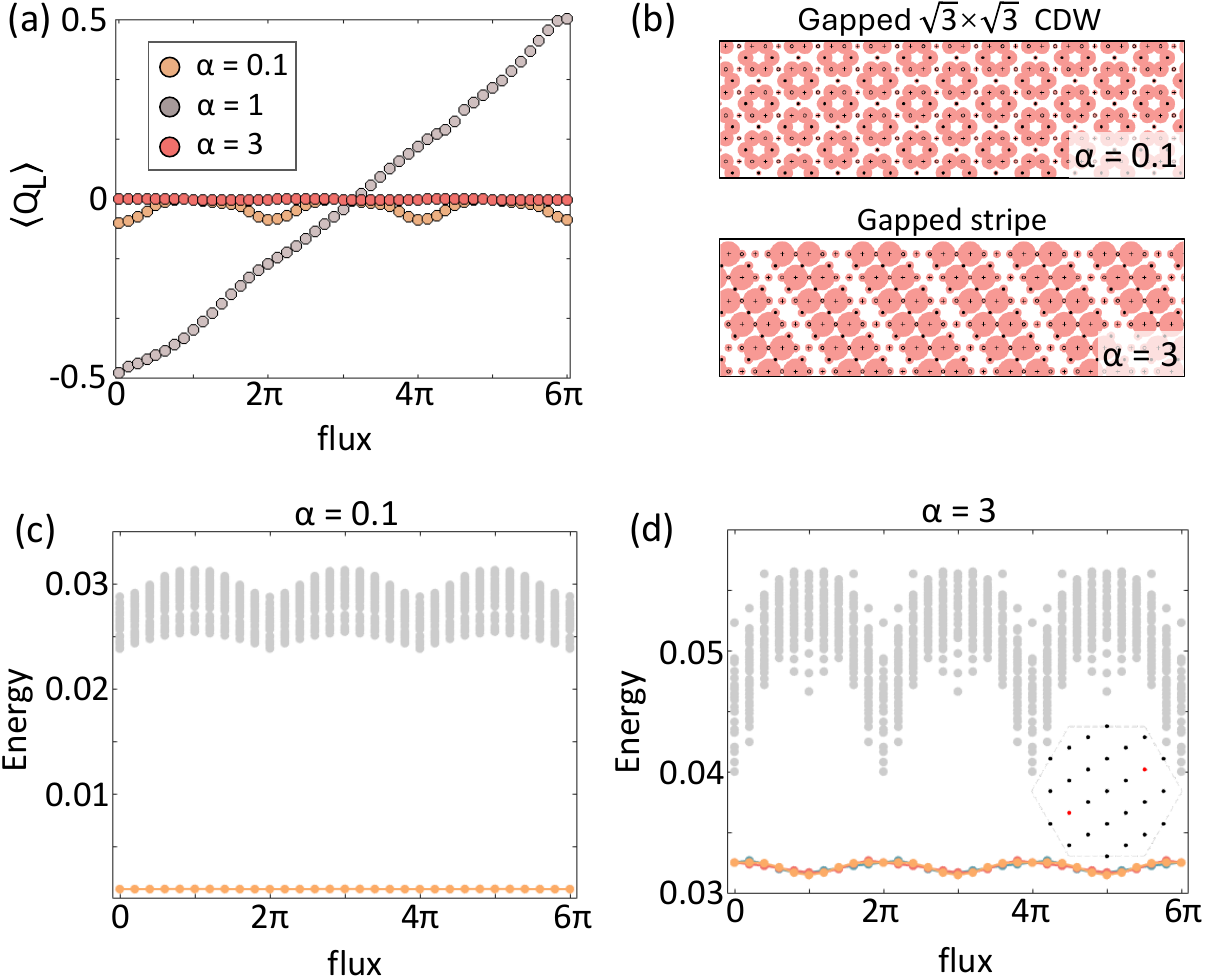}
	\caption{(a) Charge pumping simulation results for various $\alpha$ values. (b) Charge distribution pattern at $\alpha=0.1$ and $\alpha =3$. (c-d) ED spectral flow at $\alpha = 0.1$ and $\alpha = 3$, respectively. The $\bk$ grid in (c) corresponds to Fig.~\ref{FigSupp:kmesh24}(b) and the $\bk$ grid in (d) is displayed in the inset of (d). The interaction strength is fixed at $U=1$.}~\label{FigSupp:TrivialPhaseKagome}
\end{figure}

Fig.~\ref{FigSupp:TrivialPhaseKagome}(a) shows representative DMRG charge pumping simulations in the three different phases [cf. Fig.~5(f) in the main text]. Except for the FQAH phase at $\alpha=1$, the other two phases do not exhibit quantized charge transfer after $6\pi$ flux insertion, indicating that they are topologically trivial. The corresponding charge distributions of the trivial phases are visualized in Fig.~\ref{FigSupp:TrivialPhaseKagome}(b): for $\alpha = 0.1$, a $\sqrt 3\times \sqrt 3$ `donut' pattern is observed; while for $\alpha=3$, a stripe pattern emerges with a tripled periodicity along the horizontal direction. 
The DMRG results are supplemented by ED calculations with $\bk$ grids that are commensurate with the charge patterns. For $\alpha=0.1$, the tilted $\bk$ grid shown in Fig.~\ref{FigSupp:kmesh24}(b) is employed, which includes the corners of the BZ. For $\alpha=3$, the rectangular $\bk$ grid in the inset of Fig.~\ref{FigSupp:TrivialPhaseKagome}(d) is used, where the red points mark the wave vectors associated with the strip order in the triple-periodicity direction. The spectral flow in Figs.~\ref{FigSupp:TrivialPhaseKagome}(c) and (d) indicates that both trivial phases remain gapped.
We therefore conclude that the phases competing with the FQAH in the kagome model are gapped, topologically trivial CDW, characterized by a two-dimensional $\sqrt 3\times \sqrt 3$ donut pattern for $\alpha<0.35$ and a stripe pattern with tripled periodicity along the horizontal direction for $\alpha>2.46$.

\subsection{Effects of gap opening perturbation}

\begin{figure}[h!]
	\centering
	\includegraphics[width=5.5in]{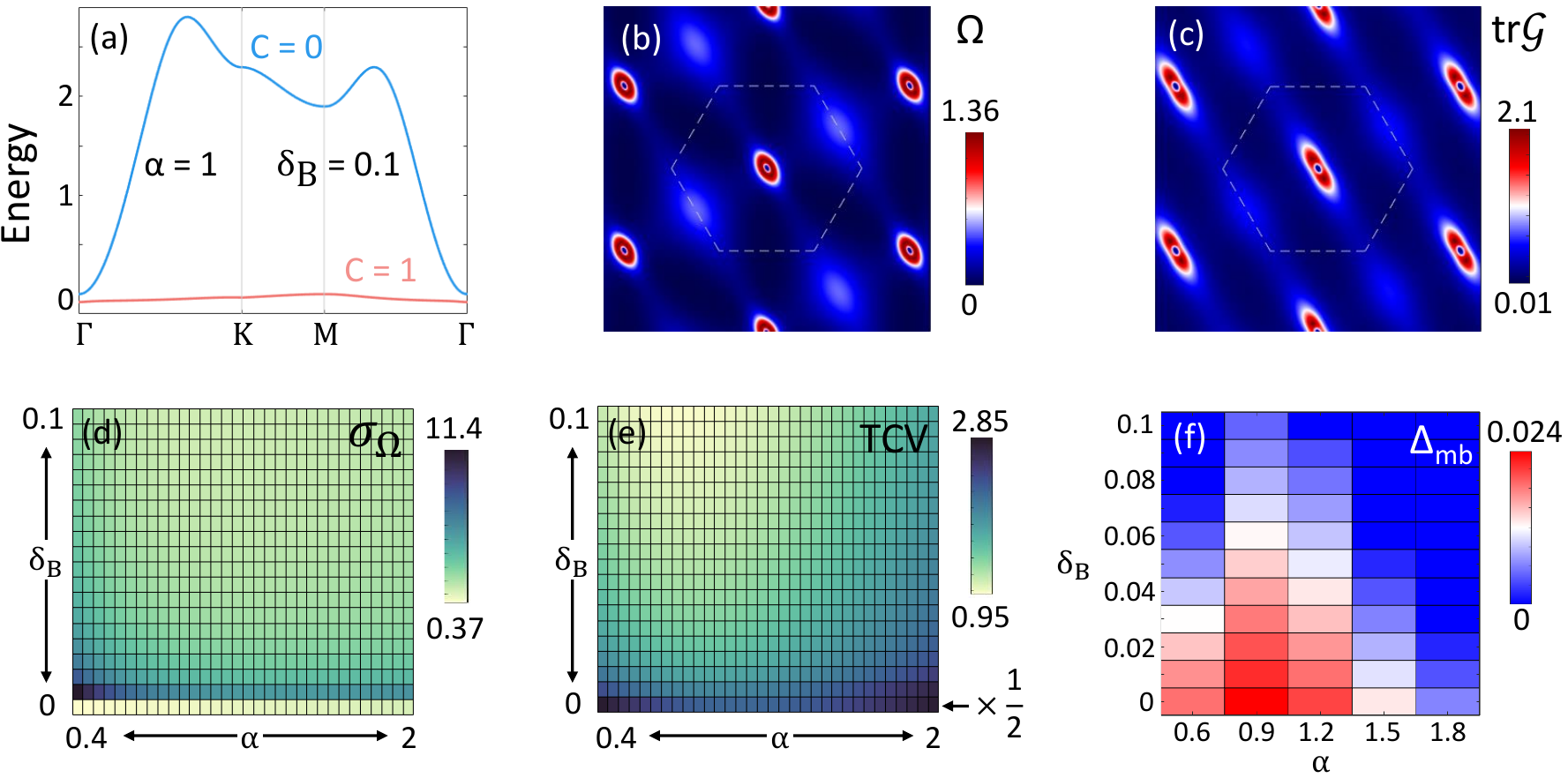}
	\caption{Effects of perturbation on quantum geometry and FQAH stability in the Kagome model: (a-c) Band structures and the quantum geometry of the lowest Chern band at $\alpha$=1 and $\delta_B=0.1$. (d-f) Analogous to (d-f) of the honeycomb model in Fig.~\ref{FigSupp:PerturbationHoney}. The lowest row of (e) is rescaled by $1/2$ for clarity. The interaction strength is  $U=1$.}~\label{FigSupp:PerturbationKagome}
\end{figure}

For the Kagome model, the dominant effective single-particle term  induced by interactions is an additional onsite energy term, $-\delta_B\hat C_{B_i}^\dagger\hat C_{B_i}$ with $\delta_B>0$.This term arises from the occupation preference on the B orbitals, which have a lower onsite energy in the original single-particle Hamiltonian. As shown in Figs.~\ref{FigSupp:PerturbationKagome}(a--c), after introducing the perturbation term, $-\delta_B\hat C_{B_i}^\dagger\hat C_{B_i}$, the SFB is gapped into an isolated Chern band with weak dispersion and strongly fluctuating quantum geometry. Figs.~\ref{FigSupp:PerturbationKagome}(d, e) characterize the idealness of the narrow Chern band: $\sigma_\Omega$ exhibits a sharp increase as $\delta_B$ is turned on, it then decreases as $\delta_B$ is further increased; while the TCV decreases monotonically with $\delta_B$. We performed two-band ED ($n_{\rm up}=2$) to evaluate the many-body gap in the $\delta_B$--$\alpha$ parameter space [Fig.~\ref{FigSupp:PerturbationKagome}(f)]. In close analogy to the honeycomb model, the FQAH states are more stable in the SFB and gapping out the band touching tends to destabilize them.

\end{document}